\documentclass[12pt,authoryear]{article}
\newcommand{\blind}{1}
\usepackage{amssymb,amsthm}
\usepackage{amsmath}
\usepackage{natbib}
\usepackage{bibunits}

\defaultbibliographystyle{plainnat}  % 或elsarticle-harv
\defaultbibliography{bibfile_main} % 默认文献库

\usepackage{graphics,enumitem,subcaption,physics}
\usepackage{tikz}
\usetikzlibrary{calc}
\usetikzlibrary{positioning} 
\usetikzlibrary{arrows.meta}
\usetikzlibrary{shadows}
\usepackage[linesnumbered,ruled,vlined]{algorithm2e} 
\usepackage{enumitem}
\usepackage{booktabs}
\usepackage{multirow}
\usepackage{setspace}
\newtheorem{theorem}{Theorem}[section]
\newtheorem{remark}{Remark}[section]
\newtheorem{definition}{Definition}[section]
\newtheorem{proposition}{Proposition}[section]
\newtheorem{assumption}{Assumption}[section]
\newtheorem{example}{Example}[section]

\usepackage{chngcntr}
\counterwithin{table}{section}
\counterwithin{figure}{section}
\numberwithin{equation}{section}
\usepackage[colorlinks=true, linkcolor=blue, urlcolor=blue, citecolor=blue]{hyperref}
\usepackage{xcolor}

\newcommand{\indep}{\perp\!\!\!\perp}
\newcommand{\notindep}{\not\perp\!\!\!\perp}
\allowdisplaybreaks[4]
\begin{document}		
	\if1\blind
	{
		\title{Identification and Debiased  Learning of Causal Effects with General Instrumental Variables}
		\author{Shuyuan Chen, Peng Zhang, Yifan Cui\thanks{The authors were partially supported by the National Key R\&D Program of China (2024YFA1015600), the National Natural Science Foundation of China (12471266 and U23A2064). Correspondence to \href{cuiyf@zju.edu.cn}{cuiyf@zju.edu.cn}.}}
		\date{Zhejiang University    \vspace{-1cm}} 
		\maketitle
	} \fi

	\if0\blind
	{
		\bigskip
		\bigskip
		\bigskip
		\begin{center}
			{\LARGE\bf Double Machine Learning of Continuous Treatment Effects with General Instrumental Variables}
		\end{center}
		\medskip
	} \fi

    \begin{abstract}
        Instrumental variable methods are fundamental to causal inference when treatment assignment is confounded by unobserved variables. In this article, we develop a general nonparametric causal framework for identification and learning with multi-categorical or continuous instrumental variables. 
		Specifically, the mean potential outcomes and the average treatment effect can be identified via a regular weighting function derived from the proposed framework.
		Leveraging semiparametric theory, we derive efficient influence functions and construct two consistent, asymptotically normal estimators via debiased machine learning. 
		The first estimator uses a prespecified weighting function, while the second estimator selects the optimal weighting function adaptively.
		Extensions to longitudinal data, dynamic treatment regimes, and multiplicative instrumental variables are further developed. We demonstrate the proposed method by employing simulation studies and analyzing real data from the Job Training Partnership Act program.
    \end{abstract}
	
	\begin{bibunit} 
	
	\section{Introduction}\label{sec: introduction}
	\subsection{Background}
	
	Observational studies are commonly employed to estimate treatment effects in biomedical and economic research \citep{hernan2020causal}. In the presence of unmeasured confounding, instrumental variable (IV) methods have been widely used to identify causal effects \citep{Robins1994SNMM,imbens1994identification,angrist1996identification,abadie2003semiparametric}. These approaches leverage exogenous variation in treatment induced by instruments that influence treatment assignment but are conditionally independent of latent confounding effects. 
	
	Under the monotonicity condition that the instrument does not decrease (or increase) the probability of receiving treatment for any individual, \citet{imbens1994identification,angrist1996identification} show that one can identify causal effects within the complier subgroup with a binary IV.
	Based on this work, \citet{heckman1999local,heckman2005structural,kennedy2019robust} propose estimators for the local IV effect curve, capturing the treatment effect among individuals who would comply when the instrument exceeds a certain threshold. Recently, \citet{tchetgen2024nudge} relaxes the monotonicity assumption, establishing the identification of the average treatment effect in the nudge subgroup, which involves mixtures of compliers and defiers.  
	
	Unlike approaches relying on the monotonicity condition, \citet{wang2018bounded, Hartwig2023} identify the average treatment effect (ATE) by imposing certain no additive interaction assumptions between the IV and the latent confounding in the treatment model; \citet{liu2025multiplicativeinstrumentalvariablemodel} propose an identifying condition that excludes any multiplicative interaction between IV and latent confounders in the treatment model, thereby enabling the identification of the ATE of the treated group. Building upon a no-interaction assumption, \citet{cui2023instrumental,wang2023instrumental,michael2024instrumental} propose marginal structural mean models and marginal structural Cox models, and \citet{fu2022offline,xu2023instrumental} propose IV approaches to confounded off-policy evaluation using a similar identification strategy. Furthermore, \citet{ye2023instrumented} propose an instrumental difference-in-difference method to identify the ATE based on no-interaction assumptions.
	In addition, \citet{cui2021semiparametric} propose to identify optimal treatment regimes under a no unmeasured common effect modifier assumption, and \citet{cui2025learning} extend the approach to learn optimal treatment regimes for censored survival data.

	In parallel, \citet{chernozhukov2005iv,chernozhukov2020instrumentalvariablequantileregression} develop IV quantile regression for estimating quantile treatment effects, accommodating non-binary instruments. 
    \cite{chen2018sequential} proposes sequential instrumental variables censored quantile regression estimators with a binary endogenous variable and a continuous instrument. More recently, \citet{chernozhukov2024estimatingcausaleffectsdiscrete} introduce a copula invariance condition to identify treatment effects for the entire population with a binary instrumental variable. 
	Collectively, these approaches, however, do not offer nonparametric identification strategies for the ATEs and mean potential outcomes in settings where the instrument is non-binary.
	
	Despite the nonparametric identification of ATEs or mean potential outcomes, several classical frameworks have been well developed to handle continuous IVs. The generalized method of moments provides a unifying and flexible parametric framework for estimation under IV settings \citep{hansen1982large}. It estimates structural parameters of interest by solving the corresponding moment conditions. It encompasses traditional methods like two-stage least squares  \citep{sargan1958estimation, wooldridge2010econometric} as a special case and extends naturally to situations with multiple instruments, heteroskedasticity, and other complexities.
	
	Nonparametric IV methods represent another important and widely adopted framework, particularly suitable for settings where the structural relationships between variables are complex and cannot be adequately captured by parametric models \citep{newey2003instrumental,darolles2011nonparametric,severini2012efficiency,AiChen2012,newey2013nonparametric,Chen_2018}. 
	By avoiding restrictive assumptions on functional forms, these methods enable flexible, data-driven estimation of treatment and outcome models. In particular, proximal causal inference methods are closely related to nonparametric IV techniques and have emerged as effective approaches to address unmeasured confounding when identifying ATEs \citep{miao2018identifying,tchetgen2024introduction,cui2024semiparametric}.

	\subsection{Contributions}
	% Inspired by existing literature, our study investigates the fundamental problem of estimating the mean potential outcomes \citep{neyman1923applications,rubin1974estimating} in settings involving discrete or continuous IVs and treatments, which broadens the scope beyond the conventional binary treatment-binary IV framework \citep{wang2018bounded}. 
	% Our method does not require more IV categories than treatments, thus enhancing its practical applicability.
	% Moreover, in certain degenerate cases, we demonstrate that the causal estimands identified by our framework are mathematically equivalent to those derived from two-stage least squares, bridging a connection with a classical econometric approach.

	We now outline the contents of the paper. 
	In Section~\ref{sec: identification}, we present the fundamental assumptions and introduce a novel concept termed regular weighting function, whose existence critically depends on the strength of the association between IV and treatment. Motivated by solving nonparametric IV problems, we further introduce a novel condition, termed additive IV, which provides a sufficient condition for the existence of nonparametric IV solutions. The additive IV condition can be viewed as a natural generalization of the no-interaction assumption. Moreover, we extend the assumption of the no unmeasured common effect modifier to the general IV setting, which also provides a sufficient condition for the existence of nonparametric IV solutions.

	In Section~\ref{sec: semiparametric theory}, we employ semiparametric theory \citep{vanderLaan2003, tsiatis2006semiparametric} to derive the efficient influence functions (EIFs) for target causal estimands defined by various regular weighting functions. For identifying the ATE, we characterize the optimal weighting function, which yields an estimator achieving the semiparametric efficiency bound under the setting of homoskedastic latent confounding. Building upon the debiased machine learning framework \citep{schick1986asymptotically, Chernozhukov2018}, we propose two novel estimators for the ATE.
	The first estimator uses a prespecified weighting function, while the second estimator does not require a prespecified weighting function and selects the optimal weighting function adaptively.

	In Section~\ref{sec: longitudinal scenario}, we extend the identification strategy from the point-exposure setting to longitudinal data. 
    We develop a cross-fitting procedure for estimation and establish that the proposed estimator is asymptotically normal, and its variance estimator is consistent.
    Further extensions to dynamic treatment regimes and multiplicative IVs are provided in the Supplementary Material, demonstrating the versatility of our framework for identification and estimation.
	% In Section~\ref{sec: simulation}, we perform simulation studies to assess the validity of the proposed estimators in both point-exposure and longitudinal settings.
	% We analyze data from the Job Training Partnership Act program in Section~\ref{sec: empirical illustration} to illustrate the practical implementation of our proposed method.
	% We conclude our paper with a discussion in Section~\ref{sec:discussion}.

	Our contributions are fourfold.
	First, we introduce a novel concept termed regular weighting function, which provides a new perspective on the IV relevance condition.
	Second, we propose a novel identification strategy in the general IV setting by solving a special class of nonparametric IV problems, and we establish sufficient conditions for the existence of such solutions. Third, we derive EIFs and construct two consistent, asymptotically normal estimators via debiased machine learning. We note that the second estimator is non-standard, which does not require a prespecified weighting function and selects the optimal weighting function adaptively.
    Fourth, we extend our identification framework to accommodate more complex settings, including longitudinal data structures, dynamic treatment regimes, and multiplicative IV models.
	Simulation and empirical studies are conducted to demonstrate the validity and applicability of the proposed methods.

	\section{Identification}\label{sec: identification}
	\subsection{Preliminaries}	
	Let \( A \in \mathcal{A} := \{0, \ldots, M\} \) denote a multi-categorical treatment variable, where \( M = 1 \) corresponds to the binary treatment setting. 
	Let \( Z \in \mathcal{Z}\subseteq\mathbb{R}^{|\mathcal{Z}|} \) denote the IV, which may be multi-categorical or continuous. Let \( U\in\mathcal{U} \subseteq\mathbb{R}^{|\mathcal{U}|}\) represent unmeasured confounders, \( L\in\mathcal{L} \subseteq\mathbb{R}^{|\mathcal{L}|}\) observable confounders, \( Y\in\mathcal{Y} \subseteq\mathbb{R}\) the observed outcome, and \( Y(a) \) the potential outcome at the treatment level \( A = a \).   
	The observed data consist of \( O = \{Z, A, Y, L\} \in \mathcal{Z}\times \mathcal{A}\times \mathcal{Y}\times\mathcal{L}\).
	We introduce four fundamental assumptions in the IV setting.
	\begin{assumption}[Consistency]\label{as: consistency}
		$Y=Y(A)$.
	\end{assumption}
	
	\begin{assumption}[Latent ignorability]\label{as: latent ignorability}
		For any $a\in\mathcal{A}$, $Y(a)\indep \{A,Z\}\mid U,L$.
	\end{assumption}
	
	\begin{assumption}[IV independence]\label{as: IV independence}
		$Z\indep U\mid L$.
	\end{assumption}
	
	\begin{assumption}[IV relevance]\label{as: IV relevance}
		For any $a\in\mathcal{A}$ and $l\in\mathcal{L}$, $Z\notindep I\{A=a\}\mid L=l$. 
		That is, there exist two distinct values $z_0,z_1\in\mathcal{Z}$ such that
		$\Pr(A=a\mid Z=z_0,L=l)\neq \Pr(A=a\mid Z=z_1,L=l).$
	\end{assumption}
	
	Assumption~\ref{as: latent ignorability} posits that, conditional on both the observed covariates and unmeasured confounders, the potential outcome \( Y(a) \) is independent of the treatment and IV. 
	Assumption~\ref{as: IV independence} states that \( Z \) is independent of the unmeasured confounders \( U \) given the observed covariates \( L \).
	Assumption~\ref{as: IV relevance} requires that \( Z \) has a nontrivial effect on the treatment \( A \), conditional on any level of \( L \). This condition is slightly stronger than \( A \notindep Z \mid L \), which only requires the existence of some \( l \in \mathcal{L} \), \( a \in \mathcal{A} \), and $z_0, z_1 \in \mathcal{Z}$ such that $\Pr(A=a\mid Z=z_0,L=l)\neq \Pr(A=a\mid Z=z_1,L=l).$
	
	\subsection{Regular weighting function}
	Next, we introduce a novel concept of a regular weighting function, which is closely related to the IV relevance and the positivity condition.
	\begin{definition}[Regular weighting function]\label{defn: regular weighting function}
		For each $a\in\mathcal{A}$, a function $\pi(Z,L)$ is a \emph{regular weighting function} (RWF) for $A = a$ if it is uniformly bounded and there exists a positive constant $\epsilon_0$ such that
		$|\mathrm{Cov}\!\{I\{A=a\}, \pi(Z,L) \mid L\} | \geq \epsilon_0$ uniformly for all $L$.
	\end{definition}
	
	The existence of an RWF for every \( a \in \mathcal{A} \) implies the IV relevance condition in Assumption~\ref{as: IV relevance}. Specifically, requiring the absolute value of the conditional covariance to be uniformly bounded below by a positive constant $\epsilon_0$ rules out scenarios where $\pi(Z,L)$ is irrelevant to $I\{A=a\}$, which would undermine the stability and validity of the identification of causal effects.
	To guarantee the existence of an RWF, the following strong IV relevance assumption is required.
	
	\begin{assumption}[Strong IV relevance]\label{as: strong IV relevance}
		There exists a positive constant $\epsilon_0$ such that for any $l\in\mathcal{L}$ and $a\in\mathcal{A}$,
		$\mathrm{Var}\!\{\Pr(A=a\mid Z,L) \mid L=l\} \geq \epsilon_0.$
	\end{assumption}
	
	\begin{remark}
		Here we directly impose a stronger version of the IV relevance condition in Assumption~\ref{as: strong IV relevance}, which is needed to derive the asymptotic properties of our proposed estimator. However, for identification or semiparametric analysis, only a weaker form is required, namely that for all \(a \in \mathcal{A}\), 
		$\mathrm{Var}\{\Pr(A=a \mid Z,L) \mid L=l\} \neq 0$.
		A similar distinction also arises in our definition of RWFs.
	\end{remark}

	Assumption~\ref{as: strong IV relevance} implies that the IV has a non-negligible effect on $A$, entailing the IV relevance condition in Assumption~\ref{as: IV relevance}. In fact, $\mathrm{Var}\!\{\Pr(A=a\mid Z,L) \mid L\}$ equals
	\begin{align*}
		\Pr(A=a\mid L)\times
		\left\{\mathbb{E}[\Pr(A=a\mid Z,L)|A=a,L]-\mathbb{E}[\Pr(A=a\mid Z,L)|L]\right\}.
	\end{align*}
	Thus, Assumption~\ref{as: strong IV relevance} also guarantees that for any $l \in \mathcal{L}$, $\Pr(A=a \mid L=l) \neq 0$, corresponding to the classical positivity condition \citep{hernan2020causal}.
	The following proposition further emphasizes the role of Assumption~\ref{as: strong IV relevance} when identifying an RWF for $A=a$.
	
	\begin{proposition}[Existence of RWFs]\label{prop: existence of RWFs}
		There exists an RWF $\pi(Z,L)$ for $A$ if and only if Assumption~\ref{as: strong IV relevance} holds. If there exists an RWF $\pi(Z,L)$ for $A=a$, then $\Pr(A=a\mid Z,L)$ must be an RWF for $A=a$. 
	\end{proposition}
	
	In particular, if there exists an RWF $\pi(Z,L)$ for $A=a$, then there actually exists a whole family of RWFs, which can be generated by multiplying $\pi(Z,L)$ by any function $f(L)$ that is uniformly bounded above and away from zero.
	In fact, according to Proposition~\ref{prop: existence of RWFs}, Assumption~\ref{as: strong IV relevance} implies that the function $\Pr(A=a \mid Z,L)$ itself is an RWF, thereby guaranteeing the existence of at least one RWF.

	% \subsection{Solving nonparametric IV models}
	\subsection{Identification of the mean potential outcome}
	In this subsection, we propose a strategy to identify the potential outcome mean \(\mathbb{E}[Y(a)]\) by formulating and solving a class of nonparametric IV problems. Throughout, we define \( A^{(a)} := I\{A = a\} \) for each \( a \in \mathcal{A} \) for convenience of notation. Our first primary goal is, for each \( a \in \mathcal{A} \), to identify a function \( f_a^o(A^{(a)}, L) \) that satisfies the conditional moment restriction
	\begin{equation}\label{eq: npiv}
		\mathbb{E}[A^{(a)} Y \mid Z, L] = \mathbb{E}[f_a^o(A^{(a)}, L) \mid Z, L].
	\end{equation}
	This conditional moment equation is common in the nonparametric IV literature \citep{newey2003instrumental}.  
	The following theorem establishes the uniqueness of the solution to Equation~\eqref{eq: npiv} and provides an explicit representation in terms of any RWF \(\pi(Z, L)\).
	
	\begin{theorem}[Uniqueness and closed form solution]\label{thm: uniqueness}
		Under Assumption~\ref{as: IV relevance}, for each $a\in\mathcal{A}$, if there is a solution \( f_a^o(A^{(a)}, L) \) to Equation~\eqref{eq: npiv}, it is unique. Moreover, for any RWF \( \pi(Z, L) \) for $A=a$, the solution \( f_a^o(A^{(a)}, L) \) satisfies
		\begin{equation}\label{eq: explicit form}
			\begin{aligned}
				f_a^o(0,L) &= -\dfrac{\mathrm{Cov}\!\{A^{(a)}Y, \pi(Z,L) \mid L\}}{\mathrm{Cov}\!\{A^{(a)}, \pi(Z,L) \mid L\}} \mathbb{E}[A^{(a)}\mid Z,L] + \mathbb{E}[A^{(a)}Y\mid Z,L],\\
				f_a^o(1,L) &= \dfrac{\mathrm{Cov}\!\{A^{(a)}Y, \pi(Z,L) \mid L\}}{\mathrm{Cov}\!\{A^{(a)}, \pi(Z,L) \mid L\}} \{1-\mathbb{E}[A^{(a)}\mid Z,L]\} + \mathbb{E}[A^{(a)}Y\mid Z,L].
			\end{aligned}
		\end{equation}
	\end{theorem}
	
	However, Theorem~\ref{thm: uniqueness} only ensures uniqueness without guaranteeing the existence of a solution to Equation~\eqref{eq: npiv}. Indeed, Equation~\eqref{eq: npiv} is over-identified when $Z$ is non-binary, meaning that a solution may not exist in general. 
	In the literature on proximal causal inference, the existence of a solution to the bridge equation is typically guaranteed under completeness conditions \citep{tchetgen2024introduction}. 
	Likewise, within the IV framework, the additive IV condition plays a central role in ensuring the existence of a solution to the nonparametric IV problem in Equation~\eqref{eq: npiv}.

	% Next, we introduce the concept of an additive IV, which serves as a key condition for the existence of solutions to nonparametric IV problems and for identifying the mean potential outcomes.
	
	\begin{definition}[Additive IV]\label{defn: additive IV}
		For each $a \in \mathcal{A}$, $Z$ is an \emph{additive IV}  for $A = a$ if there exist functions $b(U,L)$ and $c(Z,L)$ such that
		$\Pr(A = a \mid Z, U, L) = b(U, L) + c(Z, L).$
		Furthermore, $Z$ is an additive IV for $A$ if it is an additive IV for $A=a$ for all $a\in\mathcal{A}$.
	\end{definition}
	
	According to \citet{wang2018bounded}, the definition of additive IV is motivated by the no-interaction condition between $Z$ and $U$ in the treatment model.
	\citet{tchetgen2021genius,sun2022selective,ye2024geniusmawiirobustmendelianrandomization} also adopt analogous no-interaction conditions to identify the ATE with an invalid IV.
	The following proposition gives an alternative definition for additive IV.
	\begin{proposition}\label{prop: AIV equivalent form}
		For each $a \in \mathcal{A}$, $Z$ is an additive IV for $A = a$ if and only if for any $\pi(Z,L)$,
		$\mathrm{Cov}\{I\{A=a\},\pi(Z,L)\mid U,L\}=\mathrm{Cov}\{I\{A=a\},\pi(Z,L)\mid L\}.$
	\end{proposition}
	
	Specifically, when \( Z \) is binary, the condition in Definition~\ref{defn: additive IV} holds if and only if
	$$\Pr(A=a \mid Z=1, U, L)- \Pr(A=a \mid Z=0, U, L) \indep U \mid L,$$
	implying that the differential effect of the instrument on treatment $A$ is conditionally independent of the unmeasured confounders \( U \), given the observed covariates \( L \).
	Next, we establish the identification strategy of the mean potential outcome under an additive IV.

	\begin{theorem}[Identification of the mean potential outcome]\label{thm: AIV identification}
		Under Assumptions~\ref{as: consistency}--\ref{as: IV relevance}, for each \( a\in\mathcal{A} \), if $Z$ is an additive IV for $A=a$, then there exists a unique solution \( f_a^o(A^{(a)}, L) \) to Equation~\eqref{eq: npiv}, given by
		\begin{equation*}
			\begin{aligned}
				f_a^o(0,L) &= \mathrm{Cov}\!\{\mathbb{E}[Y(a)\mid U,L], \Pr(A=a\mid Z,U,L)\mid L\},\\
				f_a^o(1,L) &= \mathrm{Cov}\!\{\mathbb{E}[Y(a)\mid U,L], \Pr(A=a\mid Z,U,L)\mid L\} + \mathbb{E}[Y(a)\mid L].
			\end{aligned}
		\end{equation*}
		Furthermore, if $\pi(Z,L)$ is an RWF for $A=a$,
		% the mean potential outcome can be identified as
		\begin{align}\label{eq: identification AIV}
			\mathbb{E}[Y(a)] = \mathbb{E}\left[\dfrac{\mathrm{Cov}\!\{A^{(a)}Y, \pi(Z,L) \mid L\}}{\mathrm{Cov}\!\{A^{(a)}, \pi(Z,L) \mid L\}}\right].
		\end{align}
	\end{theorem}
	
	Notably, our identification strategy holds even when the treatment space $\mathcal{A}$ is multi-categorical. Intuitively, this is because we transform the multi-categorical treatment $A$ into a binary variable $A^{(a)}$ for each $a \in \mathcal{A}$.
	Moreover, if there are no latent confounders $U$, then \( f_a^o(0, L) \equiv 0 \) by Theorem~\ref{thm: AIV identification}. In other words, a significant deviation of \( f_a^o(0, L) \) from zero indicates the existence of latent confounding. This observation provides a novel criterion for detecting unmeasured confounder $U$, which is beyond the scope of our article. 
	
	Furthermore, when \( Z \) is a non-binary IV, Theorem~\ref{thm: AIV identification} provides a practical approach to assess whether \( Z \) qualifies as an additive IV for \( A = a \), which constitutes a relatively strong condition. Specifically, under Assumptions~\ref{as: consistency}--\ref{as: IV independence}, if \( Z \) is indeed an additive IV for \( A = a \), the choice of the RWF \( \pi(Z, L) \) does not affect the expression value on the right-hand side of Equation~\eqref{eq: identification AIV}.  
	Therefore, two distinct RWFs \( \pi_1(Z, L) \) and \( \pi_2(Z, L) \) can be selected and the corresponding values can be compared. A discrepancy between these two values would indicate a violation of the additive IV condition.

	\subsection{Identification of the average treatment effect}
	In practical applications with binary treatment \( A \), researchers are also interested 
	in identifying the ATE. The following theorem summarizes the identification results for the ATE under a weaker assumption than additive IV.
	\begin{theorem}[Identification of the ATE]\label{thm: Y(1)-Y(0)}
		Under Assumptions~\ref{as: consistency}--\ref{as: IV relevance}, and 
		\begin{align}\label{eq: no unmeasured common effect modifier}
			\mathrm{Cov}\!\{Y(1)-Y(0),\mathbb{E}[A\mid Z=z,U,L]-\mathbb{E}[A\mid U,L]\mid L\}=0,
			\text{ for any }z\in\mathcal{Z},
		\end{align}
		the nonparametric IV problem
		\begin{equation}\label{eq: npiv Y(1)-Y(0)}
			\mathbb{E}[Y\mid Z,L]=\mathbb{E}[f^o(A,L)\mid Z,L]
		\end{equation}
		has a unique solution \( f^o(A, L) \) as
		\begin{align*}
			f^o(0,L) &= \mathrm{Cov}\!\{\mathbb{E}[Y(1)-Y(0)\mid U,L], \Pr(A=1\mid U,L)\mid L\} + \mathbb{E}[Y(0)\mid L],\\
			f^o(1,L) &= \mathrm{Cov}\!\{\mathbb{E}[Y(1)-Y(0)\mid U,L], \Pr(A=1\mid U,L)\mid L\} + \mathbb{E}[Y(1)\mid L].
		\end{align*}
		Furthermore, for any RWF \( \pi(Z, L) \) for $A$, the ATE is identified as
		\begin{align}\label{eq: identification AIV Y(1)-Y(0)}
			\mathbb{E}[Y(1)-Y(0)] = \psi_{\pi}^o := \mathbb{E}\left[\dfrac{\mathrm{Cov}\!\{Y, \pi(Z,L) \mid L\}}{\mathrm{Cov}\!\{A, \pi(Z,L) \mid L\}}\right].
		\end{align}
	\end{theorem}
	
	\begin{remark}
		In fact, under Assumptions~\ref{as: consistency}--\ref{as: IV independence}, if $\pi(Z,L)$ is an RWF for $A$,
		\begin{align*}
			\psi_{\pi}^o= \mathbb{E}\left[\mathbb{E}[Y(1)-Y(0)\mid U,L]\dfrac{\mathrm{Cov}\!\{A, \pi(Z,L) \mid U,L\}}{\mathrm{Cov}\!\{A, \pi(Z,L) \mid L\}}\right].
		\end{align*}
		This result suggests that, even when $Z$ does not satisfy Equation~\eqref{eq: no unmeasured common effect modifier}, the estimand $\psi_{\pi}^o$ can still be interpreted as a weighted average of the conditional ATE $\mathbb{E}[Y(1)-Y(0)\mid U,L]$, as long as $\mathrm{Cov}\!\{A, \pi(Z,L) \mid U,L\}$ and $\mathrm{Cov}\!\{A, \pi(Z,L) \mid L\}$ have the same sign. 
		This representation is analogous to the assumption-lean inference of \citet{vansteelandt2022assumption,vansteelandt2024assumption}, in which the estimands remain meaningful and interpretable even under model misspecification.
	\end{remark}

	\begin{remark}
		For a fixed RWF $\pi(Z,L)$, Equation~\eqref{eq: identification AIV Y(1)-Y(0)} holds when 
		\begin{align}\label{eq: no unmeasured common effect modifier for pi}
			\mathrm{Cov}\!\{Y(1)-Y(0),\,\mathrm{Cov}\!\{A, \pi(Z,L) \mid U,L\}\mid L\}=0.
		\end{align}
		One can check that Equation~\eqref{eq: no unmeasured common effect modifier} holds 
		if and only if Equation~\eqref{eq: no unmeasured common effect modifier for pi} is satisfied for all $\pi(Z,L)$.
		%	Both Equation~\eqref{eq: no unmeasured common effect modifier} and Equation~\eqref{eq: no unmeasured common effect modifier for pi} satisfied for all $\pi(Z,L)$ 
		Both of them can be viewed as extensions of the ``no unmeasured common effect modifier'' assumption proposed by \citet{cui2021semiparametric}. 
	\end{remark}

	\begin{remark}
		When \( L = \emptyset \) and $Z$ are univariate, the parameter $\psi_{\pi}^o$ in Equation~\eqref{eq: identification AIV Y(1)-Y(0)} reduces to a form resembling the two-stage least squares estimator if we set \(\pi(Z,L) = Z\). Moreover, several classical works interpret $\psi_{\pi}^o$ as $\tau_0$, the solution to the conditional moment equation
		\(\mathbb{E}[Y - \tau_0 A \mid Z] = 0\)
		\citep{hansen1982large,newey1994large}, where \(\tau_0\) represents the effect of a marginal change in the endogenous variable \( A \) on the outcome. Consequently, our identification strategy can be viewed as a nonparametric generalization of the two-stage least squares and generalized method of moments, which additionally accounts for confounding effects through \( L \).    
	\end{remark}
	
	\begin{remark}
		For identifying the average dose-response curve of a continuous treatment, we provide a similar identification result in the Supplementary Material. 
	\end{remark}

	\section{Semiparametric theory for learning the average treatment effect}\label{sec: semiparametric theory}
	\subsection{Prespecified weighting function scenario}
	Recall that under Assumptions~\ref{as: consistency}--\ref{as: IV independence}, if Equation~\eqref{eq: no unmeasured common effect modifier} holds, the choice of the RWF \( \pi(Z, L) \) does not affect the value of \( \psi_{\pi}^o \) in Equation~\eqref{eq: identification AIV Y(1)-Y(0)}, as shown in Theorem~\ref{thm: Y(1)-Y(0)}. 
	However, the semiparametric efficiency bound of $\psi_{\pi}^o$ still depends on the choice of \( \pi(Z, L) \). Consequently, it is necessary to derive the EIFs corresponding to all possible RWFs \( \pi(Z, L) \).
	To this end, we define the following nuisance functions:
	\begin{align*}
		&\delta^o(L) := \mathbb{E}[A \mid L], 
		&& \eta^o(L) := \mathbb{E}[Y \mid L],\\
		&\kappa_{\pi}^o(L) := \mathrm{Cov}\!\{A, \pi(Z,L) \mid L\}, 
		&& \zeta_{\pi}^o(L) := \mathbb{E}[Y \pi(Z,L) \mid L],\\
		&\rho_{\pi}^o(L) := \mathbb{E}[\pi(Z,L) \mid L], 
		&& \gamma_{\pi}^o(L) := \dfrac{\mathrm{Cov}\!\{Y, \pi(Z,L) \mid L\}}{\mathrm{Cov}\!\{A, \pi(Z,L) \mid L\}}.
	\end{align*}
	For notational convenience, we collect them into a unified nuisance vector:
	\begin{align}\label{eq: nuisance function fixed pi}
		\alpha_{\pi}^o(L) := [\delta^o(L), \kappa_{\pi}^o(L), \rho_{\pi}^o(L), \eta^o(L), \gamma_{\pi}^o(L)].
	\end{align}	
	Note that \(\zeta_{\pi}^o(L)\) serves only as an intermediate nuisance function, which will be used in our proof and does not appear in the unified vector \(\alpha_{\pi}^o(L)\).
	Next, we derive the EIF of \(\psi_{\pi}^o\) for any choice of RWF \(\pi(Z,L)\).
	\begin{theorem}\label{thm: EIF AIV fixed pi}
		Under Assumptions~\ref{as: consistency}--\ref{as: IV independence}, 
		for any RWF $\pi(Z,L)$ for $A$, the EIF for $\psi_{\pi}^o$ in Equation~\eqref{eq: identification AIV Y(1)-Y(0)} is $\varphi_{\pi}(O;\psi_{\pi}^o,\alpha_{\pi}^o)$, where
		$\varphi_{\pi}(O;\psi_{\pi},\alpha_{\pi})$ equals
		\begin{align*}
			\dfrac{ \left\{\pi(Z,L)-\rho_{\pi}(L)\right\}}{\kappa_{\pi}(L)} \{Y-\eta(L)\} - \psi_{\pi} 
			+ \left(1 - \dfrac{\{\pi(Z,L)-\rho_{\pi}(L)\}\{A-\delta(L)\}}{\kappa_{\pi}(L)} \right) \gamma_{\pi}(L).
		\end{align*}
	\end{theorem}
	\begin{remark}
		We carry out our semiparametric analysis in the fully nonparametric model,
		%	. In this case, the tangent space coincides with $L_2(O):=\{f:\mathcal{O}\rightarrow R, \mathbb{E}[f(O)^2]^{1/2}<\infty\}$, 
		and the unique influence function corresponds to the EIF.
	\end{remark}
	This theorem underpins the construction of efficient estimators for \( \psi_{\pi}^o \). In practice, the true nuisance vector \( \alpha_{\pi}^o \) is unknown and must be estimated from the data.
	The following proposition quantifies the bias introduced by substituting an estimated \( \alpha_{\pi} \) for the true vector. This mixed bias property is well-documented in the existing literature \citep{rotnitzky2021characterization}.
	\begin{proposition}[Mixed bias property]\label{prop: mixed bias property fixed pi}
		Under Assumptions~\ref{as: consistency}--\ref{as: IV independence}, 
		for any RWF $\pi(Z,L)$ and any fixed nuisance function $\alpha_{\pi}$, $\varphi_{\pi}(O;\psi_{\pi}^o,\alpha_{\pi})$ satisfies that 
		\begin{align*}
			&\mathbb{E}[\varphi_{\pi}(O;\psi_{\pi}^o,\alpha_{\pi})]
			=\mathbb{E}\left[\dfrac{1}{\kappa_{\pi}(L)}\left\{\begin{array}{l}
				\{\kappa_{\pi}(L)-\kappa_{\pi}^o(L)\}\{\gamma_{\pi}(L)-\gamma_{\pi}^o(L)\}\\
				+\{\rho_{\pi}(L)-\rho_{\pi}^o(L)\}\{\eta(L)-\eta^o(L)\}\\
				+\{\rho_{\pi}(L)-\rho_{\pi}^o(L)\}\{\delta(L)-\delta^o(L)\}\gamma_{\pi}(L)
			\end{array}\right\}\right].
		\end{align*}
	\end{proposition}
	
	Next, we aim to select a function \( \pi(Z, L) \) from the class of RWFs that minimizes the asymptotic variance of the estimator for \( \psi_{\pi}^o \). Specifically, we seek the optimal \( \pi(Z, L) \) that minimizes the second moment of the EIF. 
	Minimizing this quantity yields the most statistically efficient estimator among all estimators based on different RWFs.
	Intuitively, Proposition~\ref{prop: existence of RWFs} suggests that 
	\(\pi_*^o(Z,L) := \Pr(A = 1 \mid Z, L)\) is a natural candidate for the optimal weighting function. 
	The following proposition characterizes the \(\pi(Z,L)\) that achieves this minimum variance bound.

	\begin{proposition}[Optimal RWF]\label{prop: lower efficiency bound}
		Under the conditions of Theorem~\ref{thm: EIF AIV fixed pi}, suppose that the solution $f^o(A,L)$ to the nonparametric IV problem in Equation~\eqref{eq: npiv Y(1)-Y(0)} satisfies
		\begin{equation}\label{eq: homoskedastic}
			\mathbb{E}\left[\{Y-f^o(A,L)\}^2\middle| Z,L\right]\indep Z\mid L.
		\end{equation}
		Then the quantity $\mathbb{E}[\varphi_{\pi}(O;\psi_{\pi}^o,\alpha_{\pi}^o)^2]$ reaches its lower bound when $\pi(Z,L)=\Pr(A=1\mid Z,L)$.
	\end{proposition}
	The condition in Equation~\eqref{eq: homoskedastic} implies that, after conditioning on $L$, the instrument $Z$ does not provide additional information about the residual variation in $Y$. \citet{wiemann2023optimal} leverage a similar assumption to derive the ``optimal instruments'' in the special case where $L=\emptyset$.
	
	\subsection{Adaptive weighting function scenario}
	As established in Proposition~\ref{prop: lower efficiency bound}, choosing  
	\(\pi_*^o(Z,L) := \Pr(A = 1 \mid Z, L)\) achieves the optimal efficiency bound for estimating \(\psi_{\pi}^o\). Moreover, Proposition~\ref{prop: existence of RWFs} implies that if \(\pi_*^o(Z,L)\) does not constitute a valid regular weighting function for \(A\), then no valid alternative exists. 
	These two observations motivate our strategy of adaptively estimating the optimal weighting function.  
	Consequently, we take \(\pi_*^o(Z,L)\) as the regular weighting function for \(A\). Since \(\pi_*^o(Z,L)\) is unknown in practice, we treat it as a nuisance function and estimate it in a data-adaptive manner.
	Concretely, we define the following nuisance functions
	\begin{align*}
		&\delta^o(L):=\mathbb{E}[A\mid L], 
		&&\eta^o(L):=\mathbb{E}[Y\mid L],\\
		&\kappa^o(L):=\mathrm{Cov}\!\{A,\pi_*^o(Z,L)\mid L\}, 
		&&\zeta^o(L):=\mathbb{E}[Y\pi_*^o(Z,L)\mid L],\\
		&\xi^o(Z,L):=\mathbb{E}[Y\mid Z,L], 
		&&\gamma^o(L):=\dfrac{\mathrm{Cov}\!\{Y, \pi_*^o(Z,L) \mid L\}}{\mathrm{Cov}\!\{A, \pi_*^o(Z,L) \mid L\}}.
	\end{align*}
	We unify these nuisance functions into a nuisance vector:
	\begin{equation}\label{eq: nuisance function}
		\beta^o(Z,L):=[\pi_*^o(Z,L),\delta^o(L),\kappa^o(L),
		\xi^o(Z,L),\eta^o(L),\gamma^o(L)].
	\end{equation}
	Note that $\zeta^o(L)$ is only an intermediate nuisance function and is not included in $\beta^o(Z,L)$.
	According to Theorem~\ref{thm: Y(1)-Y(0)}, the ATE can be identified as
	\begin{equation}\label{eq: identification AIV Y(1)-Y(0) unknown weight}
		\psi_{*}^o:=\mathbb{E}\left[\gamma^o(L)\right]
		=\mathbb{E}\left[\dfrac{\pi_*^o(Z,L)-\delta^o(L)}{\kappa^o(L)}Y\right].
	\end{equation}
	We now proceed to derive the EIF for $\psi_{*}^o$.
	\begin{theorem}\label{thm: EIF AIV}
		Under Assumptions~\ref{as: consistency}-\ref{as: IV independence} and \ref{as: strong IV relevance}, 
		the EIF for $\psi_{*}^o$ is given by $\varphi(O;\psi_{*}^o,\beta^o)$, where
		\begin{align*}
			\varphi(O;\psi_{*},\beta):=&\dfrac{\pi_*(Z,L)-\delta(L)}{\kappa(L)}Y-\psi_{*}
			+\dfrac{\gamma(L)}{\kappa(L)}\left\{
			\kappa(L) + (A-\pi_*(Z,L))^2 - (A-\delta(L))^2
			\right\}\\
			&+\dfrac{1}{\kappa(L)}
			\left\{\xi(Z,L)(A-\pi_*(Z,L))-\eta(L)(A-\delta(L))\right\}.
		\end{align*}
	\end{theorem}
	\begin{remark}
		Similarly to Theorem~\ref{thm: EIF AIV fixed pi},
		this theorem still holds even if $Z$ does not satisfy Equation~\eqref{eq: no unmeasured common effect modifier} for $A$, because the definition of $\psi_{*}^o$ does not rely on the additive IV condition.
	\end{remark}
	Next, the EIF characterization in Theorem~\ref{thm: EIF AIV} forms the foundation for analyzing the robustness of the proposed estimator, which is demonstrated in the next proposition.
	
	\begin{proposition}[Mixed bias property]\label{prop: mixed bias property}
		Under Assumptions~\ref{as: consistency}-\ref{as: IV independence} and \ref{as: strong IV relevance}, 
		for any fixed nuisance vector $\beta$ in Equation~\eqref{eq: nuisance function}, $\varphi(O;\psi_{*}^o,\beta)$ satisfies
		\begin{align*}
			\mathbb{E}[\varphi(O;\psi_{*}^o,\beta)]=
			\mathbb{E}\left[\dfrac{1}{\kappa(L)}\left\{\begin{array}{l}
				\{\gamma(L)-\gamma^o(L)\}\{\kappa(L) - \kappa^o(L)\}\\
				- \gamma(L)(\delta^o(L)-\delta(L))^2\\
				+\gamma(L)(\pi_*^o(Z,L)-\pi_*(Z,L))^2\\
				-(\xi(Z,L)-\xi^o(Z,L))(\pi_*(Z,L)-\pi_*^o(Z,L))\\
				+(\eta(L)-\eta^o(L))(\delta(L)-\delta^o(L))
			\end{array}\right\}\right].
		\end{align*}
	\end{proposition}
	
	\subsection{Cross-fitting procedure}
	In this subsection, we adopt the cross-fitting procedure \citep{Chernozhukov2018} to construct debiased estimators for $\psi_{\pi}^o$ and $\psi_{*}^o$ in Equations~\eqref{eq: identification AIV Y(1)-Y(0)} and \eqref{eq: identification AIV Y(1)-Y(0) unknown weight}. Without loss of generality, assume that the sample size \(n\) is evenly divisible by the number of folds \(K\). 
	We randomly partition the sample into \(K\) folds of equal size. Let \(I_k\) denote the set of indices belonging to the \(k\)-th fold, and let \(I_{-k}\) denote its complement. Denote by \(|I_k|\) the size of the fold \(I_k\). For any random variable \(O\), define the empirical average over fold \(k\) as 
	\(
	\mathbb{E}_{nk}[O] := \sum_{i \in I_k} O_i/|I_k|.
	\)
	We further define the $L_2$ norm of the nuisance vector $\alpha_{\pi}(L)$ from Equation~\eqref{eq: nuisance function fixed pi} as 
	\[
	\|\alpha_{\pi}(L)\|_2^2 := \|\delta(L)\|_2^2 + \|\eta(L)\|_2^2 + \|\kappa_{\pi}(L)\|_2^2 + \|\rho_{\pi}(L)\|_2^2 + \|\gamma_{\pi}(L)\|_2^2,
	\]
	and the $L_2$ norm of the nuisance vector $\beta(Z,L)$ from Equation~\eqref{eq: nuisance function} as 
	\[
	\|\beta(Z,L)\|_2^2 := \|\pi_*(Z,L)\|_2^2 + \|\xi(Z,L)\|_2^2 + \|\delta(L)\|_2^2 + \|\eta(L)\|_2^2 + \|\kappa(L)\|_2^2 + \|\gamma(L)\|_2^2.
	\]

	Next, for any fixed fold \( I_k\), the nuisance estimators \( \hat\alpha_{\pi}^{(n,k)} \) are trained using only the observations in \( I_{-k} \) with any suitable machine learning methods. By construction, \( \hat\alpha_{\pi}^{(n,k)} \) is independent of the samples in $I_k$. We derive the estimator $\hat\psi_{\pi}^{(n)}$ as the solution to the equation
	\begin{align}\label{eq: AUG estimator prespecified weight}
		\sum_{k=1}^K\mathbb{E}_{nk}\left[\varphi_{\pi}(O;\hat\psi_{\pi}^{(n)},\hat\alpha_{\pi}^{(n,k)})\right]=0.
	\end{align}
	Next, we establish the consistency and asymptotic normality of \(\hat\psi_{\pi}^{(n)}\) defined in Equation~\eqref{eq: AUG estimator prespecified weight}.
	\begin{theorem}[Asymptotic normality of $\hat\psi_{\pi}^{(n)}$]\label{thm: asymptotic normality known weight}
		Under Assumptions~\ref{as: consistency}--\ref{as: IV independence}, suppose that 
		\(\pi(Z,L)\) is an RWF for $A$. Assume further that, for any \(k=1,\ldots,K\),
		\(\mathbb{E}[\|\hat \alpha_{\pi}^{(n,k)}-\alpha_{\pi}^o\|_2^2]=o(1)\), and that
		\begin{align*}
			\left\{\begin{array}{l}
				\|\hat\kappa_{\pi}^{(n,k)}-\kappa_{\pi}^o\|_2\times \|\hat\gamma_{\pi}^{(n,k)}-\gamma_{\pi}^o\|_2\\
				+\|\hat\rho_{\pi}^{(n,k)}-\rho_{\pi}^o\|_2\times
				\|\hat\delta^{(n,k)}-\delta^o\|_2\\
				+\|\hat\rho_{\pi}^{(n,k)}-\rho_{\pi}^o\|_2\times
				\|\hat\eta^{(n,k)}-\eta^o\|_2\\
			\end{array}\right\}
			=o_p(n^{-1/2}).
		\end{align*}
		Then $\sqrt{n}\left(\hat\psi_{\pi}^{(n)}-\psi_{\pi}^o\right)/\sigma_{\pi}^o$ converges in distribution to $\mathcal{N}(0,1)$, where the asymptotic variance is defined as $(\sigma_{\pi}^o)^2:=\mathbb{E}[\varphi_{\pi}(O;\psi_{\pi}^o,\alpha_{\pi}^o)^2]$. In addition, if we define  
		$$(\hat\sigma_{\pi}^{(n)})^2:=\sum_{k=1}^K\mathbb{E}_{nk}[\varphi_{\pi}(O;\hat\psi_{\pi}^{(n)},\hat\alpha_{\pi}^{(n,k)})^2]/K,$$
		then $(\hat\sigma_{\pi}^{(n)})^2$ converges in probability to $(\sigma_{\pi}^o)^2$.
	\end{theorem}

	Analogously, for any fixed fold \(k = 1, \ldots, K\), nuisance estimators \(\hat\beta^{(n,k)}\) are trained using only the observations in \(I_{-k}\) with any suitable machine learning method. The estimator \(\hat\psi_{*}^{(n)}\) is then defined as the solution to
	\begin{align}\label{eq: AUG estimator adaptive weight}
		\sum_{k=1}^K \mathbb{E}_{nk}\!\left[\varphi\!\left(O;\hat\psi_{*}^{(n)},\hat\beta^{(n,k)}\right)\right] = 0.
	\end{align}
	Next, we establish the consistency and asymptotic normality of \(\hat\psi_{*}^{(n)}\) defined in Equation~\eqref{eq: AUG estimator adaptive weight}.
	\begin{theorem}[Asymptotic normality of $\hat\psi_{*}^{(n)}$]\label{thm: asymptotic normality unknown weight}
		Under Assumptions~\ref{as: consistency}-\ref{as: IV independence} and \ref{as: strong IV relevance}, suppose that Equation~\eqref{eq: no unmeasured common effect modifier} holds. Assume that for any $k=1,\ldots,K$, $\mathbb{E}[\|\hat \beta^{(n,k)}-\beta^o\|_2^2]=o(1)$, and that
		\begin{align*}
			\left\{
			\begin{array}{l}
				\|\hat\gamma^{(n,k)}-\gamma^o\|_2 \times \|\hat\kappa^{(n,k)}-\kappa^o\|_2
				+\|\hat\delta^{(n,k)}-\delta^o\|_2^2+\|\hat\pi_*^{(n,k)}-\pi_*^o\|_2^2\\
				+\|\hat\xi^{(n,k)}-\xi^o\|_2\times \|\hat\pi_*^{(n,k)}-\pi_*^o\|_2
				+\|\hat\eta^{(n,k)}-\eta^o\|_2\times \|\hat\delta^{(n,k)}-\delta^o\|_2
			\end{array}\right\}
			=o_p(n^{-1/2})
		\end{align*}
		Then $\sqrt{n}\left(\hat\psi_{*}^{(n)}-\psi_{*}^o\right)/\sigma_{*}^o$ converges in distribution to $\mathcal{N}(0,1)$, where the asymptotic variance is defined as $(\sigma_{*}^o)^2:=\mathbb{E}[\varphi(O;\psi_{*}^o,\beta^o)^2].$ In addition, if we define 
		$$(\hat\sigma_{*}^{(n)})^2:=\sum_{k=1}^K\mathbb{E}_{nk}[\varphi(O;\hat\psi_{*}^{(n)},\hat\beta^{(n,k)})^2]/K,$$ 
		then $(\hat\sigma_{*}^{(n)})^2$ converges in probability to $(\sigma_{*}^o)^2$.
	\end{theorem}

	\section{Extension to longitudinal data}\label{sec: longitudinal scenario}
	\subsection{Identification strategy}
	In this section, we extend the identification strategy introduced in Section~\ref{sec: identification} to a longitudinal setting with a sequence of IVs observed at each time point. Specifically, consider a longitudinal study with measurements collected at discrete time points \(T+1\), indexed by \(t = 0, 1, \ldots, T\), where \(T\) is a fixed nonnegative integer. The special case \(T = 0\) reduces to the panel data framework discussed in Section~\ref{sec: identification}. 
	
	For notation, let \(\overline{a}_t := [a_0, a_1, \ldots, a_t]\), \(\underline{a}_t := [a_t, a_{t+1}, \ldots, a_T]\), \(a_t^s := [a_t, a_{t+1}, \ldots, a_s]\), and \(\overline{a} := [a_0, a_1, \ldots, a_T]\). By convention, we set $\underline{a}_{T+1}$ and  $a_t^{t-1}$ as empty for any \(t\), and note that \(\overline{a} = \underline{a}_0 = \overline{a}_T\).
	At each time point \(t\), let \(L_t\in\mathcal{L}_t\) denote the vector of observed confounders, \(U_t\in\mathcal{U}_t\) the vector of unobserved confounders, \(Z_t\in\mathcal{Z}_t\) the IV (which may be multi-categorical or continuous), and \(A_t\in\mathcal{A}_t\) the discrete treatment assignment. The observed data are given by 
	\(
	O := [\overline{Z}_T, \overline{A}_T, \overline{L}_T, Y],
	\)
	where \(Y\) is the outcome of interest, observed only at time \(T+1\). 
	
	At each time point \(t=0,\ldots, T\), define the history \(H_t := [\overline{Z}_{t-1}, \overline{A}_{t-1}, \overline{L}_t]\in\mathcal{H}_t\), and let \(H_{T+1} := O\) denote all the observed data. Notably, the histories satisfy the recursive relation 
	\(
	H_t = [H_{t-1}, Z_{t-1}, A_{t-1}, L_t].
	\) 
	Let \(Y(\overline{a})\) denote the potential outcome in the treatment history \(\overline{A} = \overline{a}\). Next, we extend the preceding assumptions to a longitudinal setting with IVs.
	\begin{assumption}[Consistency]\label{as: consistency'}
		\(Y=Y(\overline{A})\).
	\end{assumption}
	\begin{assumption}[Latent ignorability]\label{as: latent ignorability'}
		For any fixed $t$, $\{Z_t,A_t\}\indep Y(\underline{a}_t) \mid H_t,\overline{U}_t$.
	\end{assumption}
	\begin{assumption}[IV independence]\label{as: IV independence'}
		For any fixed $t$, $Z_t\indep \overline{U}_t\mid H_t$.
	\end{assumption}
	\begin{assumption}[Strong IV relevance]\label{as: strong IV relevance'}
		For each time $t=0,\ldots,T$, there exists a positive constant $\epsilon_0$ such that for any $h_t\in\mathcal{H}_t$ and $a_t\in\mathcal{A}_t$,
		$\mathrm{Var}\!\{\Pr(A_t=a_t\mid Z_t,H_t) \mid H_t=h_t\} \geq \epsilon_0.$
	\end{assumption}

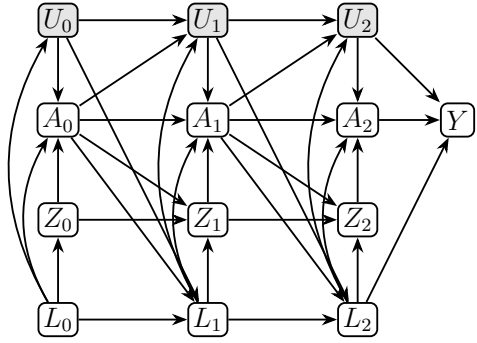
\begin{figure}
		\centering
		\resizebox{0.4\textwidth}{!}{
			\begin{tikzpicture}[
				->,
				>=Stealth,
				node distance=1.5cm,
				thick,
				every node/.style={
					draw=black,
					rounded corners=3pt,
					minimum size=0.5cm,
					inner sep=2pt,
					font=\normalsize\sffamily,
					fill=white
				},
				every edge/.style={
					draw=black,
					->,
					shorten >=1pt,
					shorten <=1pt
				}
				]
				
				\node (Z0) at (0,-1.5) {$Z_0$};
				\node (Z1) at (2.25,-1.5) {$Z_1$};
				\node (Z2) at (4.5,-1.5) {$Z_2$};
				
				\node (A0) at (0,0) {$A_0$};
				\node (A1) at (2.25,0) {$A_1$};
				\node (A2) at (4.5,0) {$A_2$};
				
				\node (U0)[fill=gray!20] at (0,1.5) {$U_0$};
				\node (U1)[fill=gray!20] at (2.25,1.5) {$U_1$};
				\node (U2)[fill=gray!20] at (4.5,1.5) {$U_2$};
				
				\node (L0) at (0,-3) {$L_0$};
				\node (L1) at (2.25,-3) {$L_1$};
				\node (L2) at (4.5,-3) {$L_2$};
				
				\node (Y) at (6,0) {$Y$};
				
				\draw (Z0) -> (A0);
				\draw (Z0) -> (Z1);
				
				\draw (A0) -> (A1);
				\draw (A0) -> (Z1);
				\draw (A0) -> (U1);
				\draw (A0) -> (L1);
				
				\draw (U0) -> (A0);
				\draw (U0) -> (U1);			
				\draw (U0) -> (L1);
				
				\draw (Z1) -> (A1);
				\draw (Z1) -> (Z2);

				\draw (A1) -> (A2);
				\draw (A1) -> (Z2);
				\draw (A1) -> (U2);
				\draw (A1) -> (L2);
				
				\draw (U1) -> (A1);
				\draw (U1) -> (U2);
				\draw (U1) -> (L2);

				\draw (Z2) -> (A2);
				\draw (U2) -> (A2);
				
				\draw (L0) to [bend left=30] (A0);
				\draw (L0) -> (Z0);
				\draw (L0) -> (L1);
				\draw (L0) to [bend left=30] (U0);
				
				\draw (L1) to [bend left=30] (A1);
				\draw (L1) -> (Z1);
				\draw (L1) -> (L2);
				\draw (L1) to [bend left=30] (U1);
				
				\draw (L2) to [bend left=30] (A2);
				\draw (L2) -> (Z2);
				\draw (L2) -> (Y);
				\draw (L2) to [bend left=30] (U2);
				
				\draw (U2) -> (Y);
				\draw (A2) -> (Y);
		\end{tikzpicture}}
		\caption{
			Sequential DAG when $T=2$. 
			Gray nodes indicate unobserved variables, while white nodes indicate observed ones. 
		}
		
		\label{fig: sequential DAG}
	\end{figure}

	Assumptions~\ref{as: consistency'}--\ref{as: strong IV relevance'} can be regarded as a longitudinal extension of Assumptions~\ref{as: consistency}-\ref{as: IV independence} and \ref{as: strong IV relevance}.
	For illustration, Figure~\ref{fig: sequential DAG} displays a sequential directed acyclic graph (DAG) for the IV setting with \(T = 2\) under the one-step Markov property, where Assumptions~\ref{as: latent ignorability'},~\ref{as: IV independence'} hold.

	Next, we generalize the definitions of RWF and additive IV from the point-exposure setting to accommodate longitudinal data.
	
	\begin{definition}[Longitudinal RWF]\label{defn: RWF'}
		A function \(\pi_t(Z_t, H_t)\) is said to be an RWF for \(A_t\) if it is uniformly bounded, and  for each \(a_t \in \mathcal{A}_t\), there exists a constant \(\epsilon_0 > 0\) such that
		\[
		\left| \mathrm{Cov}\!\left\{ I\{A_t = a_t\}, \pi_t(Z_t, H_t) \mid H_t =h_t\right\} \right|
		\geq \epsilon_0, 
		\quad \text{for all } h_t\in\mathcal{H}_t.
		\]
	\end{definition}  
	\begin{definition}[Longitudinal additive IV]\label{defn: AIV'}
		For any fixed \(t\), we say that \(Z_t\) is an additive IV for \(A_t\) if there exist functions \(b_{t,a_t}(\overline{U}_t,H_t)\) and \(c_{t,a_t}(Z_t,H_t)\) such that, for all \(a_t \in \mathcal{A}_t\),
		\[
		\Pr(A_t = a_t \mid Z_t, \overline{U}_t, H_t)
		= b_{t,a_t}(\overline{U}_t, H_t) + c_{t,a_t}(Z_t, H_t).
		\]
	\end{definition}

	These two definitions are required for identifying the longitudinal mean potential outcomes. For a fixed sequence of RWFs \(\pi_t(Z_t,H_t)\), define \(\gamma_{T+1,\underline{a}_{T+1}}^o(H_{T+1}) := Y.\)
	Then, for \(t = T, \ldots, 0\), recursively define the nuisance function
	\[
	\gamma_{t,\underline{a}_t}^o(H_t) 
	:= \frac{\mathrm{Cov}\!\left\{ I\{A_t = a_t\} \, \gamma_{t+1,\underline{a}_{t+1}}^o(H_{t+1}), \pi_t(Z_t, H_t) \mid H_t \right\}}
	{\mathrm{Cov}\!\left\{ I\{A_t = a_t\}, \pi_t(Z_t, H_t) \mid H_t \right\}}.
	\]
	Intuitively, \(\gamma_{t,\underline{a}_t}^o(H_t)\) can be interpreted as an estimator of the conditional mean potential outcomes \(\mathbb{E}[Y(\underline{a}_t) \mid H_t]\). 
	Building on this intuition, we derive the following identification result for longitudinal additive IVs.

	\begin{theorem}[Longitudinal additive IV identification]\label{thm: longitudinal AIV}
		Under Assumptions~\ref{as: consistency'}--\ref{as: IV independence'}, let $s$, $r$ be two integers with \(0 \le s \le T+1\), \(r \ge 0\), and \(s + r \le T+1\). Suppose that for each \(t=0,\ldots,T\), \(Z_t\) serves as an additive IV for \(A_t\), and that \(\pi_t(Z_t, H_t)\) is an RWF for \(A_t\). Then, the mean potential outcomes \(\mathbb{E}\bigl[Y(\underline{a}_{s})\bigr]\) can be expressed as
		\begin{equation}\label{eq: identification AIV longitudinal}
			\begin{aligned}
				\mathbb{E}\left[
				\prod_{t=s}^{T-r}
				\dfrac{\left(\pi_t(Z_t,H_t)-\mathbb{E}[\pi_t(Z_t,H_t)\mid H_t]\right)I\{A_t=a_t\}}{\mathrm{Cov}\!\{I\{A_t=a_t\},\pi_t(Z_t,H_t)\mid H_t\}}
				\gamma_{T+1-r,\underline{a}_{T+1-r}}^o(H_{T+1-r})
				\right].
			\end{aligned}
		\end{equation}
	\end{theorem}
	In particular, consider the special case of Theorem~\ref{thm: longitudinal AIV} with \(s = 0\) and \(r = 0\), which corresponds to identifying the mean of potential outcomes from the initial time point to the final time \(T\). Without loss of generality, if we set \(\pi(Z_t,H_t) = Z_t\), the identification formula simplifies to
	\begin{align}
		\mathbb{E}[Y(\overline{a})] = \psi_{\overline{a}}^o :=
		\mathbb{E}\Biggl[
		\prod_{t=0}^{T}
		\frac{(Z_t - \mathbb{E}[Z_t \mid H_t]) \, I\{A_t = a_t\}}{\mathrm{Cov}\!\{ I\{A_t = a_t\}, Z_t \mid H_t \}}
		\times Y
		\Biggr].\label{eq: identification AIV longitudinal 2}
	\end{align}
	This expression is directly analogous to the inverse probability weighting approach in settings without unmeasured confounding.
	Alternatively, setting \(s = 0\) and \(r = T+1\), the identification formula reduces to
	\(
	\psi_{\overline{a}}^o = \mathbb{E}\bigl[\gamma_{0,\overline{a}}^o(H_0)\bigr],
	\)
	which corresponds to the outcome regression approach or the g-formula in longitudinal causal inference \citep{hernan2020causal}.
	
	\subsection{Semiparametric theory}
	In this subsection, without loss of generality, we focus on the case where $Z_t$ is univariate and \(\pi_t(Z_t, H_t)=Z_t\) is an RWF for \(A_t\). We derive the EIFs for \(\psi_{\overline{a}}^o\) in Equation~\eqref{eq: identification AIV longitudinal 2} when \(\pi_t(Z_t, H_t) = Z_t\); for a general RWF \(\pi_t(Z_t, H_t)\), we can define \(Z_t^{\pi} := \pi_t(Z_t, H_t)\) and replace \(Z_t\) with \(Z_t^{\pi}\) in the subsequent analysis. 
	For notational convenience, define \(\gamma_{T+1,\underline{a}_{T+1}}^o(H_{T+1}) := Y\) and \(A_t^{(a_t)} := I\{A_t = a_t\}\).
	Next, for \(t = T, \ldots, 0\), recursively define nuisance functions:
	\begin{align*}
		&\kappa_{t,a_t}^o(H_t) := \mathrm{Cov}\!\{A_t^{(a_t)}, Z_t \mid H_t\},\\ 
		&\delta_{t,a_t}^o(H_t) := \mathbb{E}[A_t^{(a_t)} \mid H_t], 
		&& \eta_{t,\underline{a}_t}^o(H_t) := \mathbb{E}\bigl[A_t^{(a_t)} \gamma_{t+1,\underline{a}_{t+1}}^o(H_{t+1}) \mid H_t\bigr],\\
		&\rho_t^o(H_t) := \mathbb{E}[Z_t \mid H_t], 
		&& \gamma_{t,\underline{a}_t}^o(H_t) := \frac{\mathrm{Cov}\!\{ A_t^{(a_t)} \gamma_{t+1,\underline{a}_{t+1}}^o(H_{t+1}), Z_t \mid H_t \}}
		{\mathrm{Cov}\!\{ A_t^{(a_t)}, Z_t \mid H_t \}}.
	\end{align*}
	We unify these nuisance functions into one nuisance vector:
	\begin{align}\label{eq: nuisance function longitudinal}
		\alpha_{\overline{a}}^o := \{\alpha_{t,\underline{a}_t}^o\}_{t=0}^T, \qquad
		\alpha_{t,\underline{a}_t}^o := [
		\delta_{t,a_t}^o,
		\kappa_{t,a_t}^o,
		\rho_t^o,
		\eta_{t,\underline{a}_t}^o,
		\gamma_{t,\underline{a}_t}^o].
	\end{align}
	We now proceed to derive the EIF for \(\psi_{\overline{a}}^o\) in Equation~\eqref{eq: identification AIV longitudinal 2} in the following theorem.
	
	\begin{theorem}\label{thm: EIF AIV longitudinal}
		Under Assumptions~\ref{as: consistency'}--\ref{as: IV independence'}, suppose that for each \(t=0,\ldots,T\), 
		\(\pi_t(Z_t, H_t)=Z_t\) is an RWF for \(A_t=a_t\). Then, the EIF for $\psi_{\overline{a}}^o$ is $\varphi_{\overline{a}}(O;\psi_{\overline{a}}^o,\alpha_{\overline{a}}^o)$, where 
		\begin{align*}
			&\varphi_{\overline{a}}(O;\psi_{\overline{a}},\alpha_{\overline{a}}):=
			\prod_{t=0}^{T}
			\dfrac{\left\{Z_t-\rho_t(H_t)\right\}A_t^{(a_t)}}
			{\kappa_{t,a_t}(H_t)}
			Y-\psi_{\overline{a}}
			+\sum_{t=0}^T\left(\displaystyle\prod_{s=0}^{t-1}\dfrac{\left\{Z_s-\rho_s(H_s)\right\}A_s^{(a_s)}}
			{\kappa_{s,a_s}(H_s)}\right)
			\\&\times \left\{
			\left(1-\dfrac{\{Z_t-\rho_t(H_t)\}\{A_t^{(a_t)}-\delta_{t,a_t}(H_t)\}}{\kappa_{t,a_t}(H_t)} \right)\gamma_{t,\underline{a}_t}(H_t) - 
			\dfrac{(Z_t-\rho_t(H_t))\eta_{t,\underline{a}_t}(H_t)}{\kappa_{t,a_t}(H_t)}\right\}.
		\end{align*}
	\end{theorem}
	Notably, when $t=0$, the EIF in Theorem~\ref{thm: EIF AIV longitudinal} 
	has a structure similar to that derived in Theorem~\ref{thm: EIF AIV fixed pi}.
	The next proposition derives the mixed bias property for the EIF in Theorem~\ref{thm: EIF AIV longitudinal}.
	\begin{proposition}[Mixed bias property]\label{prop: mixed bias property longitudinal}
		Under the conditions of Theorem~\ref{thm: EIF AIV longitudinal}, for any fixed nuisance vector $\alpha_{\overline{a}}$ in Equation~\eqref{eq: nuisance function longitudinal}, $\mathbb{E}[\varphi_{\overline{a}}(O;\psi_{\overline{a}}^o,\alpha_{\overline{a}})]$ equals
		\begin{align*}
			&\mathbb{E}\left[\begin{array}{l}
				\displaystyle\sum_{t=0}^T\left(\displaystyle\prod_{s=0}^{t-1}\dfrac{\left\{Z_s-\rho_s(H_s)\right\}A_s^{(a_s)}}
				{\kappa_{s,a_s}(H_s)}\right)
				\times \dfrac{1}{\kappa_{t,a_t}(H_t)}\\
				\times\left(\begin{array}{l}
					\left\{\kappa_{t,a_t}(H_t)- \kappa_{t,a_t}^o(H_t)\right\} \left\{\gamma_{t,\underline{a}_t}(H_t)-\gamma_{t,\underline{a}_t}^o(H_t)\right\}\\
					+\left\{\rho_t(H_t) - \rho_t^o(H_t)\right\}\left\{\eta_{t,\underline{a}_t}(H_t)-\eta_{t,\underline{a}_t}^o(H_t)\right\}\\
					+\left\{\rho_t(H_t) - \rho_t^o(H_t)\right\}\left\{\delta_{t,a_t}(H_t)-\delta_{t,a_t}^o(H_t)\right\} \gamma_{t,\underline{a}_t}(H_t)
				\end{array}\right)
			\end{array}\right].
		\end{align*}
	\end{proposition}

	\subsection{Cross-fitting procedure}
	We develop a cross-fitting procedure for estimating $\psi_{\overline{a}}^o$. This estimator can be intuitively understood as resulting from a backward fitting strategy. Let 
	\(\hat{\mathbb{E}}^{(n,k)}[\phi(O) \mid H_t]\) denote an estimate of the conditional expectation \(\mathbb{E}[\phi(O)\mid H_t]\), and 
	\(\widehat{\text{Cov}}^{(n,k)}\{\phi_1(O), \phi_2(O)\mid H_t\}\) denote an estimate of the conditional covariance 
	\(\text{Cov}\{\phi_1(O),\phi_2(O)\mid H_t\}\). Both estimates are obtained using only the observations in \(I_{-k}\) and fitted using an appropriate machine learning method. The Algorithm~\ref{alg: longitudinal} summarizes the backward cross-fitting procedure.
	
	\begin{algorithm}[htbp]
		\caption{Backward Cross-Fitting Procedure for Longitudinal Data}
		\label{alg: longitudinal}
		
		Randomly divide the samples evenly into \(K\) folds \(\{I_k\}_{k=1}^K\)\;
		
		\For{$k = 1$ \KwTo $K$}{
			Initialize \(t \gets T\) and set \(\hat\Psi_{T+1,\underline{a}_{T+1}}^{(n,k)} := Y\)\;
			
			\While{$t \ge 0$}{
				Fit the nuisance functions \(\hat \alpha_{t,\overline{a}}^{(n,k)}(H_t)\) using samples in \(I_{-k}\):
				\begin{align*}
					&\hat\eta_{t,\underline{a}_t}^{(n,k)}(H_t) := \hat{\mathbb{E}}^{(n,k)}[A_t^{(a_t)} \hat\Psi_{t+1,\underline{a}_{t+1}}^{(n,k)} \mid H_t], &
					\hat \delta_{t,a_t}^{(n,k)}(H_t) := \hat{\mathbb{E}}^{(n,k)}[A_t^{(a_t)} \mid H_t],\\
					&\hat\gamma_{t,\underline{a}_t}^{(n,k)}(H_t) := 
					\frac{\widehat{\mathrm{Cov}}^{(n,k)}\{ A_t^{(a_t)} \hat\Psi_{t+1,\underline{a}_{t+1}}^{(n,k)}, Z_t \mid H_t\}}
					{\widehat{\mathrm{Cov}}^{(n,k)}\{ A_t^{(a_t)}, Z_t \mid H_t \}}, &
					\hat \rho_t^{(n,k)}(H_t) := \hat{\mathbb{E}}^{(n,k)}[Z_t \mid H_t],\\
					&\hat\kappa_{t,a_t}^{(n,k)}(H_t) := \widehat{\mathrm{Cov}}^{(n,k)}\{A_t^{(a_t)},Z_t \mid H_t\}.
				\end{align*}
				
				Update for all samples:
				
				\begin{align*}
					\hat\Psi_{t,\underline{a}_{t}}^{(n,k)} :=&
					\frac{1}{\hat\kappa_{t,a_t}^{(n,k)}(H_t)}
					\left(Z_t - \hat\rho_t^{(n,k)}(H_t)\right)
					\left(A_t^{(a_t)} \hat\Psi_{t+1,\underline{a}_{t+1}}^{(n,k)} - \hat\eta_{t,\underline{a}_t}^{(n,k)}(H_t)\right)
					+ \\&\left(1 - 
					\frac{(Z_t-\hat\rho_t^{(n,k)}(H_t))(A_t^{(a_t)} - \hat\delta_{t,a_t}^{(n,k)}(H_t))}{\hat\kappa_{t,a_t}^{(n,k)}(H_t)}\right)
					\hat\gamma_{t,\underline{a}_t}^{(n,k)}(H_t).
				\end{align*}
				
				Decrement \(t \gets t-1\)\;
			}
		}
		
		Output the estimator and its variance:
		\begin{equation}\label{eq: AUG estimator longitudinal}
			\hat\psi_{\overline{a}}^{(n)} := \frac{1}{K} \sum_{k=1}^K \mathbb{E}_{nk}[\hat\Psi_{0,\overline{a}}^{(n,k)}],\quad
			(\hat\sigma_{\overline{a}}^{(n)})^2 := \frac{1}{K} \sum_{k=1}^K \mathbb{E}_{nk} \Big[(\hat\Psi_{0,\overline{a}}^{(n,k)} - \mathbb{E}_{nk}[\hat\Psi_{0,\overline{a}}^{(n,k)}])^2 \Big].
		\end{equation}
		
	\end{algorithm}
	
	\begin{figure}[htbp]
		\centering
		\resizebox{0.8\textwidth}{!}{
			\begin{tikzpicture}[
				arrow/.style={-{Stealth[length=2mm, width=1.2mm]}, thick, gray!70},
				every node/.style={minimum width=1.6cm, minimum height=0.9cm, align=center, font=\small},
				state/.style={
					draw=gray!70,
					rounded corners=3pt,
					top color=white,
					bottom color=blue!5,
					very thick,
					drop shadow
				},
				est/.style={
					circle,
					draw=orange!60!black,
					top color=white,
					bottom color=orange!20,
					thick,
					drop shadow
				}
				]
				
				% spacing
				\def\xsep{3}
				\def\ysep{1.8}
				
				% === Top row: Validation set ===
				\node[state] (T3) at (0, 0) {$\hat\Psi_{3,\underline{a}_3}^{(n,k)},H_3$};
				\node[state] (T2) at (\xsep, 0) {$\hat\Psi_{2,\underline{a}_2}^{(n,k)},H_2$};
				\node[state] (T1) at (2*\xsep, 0) {$\hat\Psi_{1,\underline{a}_1}^{(n,k)},H_1$};
				\node[state] (T0) at (3*\xsep, 0) {$\hat\Psi_{0,\underline{a}_0}^{(n,k)},H_0$};
				
				% === Middle row: alpha estimators ===
				\node[est] (alpha2) at (\xsep, -\ysep) {$\hat\alpha_{2,\overline{a}}^{(n,k)}$};
				\node[est] (alpha1) at (2*\xsep, -\ysep) {$\hat\alpha_{1,\overline{a}}^{(n,k)}$};
				\node[est] (alpha0) at (3*\xsep, -\ysep) {$\hat\alpha_{0,\overline{a}}^{(n,k)}$};
				
				% === Bottom row: Training set ===
				\node[state] (T3') at (0, -2*\ysep) {$\hat\Psi_{3,\underline{a}_3}^{(n,k)},H_3$};
				\node[state] (T2') at (\xsep, -2*\ysep) {$\hat\Psi_{2,\underline{a}_2}^{(n,k)},H_2$};
				\node[state] (T1') at (2*\xsep, -2*\ysep) {$\hat\Psi_{1,\underline{a}_1}^{(n,k)},H_1$};
				\node[state] (T0') at (3*\xsep, -2*\ysep) {$\hat\Psi_{0,\underline{a}_0}^{(n,k)},H_0$};
				
				% === Time labels ===
				\node[draw=none] at (0, 0.9) {\scriptsize \textbf{$t=3$}};
				\node[draw=none] at (\xsep, 0.9) {\scriptsize \textbf{$t=2$}};
				\node[draw=none] at (2*\xsep, 0.9) {\scriptsize \textbf{$t=1$}};
				\node[draw=none] at (3*\xsep, 0.9) {\scriptsize \textbf{$t=0$}};
				
				% === Labels: evaluation / training ===
				\node[draw=none, anchor=east, text=blue!50!black] at (-1, 0) {\textbf{\small Evaluation set} $I_k$};
				\node[draw=none, anchor=east, text=orange!60!black] at (-1, -2*\ysep) {\textbf{\small Training set} $I_{-k}$};
				
				% === Arrows ===
				% Top row arrows
				\draw[arrow] (T3) -- (T2);
				\draw[arrow] (T2) -- (T1);
				\draw[arrow] (T1) -- (T0);
				
				% Bottom row arrows
				\draw[arrow] (T3') -- (T2');
				\draw[arrow] (T2') -- (T1');
				\draw[arrow] (T1') -- (T0');
				
				% eta' to alpha
				\draw[arrow, orange!70!black] (T3') -- (alpha2);
				\draw[arrow, orange!70!black] (T2') -- (alpha1);
				\draw[arrow, orange!70!black] (T1') -- (alpha0);
				
				% alpha to eta' and eta
				\draw[arrow, dashed, blue!60!black] (alpha2) -- (T2');
				\draw[arrow, dashed, blue!60!black] (alpha2) -- (T2);
				\draw[arrow, dashed, blue!60!black] (alpha1) -- (T1');
				\draw[arrow, dashed, blue!60!black] (alpha1) -- (T1);
				\draw[arrow, dashed, blue!60!black] (alpha0) -- (T0');
				\draw[arrow, dashed, blue!60!black] (alpha0) -- (T0);
				
			\end{tikzpicture}
		}
		\caption{Illustration of the backward cross-fitting procedure in the $k$-th fold for $T=2$.}
		\label{fig: cross-fitting procedure}
	\end{figure}
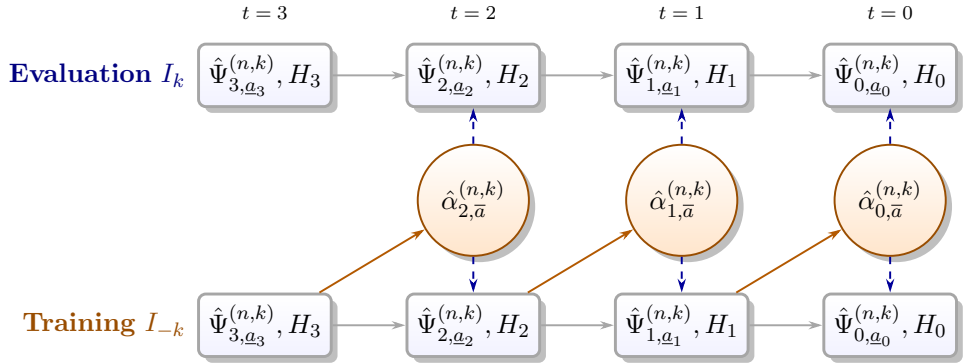
	Figure~\ref{fig: cross-fitting procedure} provides a graphical illustration of
	Algorithm~\ref{alg: longitudinal} for \(T=2\). The upper row corresponds to the
	evaluation set \(I_k\), while the bottom row represents the training set
	\(I_{-k}\). The middle row displays the nuisance estimators
	\(\hat\alpha_{t,\overline{a}}^{(n,k)}\), which are fitted using the training data and
	adopted to form the final predictions. Importantly,
	\(\hat\alpha_{t,\overline{a}}^{(n,k)}\) is constructed exclusively from samples in
	\(I_{-k}\), thereby ensuring its independence from the evaluation set \(I_k\).
	Moreover, in Algorithm~\ref{alg: longitudinal}, we introduce the random variable $\hat\Psi_{t,\underline{a}_{t}}^{(n,k)}$. By induction, one can verify that for any $t=0,\ldots,T$,
	\begin{align*} 
		&\hat\Psi_{t,\underline{a}_{t}}^{(n,k)}=\prod_{s=t}^{T} \dfrac{\left\{Z_s-\hat\rho_s^{(n,k)}(H_s)\right\}A_s^{(a_s)}} {\hat\kappa_{s,a_s}^{(n,k)}(H_s)} Y +\sum_{s=t}^T\left(\displaystyle\prod_{r=t}^{s-1}\dfrac{\left\{Z_r-\hat\rho_r^{(n,k)}(H_r)\right\}A_r^{(a_r)}} {\hat\kappa_{r,a_r}^{(n,k)}(H_r)}\right) \\&\times \left\{\begin{array}{c}
			\left(1-\dfrac{\{Z_s-\hat\rho_s^{(n,k)}(H_s)\}\{A_s^{(a_s)}-\hat\delta_{s,a_s}^{(n,k)}(H_s)\}}{\hat\kappa_{s,a_s}^{(n,k)}(H_s)} \right)\hat\gamma_{s,\underline{a}_s}^{(n,k)}(H_s) \\- \dfrac{(Z_s-\hat\rho_s^{(n,k)}(H_s))\times\hat\eta_{s,\underline{a}_s}^{(n,k)}(H_s)}{\hat\kappa_{s,a_s}^{(n,k)}(H_s)}
		\end{array}\right\}. 
	\end{align*}
	Intuitively, \(\hat\Psi_{t,\underline{a}_t}^{(n,k)}\) serves as an augmented estimator of the
	true conditional mean potential outcome
	\(\gamma_{t,\underline{a}_t}^o(H_t)\). In particular, if all nuisance functions are
	correctly specified, then
	$\mathbb{E}[\hat\Psi_{t,\underline{a}_t}^{(n,k)} \mid H_t]
	= \gamma_{t,\underline{a}_t}^o(H_t).$
	In addition, one can verify that 
	\(
	\varphi_{\overline{a}}(O;\psi_{\overline{a}},\hat\alpha_{\overline{a}}^{(n,k)}) = 
	\hat\Psi_{0,\overline{a}}^{(n,k)} - \psi_{\overline{a}}.
	\)
	This representation motivates the construction of the estimator $\hat\psi_{\overline{a}}^{(n)}$ in Equation~\eqref{eq: AUG estimator longitudinal}.
	
	An estimator of nuisance function $\hat{\alpha}_{t,\overline{a}}^{(n,k)}$ is referred to as a DR-Learner(or IF-Learner), and the theoretical properties of the nuisance vector $\hat{\alpha}_{t,\overline{a}}^{(n,k)}$ have been well-established in \citet{curth2021estimatingstructuraltargetfunctions,nie2020quasioracleestimationheterogeneoustreatment,foster2023orthogonalstatisticallearning, morzywolek2024weightedorthogonallearnersheterogeneous}.
	Next, we establish that, under the convergence rates required
	for nuisance estimators, $\hat{\psi}_{\overline{a}}^{(n)}$ is asymptotically normal, 
	and its variance estimator $\hat{\sigma}_{\overline{a}}^{(n)}$ is consistent.
	\begin{theorem}[Asymptotic normality of $\hat\psi_{\overline{a}}^{(n)}$]\label{thm: asymptotic normality longitudinal}
		Under Assumptions~\ref{as: consistency'}--\ref{as: IV independence'}, suppose that for each \(t = 0, \ldots, T\), \(Z_t\) serves as an additive IV for \(A_t\), and that \(\pi_t(Z_t, H_t) = Z_t\) is an RWF for \(A_t\). 
		Further, for each \(t = 0, \ldots, T\) and \(k = 1, \ldots, K\), suppose that the following rate condition holds for the nuisance functions $\hat\alpha_{t,\overline{a}}^{(n,k)}$ defined in Algorithm~\ref{alg: longitudinal}:
		\begin{align*}
			\left\{\begin{array}{l}s
				\|\hat\kappa_{t,a_t}^{(n,k)}- \kappa_{t,a_t}^{o}\|_2
				\times\|\hat\gamma_{t,\underline{a}_t}^{(n,k)}-\gamma_{t,\underline{a}_t}^o\|_2\\
				+\|\hat\rho_t^{(n,k)} - \rho_t^o\|_2\times
				\|\hat\eta_{t,\underline{a}_t}^{(n,k)}-\eta_{t,\underline{a}_t}^o\|_2\\
				+\|\hat\rho_t^{(n,k)} - \rho_t^o\|_2\times
				\|\hat\delta_{t,a_t}^{(n,k)}- \delta_{t,a_t}^{o}\|_2
			\end{array}\right\}=o_p(n^{-1/2}).
		\end{align*}
		Furthermore, assume that for any fixed $k$ and time $t$,
		\(\mathbb{E}[\|\hat\alpha_{t,\overline{a}}^{(n,k)}- \alpha_{t,\overline{a}}^o\|_2^2]=o(1)\).
		Then $\sqrt{n}\{\hat\psi_{\overline{a}}^{(n)}-\psi_{\overline{a}}^o\}/\sigma_{\overline{a}}^o$ converges in distribution to $\mathcal{N}(0,1)$, where $\hat\psi_{\overline{a}}^{(n)}$ is defined in Equation~\eqref{eq: AUG estimator longitudinal}, and $(\sigma_{\overline{a}}^o)^2:=\mathbb{E}[\varphi_{\overline{a}}(O;\psi_{\overline{a}}^o,\alpha_{\overline{a}}^o)^2]$.
		In addition, $\hat \sigma_{\overline{a}}^{(n)}$ converges in probability to $\sigma_{\overline{a}}^o$.
	\end{theorem}
	
	Notably, the RWF \(\pi_t(Z_t,H_t)\) can be can adaptively selected as demonstrated in Section~\ref{sec: semiparametric theory} to obtain an efficient estimator of \(\psi_{\overline{a}}^o\). A detailed discussion of this adaptive selection is provided in the Supplementary Material. 
	Intuitively, a reliable candidate for adaptive RWF is the conditional probability function \(\pi_t^o(Z_t,H_t) = \Pr(A_t = a_t \mid Z_t,H_t)\), which directly characterizes the treatment assignment mechanism.
	
	Finally, although we focus on the evaluation of static treatment rules, the proposed methods can be extended to facilitate the evaluation of dynamic treatment rules, which is also demonstrated in the Supplementary Material. This extension would allow the incorporation of time-varying treatment strategies and enable more comprehensive assessments of treatment effects over time, enhancing the applicability of our approach to dynamic decision-making processes.

	\section{Simulations}\label{sec: simulation}
	\subsection{Point exposure scenario}
	In this section, we conduct simulation studies with a binary treatment and a continuous IV to illustrate the asymptotic results established in Section~\ref{sec: semiparametric theory}. 
	We generate $U$ and $L$ from the uniform distribution $U(-1,1)$ independently. The continuous IV is constructed as $Z = L + \sin(3L) + 2\epsilon_Z$, where $\epsilon_Z$ is an exogenous error term drawn from the standard normal distribution. Let ``Ber'' denote the binomial distribution. The binary treatment $A$ is generated under the following two designs:
	\begin{enumerate}[label=(A\arabic*),leftmargin=1cm]
		\item $A \sim \mathrm{Ber}\big(1,0.7\Phi(-2Z + 2L) + 0.3\Phi(3U - L)\big)$, where $\Phi$ denotes the cumulative distribution function of the standard normal distribution. \label{A1}
		\item $A \sim \mathrm{Ber}\big(1,\{1 + \exp(-(Z - L + U))\}^{-1}\big)$. \label{A2}
	\end{enumerate}
	
	It is straightforward to verify that $Z$ is an additive IV for $A$ in \ref{A1}, while in \ref{A2} the additive IV conditions are violated.  
	Next, we independently generate $\epsilon_Y \sim N(0,1)$. The outcome $Y$ is then generated according to the following two models:
	\begin{enumerate}[label=(Y\arabic*),leftmargin=1cm]
		\item $Y = Y(A) = 2U - 2L + 4AL + \epsilon_Y$. \label{Y1}
		\item $Y = Y(A) = (1 - A)\{3\cos(2U) - 3\cos(2L)\} + A\{3\sin(2U) + 2L\} + \epsilon_Y$. \label{Y2}
	\end{enumerate}
	The condition $Y(1) - Y(0)\indep U \mid L$ is satisfied in \ref{Y1} but violated in \ref{Y2}. It is straightforward to verify that the ATEs in \ref{Y1} and \ref{Y2} are both zero.
	We set the sample size $n=5000$ and use $K=2$ folds for cross-fitting. Nuisance functions are estimated via spline methods implemented in the \texttt{R} package \texttt{mgcv} \citep{wood2011fast}.
	
	The results are summarized in Table~\ref{tbl: cross-sectional}.
	We report the finite sample behavior of the ATE estimators
	\(\hat\psi_{\pi}^{(n)}\) and \(\hat\psi_{*}^{(n)}\), defined in
	Equations~\eqref{eq: AUG estimator prespecified weight} and
	\eqref{eq: AUG estimator adaptive weight}, respectively.
	The corresponding results are also presented for \(\mathbb{E}[Y(1)]\) (treated) and \(\mathbb{E}[Y(0)]\) (control).
	Under the model misspecification settings \ref{Y2}, \ref{A2}, the ATE estimator tends to exhibit larger bias due to the violation of Equation~\eqref{eq: no unmeasured common effect modifier}. We also find that the adaptive weighting method generally yields a smaller variance, suggesting superior efficiency.
	
	\renewcommand{\arraystretch}{0.75}
	\begin{table}
		\centering
		\caption{Simulation results under four DGPs. In each setting, the estimation is repeated 1000 times to calculate the average bias (Bias), empirical standard deviation (SD), average estimated standard error (SE), and the empirical 95\% coverage rate (CR).}
		\label{tbl: cross-sectional}
		% \resizebox{0.5\textwidth}{!}{
			\begin{tabular}{cc| ccc |ccc}
				\toprule
				\multirow{2}{*}{DGP}&\multirow{2}{*}{Metric}&
				\multicolumn{3}{c}{Adaptive weight}&\multicolumn{3}{|c}{Prespecified ($\pi(Z,L)=Z$)}\\
				&& Treated & Control & ATE & Treated & Control & ATE\\
				\midrule
				\multirow{4}{*}{\ref{Y1}, \ref{A1}}
				&Bias &.0011&.0011&.0019& .0018 &.0032&.0028  \\
				&SE   &.0543&.0545&.0807& .0626 &.0626&.0920 \\
				&SD   &.0577&.0581&.0844& .0612 &.0648&.0922 \\
				&CR   &.927&.938&.939& .947 &.935&.950 \\
				[5pt]
				\multirow{4}{*}{\ref{Y2}, \ref{A1}}
				&Bias &.0023&.0078&.0064& .0043 &.0044&.0012  \\
				&SE   &.0864&.0611&.1060& .1006 &.0691&.1220  \\
				&SD   &.0906&.0637&.1092& .0987 &.0703&.1200  \\
				&CR   &.926&.946&.941& .951 &.951&.952  \\
				[5pt]
				\multirow{4}{*}{\ref{Y1}, \ref{A2}}
				&Bias &.0006&.0020&.0013& .0006 &.0013&.0003  \\
				&SE   &.0554&.0561&.0820& .0565 &.0569&.0833  \\
				&SD   &.0547&.0575&.0830& .0563 &.0557&.0828  \\
				&CR   &.954&.934&.938& .956 &.956&.944  \\
				[5pt]
				\multirow{4}{*}{\ref{Y2}, \ref{A2}}
				&Bias &.0000&.0318&.0315& .0027 &.0269& .0301 \\
				&SE   &.0886&.0615&.1079& .0908 &.0621& .1100 \\
				&SD   &.0891&.0619&.1089& .0915 &.0618& .1110 \\
				&CR   &.948&.912&.936& .955 &.924& .941 \\
				\bottomrule
			\end{tabular}
			% }
	\end{table}

	\subsection{Longitudinal scenario}
	To simplify the setting, we fix the time horizon at \( T = 1 \). For each time point \( t = 0,\,1 \), we independently generate noise terms \( \epsilon_{U_t} \), \( \epsilon_{L_t} \), \( \epsilon_{Z_t} \), and \( \epsilon_Y \) from standard normal distributions. Based on these, the variables are simulated according to the following DGP:
	\begin{align*}
		&L_0 \sim 1.5\epsilon_{L_0},\quad U_0 \mid H_0 \sim 1.5\epsilon_{U_0};\\
		&Z_0 \mid H_0,U_0 \sim 0.3 L_0 + \sin(1.5 L_0)+ 2\epsilon_{Z_0};\\
		&A_0 \mid H_0,U_0,Z_0 \sim \text{Ber}(1,\,0.7\Phi(-2Z_0+0.6L_0) + 0.3\Phi(3U_0-L_0));\\
		&L_1 \mid H_0,U_0,Z_0,A_0 \sim (A_0-0.5)+0.5L_0+0.3U_0+0.5 \epsilon_{L_1};\\
		&U_1 \mid H_1,U_0 \sim (A_0-0.5)+0.5U_0+0.3L_1+0.5 \epsilon_{U_1};\\
		&Z_1 \mid H_1,\overline{U}_1 \sim 0.5L_1-0.5(A_0-0.5)-0.3Z_0+2\epsilon_{Z_1};\\
		&A_1 \mid H_1,\overline{U}_1,Z_1 \sim \text{Ber}(1,\,0.7\Phi(-2Z_1+L_1) + 0.3\Phi(3U_1-L_1));\\
		&Y \mid H_1,\overline{U}_1,Z_1,A_1 \sim (A_1-0.5)+2L_1+U_1+0.5\epsilon_{Y}.
	\end{align*}
	Notably, $Z_0$ and $Z_1$ serve as additive IVs for $A_0$ and $A_1$, respectively. 
	Data are generated using the \texttt{R} package \texttt{simcausal} \citep{sofrygin2017simcausal}. The sample sizes are set to $2000$ and $5000$, with cross-fitting performed using $K=2$ folds. Nuisance functions are estimated via spline methods implemented in the R package \texttt{mgcv}.
	Because the true treatment effects are not analytically available under the constructed DGPs, we estimate them using 100{,}000 samples by simulating potential outcomes under modified data-generating processes, where pairs of treatment assignments, $(A_0,A_1)$, are set deterministically to \((0,0)\), \((1,1)\), \((0,1)\), \((1,0)\), \((A_0,1),\) and \((A_0,0)\) (We use the notation \((A_0,0)\) to denote a dynamic treatment regime that follows the natural treatment rule for \(A_0\) while fixing \(A_1 = 0\)).
	
	Table~\ref{tbl: longitudinal data} summarizes the simulation results adopting Algorithm~\ref{alg: longitudinal} under different intervention strategies and sample sizes. Across all scenarios, the estimators show small bias, and the estimated standard errors closely match the empirical standard deviations, indicating accurate variance estimation. As expected, increasing the sample size from 2000 to 5000 reduces variability and tightens confidence intervals. The empirical coverage rates are generally close to the nominal 95\% level, although a modest decline is observed in certain intervention settings. Overall, the results demonstrate that the proposed method performs well and provides reliable inference in most cases.
	
	\begin{table}
		\centering
		\caption{Simulation results under three DGPs. In each setting, the estimation is repeated 500 times to calculate the average bias (Bias), average estimated standard error (SE), empirical standard deviation (SD), and the empirical 95\% coverage rate (CR).}
		\label{tbl: longitudinal data}
		%		\resizebox{\textwidth}{!}{%
			\begin{tabular}{ccc|cccccc}
				\toprule
				&\multirow{2}{*}{Size}&
				\multirow{2}{*}{Metric}&\multicolumn{6}{c}{Intervention}\\
				&&& $(0,0)$ & $(0,1)$ & $(1,0)$ & $(1,1)$ & $(A_0,0)$ & $(A_0,1)$ \\
				\midrule
				& \multirow{4}{*}{2000}
				& Bias &.0234&.0071&.0147&.0189&.0016&.0033\\
				& & SE &.6090&.5267&.5266&.5960&.1401&.1403\\
				& & SD &.6102&.5382&.5203&.6343&.1395&.1460\\
				& & CR &.955&.959&.960&.938&.938&.929\\
				[5pt]
				& \multirow{4}{*}{5000}
				& Bias &.0134&.0078&.0057&.0155&.0057&.0037\\
				& & SE &.2143&.1882&.1962&.2115&.0658&.0651\\
				& & SD &.2211&.1949&.1870&.2004&.0607&.0612\\
				& & CR &.940&.930&.942&.956&.956&.950\\
				\bottomrule
			\end{tabular}
			%	}
	\end{table}
	
	\section{Empirical illustration}\label{sec: empirical illustration}
	
	We perform a policy analysis using the dataset provided in the Supplementary Material of \citet{han2024optimal}.
	Schooling and post-school training are two central interventions that influence labor market outcomes such as earnings and employment \citep{ashenfelter2010handbook}. 
	To enable this analysis, \citet{han2024optimal} merged data from the Job Training Partnership Act (JTPA) Title II with additional sources on high school (HS) diploma, thereby constructing a dataset suitable for evaluating the effects of HS diplomas and subsidized job training as sequential treatments. 
	The final sample comprises 9,223 individuals.

	We now describe the key features of this dataset. 
	Let $A_0$ denote whether an individual obtains an HS diploma, and let $A_1$ indicate participation in the job training program. 
	We define $L_0$ as sex (a baseline covariate) and $L_1$ as pre-program earnings, which act as two time-varying confounders. 
	The initial treatment $A_0$ affects subsequent pre-program earnings $L_1$, and the assignment to the job training program $A_1$ may depend on $L_1$. 
	The instruments are $Z_0$, the number of high schools per square mile, and $Z_1$, a random assignment to job training program. 
	Our target outcome $Y$ indicates whether an individual's terminal earnings exceed the empirical median.

	We consider the dynamic treatment regime (DTR) $[g_0, g_1] \in \{0,1,\text{x}\} \times \{0,1,d^+,d^-\}$. 
	For DTR $g_0$, the value `0' assigns $A_0=0$, `1' assigns $A_0=1$, and `x' assigns the natural selection rule (i.e., the observed assignment). For the treatment rule $g_1$, the value `0' assigns $A_1=0$, `1' assigns $A_1=1$, `$d^+$' assigns $A_1=1$ only when $L_1$ is below the 80\% quantile, and `$d^-$' assigns $A_1=1$ only when $L_1$ is above the 80\% quantile.
	The target is to estimate $\mathbb{E}[Y(g_0(H_0),g_1(H_1))]$.
	We use the spline methods in the R package \texttt{mgcv} to estimate the nuisance parameters and calculate the bootstrapped estimated mean and standard deviation based on 1,000 replications.

	\begin{table}
		\centering
		\caption{
			Estimations for different types of DTRs are repeated 1,000 times using the bootstrap method. 
		}
		\label{tbl: real study}
		
		\begin{tabular}{c| cccc| cccc| cccc}
			\toprule
			DTRs 
			& \texttt{00} & \texttt{01} & \texttt{0$d^+$} & \texttt{0$d^-$} 
			& \texttt{10} & \texttt{11} & \texttt{1$d^+$} & \texttt{1$d^-$} 
			& \texttt{x0} & \texttt{x1} & \texttt{x$d^+$} & \texttt{x$d^-$} \\
			\midrule
			EST &.292 &.321 &.343 &.264 &.653 &.668 &.732 &.589 &.485 &.550 &.548 &.487 \\
			SE  &.124 &.089 &.094 &.120 &.155 &.122 &.132 &.147 &.017 &.012 &.013 &.016 \\
			SD  &.108 &.074 &.081 &.103 &.129 &.089 &.102 &.126 &.017 &.011 &.012 &.015 \\
			\bottomrule
		\end{tabular}
		
	\end{table}
	
	The results are demonstrated in Table~\ref{tbl: real study}. 
	The table presents estimated mean of mean potential terminal income (EST), mean estimated standard errors (SE), and empirical standard deviations (SD) for various DTRs. 
	Specifically, the DTRs are denoted by combinations of values in the set $\{0,1,\text{x}\}$ for the first treatment $g_0$ and $\{0,1,d^+,d^-\}$ for the second treatment $g_1$. 
	The treatment values and their respective effects are summarized for 12 distinct combinations.
	
	We observe that the SE and SD for \{\texttt{x0}, \texttt{x1}, \texttt{x$d^+$}, \texttt{x$d^-$}\} are smaller compared to the other DTRs. 
	This is likely due to the relatively weak correlation between \( Z_0 \) and \( A_0 \). 
	Furthermore, the SE and SD values for the same DTR are fairly consistent with each other, demonstrating the validity of the variance estimator in our algorithm.
	
	Considering the EST, the DTRs $\{\texttt{10}, \texttt{11}, \texttt{1d}^+, \texttt{1d}^-\}$ generally yield higher estimates than their counterparts $\{\texttt{00}, \texttt{01}, \texttt{0d}^+, \texttt{0d}^-\}$. The natural treatment rules $\{\texttt{x0}, \texttt{x1}, \texttt{xd}^+, \texttt{xd}^-\}$ produce estimates that lie between these two groups. This pattern suggests that obtaining an HS diploma has a positive effect on income.
	
	Meanwhile, the estimated average terminal income is higher for DTRs $\{\texttt{0d}^+, \texttt{1d}^+, \texttt{xd}^+\}$, which assign the program only to low-earning individuals, than for DTRs $\{\texttt{00}, \texttt{10}, \texttt{x0}\}$ or $\{\texttt{01}, \texttt{11}, \texttt{x1}\}$. In contrast, DTRs $\{\texttt{0d}^-, \texttt{1d}^-, \texttt{xd}^-\}$, which assign the program only to high-earning individuals, yield lower estimates. This pattern aligns with the findings of \citet{han2024optimal}, suggesting that restricting the job-training program to low-earners positively affects terminal income.
	
	% \section{Discussion}\label{sec:discussion}	
	% In this article, we develop a framework for identifying causal effects for multi-categorical or continuous IVs and treatment variables. We elucidate the connection between classical two-stage least square estimators and the identification of causal estimands under the IV framework. Our methods are further generalized to identify mean potential outcomes in longitudinal settings.
	% Furthermore, we analyze the efficiency, asymptotic normality, and asymptotic unbiasedness of the proposed estimator when different RWFs are employed to estimate the ATE.
	
	% Looking ahead, a promising direction is to identify other conditions analogous to additive IV or no unmeasured common effect modifier that guarantee the existence of a solution to the nonparametric IV problem in formula~\eqref{eq: npiv}. Additionally, accommodating right censoring and estimating counterfactual survival curves under a general IV framework constitute important directions for future research \citep{westling2024inference}. 
	% Furthermore, given the proposed identification result for continuous treatments, it would be of interest to propose a debiased learning approach by leveraging \cite{kennedy2017non}. 
	% Finally, identifying the optimal weighting function under more general settings within the proposed framework represents another avenue for further investigation.
	
	\section{Supplementary Material}
	The supplementary material contains the additional results on dynamic treatment regimes, multiplicative IVs, connection to handling continuous instruments via discretization, identification with continuous treatments, and proofs. 
	The codes of the simulation results and real data analysis are publicly available at \url{https://github.com/chensy123-sys/Additive-IV}.

	\putbib[bibfile_main]
    \end{bibunit}

	\clearpage
    \begin{bibunit}

	\appendix
    \section*{Supplementary Material}
	\section{Additional extensions under longitudinal setting}
	\subsection{Dynamic treatment regime}
	In this subsection, we illustrate the identification strategy under a dynamic treatment regime. 
	Let $\overline{g} := [g_0, \ldots, g_T]$ denote a sequence of deterministic DTRs, where each mapping satisfies $g_t: \mathcal{H}_t \rightarrow \mathcal{A}_t$. 
	For $s = 0, \ldots, T$, define the potential outcome under regime $\overline{g}$ as
	\[
	Y(\overline{g}) := Y\{g_0(H_0), \ldots, g_T(H_T)\},
	\]
	that is, the outcome observed when the subject follows the treatment rule $A_t = g_t(H_t)$ at each time $t$. 
	Similarly, define
	\[
	Y(\overline{g}_s) := Y(A_0, \ldots, A_{s-1}, g_s(H_s), \ldots, g_T(H_T))
	\]
	as the potential outcome when the subject follows the rule $A_t = g_t(H_t)$ for all $t \ge s$. 
	
	Without loss of generality, we assume that $\pi_t(Z_t, H_t) = Z_t$, which is taken to be an RWF.
	Now we illustrate the nuisance functions used in the dynamic treatment setting.
	Define \(\gamma_{T+1,\underline{g}_{T+1}}^o(H_{T+1}) := Y\). 
	For \(t = T, \ldots, 0\), recursively define the nuisance functions:
	\begin{align*}
		&\delta_{t,g_t}^o(H_t) := \mathbb{E}[I\{A_t=g_t(H_t)\} \mid H_t], \\
		& \eta_{t,\underline{g}_t}^o(H_t) := \mathbb{E}\bigl[I\{A_t=g_t(H_t)\} \gamma_{t+1,\underline{a}_{t+1}}^o(H_{t+1}) \mid H_t\bigr],\\
		&\rho_t^o(H_t) := \mathbb{E}[Z_t \mid H_t], \\
		&\kappa_{t,g_t}^o(H_t) := \mathrm{Cov}\!\{I\{A_t=g_t(H_t)\}, Z_t \mid H_t\}, \\
		& \gamma_{t,\underline{g}_t}^o(H_t) := \frac{\mathrm{Cov}\!\{ I\{A_t=g_t(H_t)\} \gamma_{t+1,\underline{g}_{t+1}}^o(H_{t+1}), Z_t \mid H_t \}}{\mathrm{Cov}\!\{ I\{A_t=g_t(H_t)\}, Z_t \mid H_t \}}.
	\end{align*}
	Denote the nuisance vector as
	\begin{align*}
		\alpha_{\overline{g}}^o := \{\alpha_{t,\underline{g}_t}^o\}_{t=0}^T, \qquad
		\alpha_{t,\underline{g}_t}^o := [
		\delta_{t,g_t}^o,
		\kappa_{t,g_t}^o,
		\rho_t^o,
		\eta_{t,\underline{g}_t}^o,
		\gamma_{t,\underline{g}_t}^o].
	\end{align*}

	Then the identification strategy is derived in the next theorem.
	\begin{theorem}\label{thm: longitudinal AIV DTR}
		Under Assumptions~\ref{as: consistency'}--\ref{as: IV independence'}, let \(0 \le s \le T+1\), \(r \ge 0\), and \(s + r \le T+1\). Suppose that, for each \(t=0,\ldots,T\), \(Z_t\) serves as an AIV for \(A_t\), and that \(\pi_t(Z_t, H_t)\) is an RWF for \(A_t\). Then,
		\begin{align*}
			\mathbb{E}\left[Y(\underline{g}_{s})\right]=
			\mathbb{E}\left[
			\prod_{t=s}^{T-r}
			\dfrac{\left(Z_t-\rho_t^o(H_t)\right)I\{A_t=g_t(H_t)\}}{
				\kappa_{t,g_t}^o(H_t)
			}
			\gamma_{T+1-r,\underline{g}_{T+1-r}}^o(H_{T+1-r})
			\right].
		\end{align*}
	\end{theorem}
	
	The proof for Theorem~\ref{thm: longitudinal AIV DTR} is similar to that of Theorem~\ref{thm: longitudinal AIV}, thus omitted.
	Set $s=0$ and $r=0$, we know that the mean potential outcomes can be identified as
	
	\begin{align*}
		\mathbb{E}\bigl[Y(\overline{g})\bigr]=\psi_{\overline{g}}^o:=
		\mathbb{E}\left[
		\prod_{t=0}^{T}
		\dfrac{\left(Z_t-\rho_t^o(H_t)\right)I\{A_t=g_t(H_t)\}}
		{
			\kappa_{t,g_t}^o(H_t)
		}Y
		\right].
	\end{align*}
	Next, we derive the EIF for $\psi_{\overline{g}}$.
	\begin{theorem}\label{thm: EIF AIV longitudinal DTR}
		Under Assumptions~\ref{as: consistency'}--\ref{as: IV independence'}, suppose that for each \(t=0,\ldots,T\),  \(\pi_t(Z_t, H_t)=Z_t\) is an RWF for \(A_t\). Then, if we define $A_t^{(g_t)}:=I\{A_t = g_t(H_t)\}$, the EIF for $\psi_{\overline{g}}^o$ is $\varphi_{\overline{g}}(O;\psi_{\overline{g}}^o,\alpha_{\overline{g}}^o)$, where 
		\begin{align*}
			&\varphi_{\overline{g}}(O;\psi_{\overline{g}},\alpha_{\overline{g}}):=
			\prod_{t=0}^{T}
			\dfrac{\left\{Z_t-\rho_t(H_t)\right\}A_t^{(g_t)}}
			{\kappa_{t,g_t}(H_t)}
			Y-\psi_{\overline{g}}
			+\sum_{t=0}^T\left(\displaystyle\prod_{s=0}^{t-1}\dfrac{\left\{Z_s-\rho_s(H_s)\right\}A_s^{(g_s)}}
			{\kappa_{s,g_s}(H_s)}\right)
			\\&\times \left\{
			\left(1-\dfrac{\{Z_t-\rho_t(H_t)\}\{A_t^{(g_t)}-\delta_{t,g_t}(H_t)\}}{\kappa_{t,g_t}(H_t)} \right)\gamma_{t,\underline{g}_t}(H_t) - 
			\dfrac{(Z_t-\rho_t(H_t))\times\eta_{t,\underline{g}_t}(H_t)}{\kappa_{t,g_t}(H_t)}\right\}.
		\end{align*}
	\end{theorem}
	
	For estimation, we can replace $A_t^{(a_t)}$ with $A_{t}^{(g_t)}=I(A_t=g_t(H_t))$, and $\underline{a}_t$ with $\underline{g}_t$ in Algorithm~\ref{alg: longitudinal}, and construct the corresponding backward cross-fitting algorithm for dynamic treatment regimes.
	Concretely, Algorithm~\ref{alg: longitudinal DTR} can be viewed as a generalization of Algorithm~\ref{alg: longitudinal} to estimating the mean potential outcomes under the DTRs $\overline{g}$.
	An analogous version of Theorem~\ref{thm: asymptotic normality longitudinal} can be derived for potential outcome means $\hat\psi_{\overline{g}}^{(n)}$ in Algorithm~\ref{alg: longitudinal DTR}, deriving the asymptotic consistency and normality of our proposed estimator.
	
	\begin{theorem}[Asymptotic normality of $\hat\psi_{\overline{g}}^{(n)}$ in Algorithm~\ref{alg: longitudinal DTR}]\label{thm: asymptotic normality longitudinal DTR}
		Under Assumptions~\ref{as: consistency'}--\ref{as: strong IV relevance'}, suppose that for each \(t = 0, \ldots, T\), \(Z_t\) serves as an AIV for \(A_t\). 
		Further, for each \(t = 0, \ldots, T\) and \(k = 1, \ldots, K\), suppose that the following rate condition holds for the nuisance functions $\hat\alpha_{t,\overline{g}}^{(n,k)}$ defined in Algorithm~\ref{alg: longitudinal DTR}:
		\begin{align*}
			\left\{\begin{array}{l}
				\|\hat\kappa_{t,g_t}^{(n,k)}- \kappa_{t,g_t}^{o}\|_2
				\times\|\hat\gamma_{t,\underline{g}_t}^{(n,k)}-\gamma_{t,\underline{g}_t}^o\|_2\\
				+\|\hat\rho_t^{(n,k)} - \rho_t^o\|_2\times
				\|\hat\eta_{t,\underline{g}_t}^{(n,k)}-\eta_{t,\underline{g}_t}^o\|_2\\
				+\|\hat\rho_t^{(n,k)} - \rho_t^o\|_2\times
				\|\hat\delta_{t,g_t}^{(n,k)}- \delta_{t,g_t}^{o}\|_2
			\end{array}\right\}=o_p(n^{-1/2}).
		\end{align*}
		Furthermore, assume that for any fixed $k$ and time $t$,
		\(\mathbb{E}[\|\hat\alpha_{t,\overline{g}}^{(n,k)}- \alpha_{t,\overline{g}}^o\|_2^2]=o(1)\).
		Then $\sqrt{n}\{\hat\psi_{\overline{g}}^{(n)}-\psi_{\overline{g}}^o\}/\sigma_{\overline{g}}^o$ converges in distribution to $\mathcal{N}(0,1)$, where $\hat\psi_{\overline{g}}^{(n)}$ is defined in Equation~\eqref{eq: AUG estimator longitudinal DTR}, and $(\sigma_{\overline{g}}^o)^2:=\mathbb{E}[\varphi_{\overline{g}}(O;\psi_{\overline{g}}^o,\alpha_{\overline{g}}^o)^2]$.
		In addition, $\hat \sigma_{\overline{g}}^{(n)}$ converges in probability to $\sigma_{\overline{g}}^o$.
	\end{theorem}
	The proofs for the two theorems above are omitted, since their proofs are similar to those of Theorems~\ref{thm: asymptotic normality known weight}~and~\ref{thm: EIF AIV longitudinal}.

	\subsection{Adaptive choosing weighting function}
	In Section~\ref{sec: semiparametric theory}, we discuss that one can adaptively choose the weight by setting $\pi_*^o(Z,L)=\mathbb{E}[A\mid Z,L]$ when estimating the ATE of interest. In this subsection, we generalize this strategy to the longitudinal data. 
	First, we define the nuisance functions as
	\begin{align*}
		&\pi_{t,a_t}^o(Z_t, H_t):=\mathbb{E}[A_t^{(a_t)}\mid Z_t, H_t],\\
		&\delta_{t,a_t}^o(H_t):=\mathbb{E}[A_t^{(a_t)}\mid H_t],\\
		&\kappa_{t,a_t}^o(H_t):=\mathrm{Var}\!\{\pi_{t,a_t}^o(Z_t,H_t)\mid H_t\}, \\
		&\xi_{t,\underline{a}_t}^o(Z_t,H_t):=\mathbb{E}[A_t^{(a_t)}\gamma_{t+1,\underline{a}_{t+1}}^o(H_{t+1})\mid Z_t,H_t], \\
		&\eta_{t,\underline{a}_t}^o(H_t):=\mathbb{E}[A_t^{(a_t)}\gamma_{t+1,\underline{a}_{t+1}}^o(H_{t+1})\mid H_t],\\
		&\gamma_{t,\underline{a}_t}^o(H_t):=\dfrac{\mathrm{Cov}\!\{A_t^{(a_t)}\gamma_{t+1,\underline{a}_{t+1}}^o(H_{t+1}), \pi_{t,a_t}^o(Z_t,H_t) \mid H_t\}}{\mathrm{Var}\!\{ \pi_{t,a_t}^o(Z_t,H_t) \mid H_t\}}.
	\end{align*}
	Define the nuisance vector as
	\begin{align*}
		\beta_{\overline{a}} := \{\beta_{t,\underline{a}_t}\}_{t=0}^T,\quad\beta_{t,\underline{a}_t} := [\pi_{t,a_t}, \delta_{t,a_t}, \kappa_{t,a_t}, \xi_{t,\underline{a}_t}, 
		\eta_{t,\underline{a}_t}, \gamma_{t,\underline{a}_t}].
	\end{align*}
	
	\begin{algorithm}[h]
		\caption{Backward Cross-Fitting Procedure for DTRs}
		\label{alg: longitudinal DTR}
		
		Randomly divide the samples evenly into \(K\) folds \(\{I_k\}_{k=1}^K\)\;
		
		\For{$k = 1$ \KwTo $K$}{
			Initialize \(t \gets T\) and set \(\hat\Psi_{T+1,\underline{g}_{T+1}}^{(n,k)} \gets Y\)\;
			
			\While{$t \ge 0$}{
				Fit the nuisance functions \(\hat \alpha_{t,\overline{a}}^{(n,k)}(H_t)\) using samples in \(I_{-k}\):
				\begin{align*}
					&\hat\eta_{t,\underline{g}_t}^{(n,k)}(H_t) := \hat{\mathbb{E}}^{(n,k)}[A_t^{(g_t)} \hat\Psi_{t+1,\underline{g}_{t+1}}^{(n,k)} \mid H_t], &
					\hat \delta_{t,g_t}^{(n,k)}(H_t) := \hat{\mathbb{E}}^{(n,k)}[A_t^{(g_t)} \mid H_t],\\
					&\hat\gamma_{t,\underline{g}_t}^{(n,k)}(H_t) := 
					\frac{\widehat{\mathrm{Cov}}^{(n,k)}(A_t^{(g_t)} \hat\Psi_{t+1,\underline{g}_{t+1}}^{(n,k)}, Z_t \mid H_t)}
					{\widehat{\mathrm{Cov}}^{(n,k)}(A_t^{(g_t)}, Z_t \mid H_t)}, &
					\hat \rho_t^{(n,k)}(H_t) := \hat{\mathbb{E}}^{(n,k)}[Z_t \mid H_t],\\
					&\hat\kappa_{t,g_t}^{(n,k)}(H_t) := \widehat{\mathrm{Cov}}^{(n,k)}(A_t^{(g_t)},Z_t \mid H_t).
				\end{align*}
				
				Update for all samples:
				\begin{align*}
					\hat\Psi_{t,\underline{g}_{t}}^{(n,k)} :=&
					\frac{1}{\hat\kappa_{t,g_t}^{(n,k)}(H_t)}
					(Z_t - \hat\rho_t^{(n,k)}(H_t))
					(A_t^{(g_t)} \hat\Psi_{t+1,\underline{g}_{t+1}}^{(n,k)} - \hat\eta_{t,\underline{g}_t}^{(n,k)}(H_t))
					+\\&
					\left(1 - 
					\frac{(Z_t - \hat\rho_t^{(n,k)}(H_t))(A_t^{(g_t)} - \hat\delta_{t,g_t}^{(n,k)}(H_t))}{\hat\kappa_{t,g_t}^{(n,k)}(H_t)}\right)
					\hat\gamma_{t,\underline{g}_t}^{(n,k)}(H_t).
				\end{align*}
				
				Decrement \(t \gets t-1\)\;
			}
		}
		
		Output the estimator and its variance:
		\begin{equation}\label{eq: AUG estimator longitudinal DTR}
			\hat\psi_{\overline{g}}^{(n)} := \frac{1}{K} \sum_{k=1}^K \mathbb{E}_{nk}[\hat\Psi_{0,\overline{g}}^{(n,k)}],\quad
			(\hat\sigma_{\overline{g}}^{(n)})^2 := \frac{1}{K} \sum_{k=1}^K \mathbb{E}_{nk} \Big[(\hat\Psi_{0,\overline{g}}^{(n,k)} - \mathbb{E}_{nk}[\hat\Psi_{0,\overline{g}}^{(n,k)}])^2 \Big].
		\end{equation}
		
	\end{algorithm}

	\begin{algorithm}[htbp]
		\caption{Backward Cross-Fitting Procedure under the Adaptive Weighting Scenario}
		\label{alg: longitudinal'}
		
		Randomly divide the samples evenly into \(K\) folds \(\{I_k\}_{k=1}^K\)\;
		
		\For{$k = 1$ \KwTo $K$}{
			Initialize \(t \gets T\) and set \(\hat\Psi_{T+1,\underline{a}_{T+1}}^{(n,k)} \gets Y\)\;
			
			\While{$t \ge 0$}{
				Fit the nuisance functions \(\hat \alpha_{t,\overline{a}}^{(n,k)}(H_t)\) using samples in \(I_{-k}\):
				\[
				\begin{split}
					\hat\pi_{t,a_t}^{(n,k)}(Z_t, H_t) &:= \hat{\mathbb{E}}^{(n,k)}\big[A_t^{(a_t)} \mid Z_t, H_t\big],\\
					\hat\delta_{t,a_t}^{(n,k)}(H_t) &:= \hat{\mathbb{E}}^{(n,k)}\big[A_t^{(a_t)} \mid H_t\big],\\
					\hat\kappa_{t,a_t}^{(n,k)}(H_t) &:= \hat{\mathbb{E}}^{(n,k)}\big[(\hat\pi_{t,a_t}^{(n,k)}(Z_t,H_t)-\hat\delta_{t,a_t}^{(n,k)}(H_t))^2 \mid H_t\big],\\
					\hat\xi_{t,\underline{a}_t}^{(n,k)}(Z_t,H_t) &:= \hat{\mathbb{E}}^{(n,k)}[A_t^{(a_t)} \hat\Psi_{t+1,\underline{a}_{t+1}}^{(n,k)} \mid Z_t,H_t],\\
					\hat\eta_{t,\underline{a}_t}^{(n,k)}(H_t) &:= \hat{\mathbb{E}}^{(n,k)}[A_t^{(a_t)} \hat\Psi_{t+1,\underline{a}_{t+1}}^{(n,k)} \mid H_t],\\
					\hat\gamma_{t,\underline{a}_t}^{(n,k)}(H_t) &:= 
					\frac{\widehat{\mathrm{Cov}}^{(n,k)}\{A_t^{(a_t)}\hat\Psi_{t+1,\underline{a}_{t+1}}^{(n,k)}, \hat\pi_{t,a_t}^{(n,k)}(Z_t,H_t) \mid H_t\}}
					{\hat\kappa_{t,a_t}^{(n,k)}(H_t)}.
				\end{split}
				\]
				
				Update for all samples:
				\[
				\begin{split}
					\hat\Psi_{t,\underline{a}_{t}}^{(n,k)} :=&
					\frac{\hat\pi_{t,a_t}^{(n,k)}(Z_t,H_t) - \hat\delta_{t,a_t}^{(n,k)}(H_t)}{\hat\kappa_{t,a_t}^{(n,k)}(H_t)}
					A_t^{(a_t)}\hat\Psi_{t+1,\underline{a}_{t+1}}^{(n,k)}\\
					&+\frac{A_t^{(a_t)} - \hat\pi_{t,a_t}^{(n,k)}(Z_t,H_t)}{\hat\kappa_{t,a_t}^{(n,k)}(H_t)}
					\hat\xi_{t,\underline{a}_t}^{(n,k)}(Z_t,H_t)\\
					&-\frac{A_t^{(a_t)} - \hat\delta_{t,a_t}^{(n,k)}(H_t)}{\hat\kappa_{t,a_t}^{(n,k)}(H_t)}
					\hat\eta_{t,\underline{a}_t}^{(n,k)}(H_t)\\
					&+\left(1+\frac{\{A_t^{(a_t)} - \hat\pi_{t,a_t}^{(n,k)}(Z_t,H_t)\}^2-\{A_t^{(a_t)} - \hat\delta_{t,a_t}^{(n,k)}(H_t)\}^2}
					{\hat\kappa_{t,a_t}^{(n,k)}(H_t)} \right)
					\hat\gamma_{t,\underline{a}_t}^{(n,k)}(H_t).
				\end{split}
				\]
				
				Decrement \(t \gets t-1\)\;
			}
		}
		
		Output the estimator and its variance:
		
		\begin{equation}\label{eq: AUG estimator longitudinal'}
			\hat\psi_{\overline{a},*}^{(n)} := \frac{1}{K} \sum_{k=1}^K \mathbb{E}_{nk}[\hat\Psi_{0,\overline{a}}^{(n,k)}],\quad
			(\hat\sigma_{\overline{a},*}^{(n)})^2 := \frac{1}{K} \sum_{k=1}^K \mathbb{E}_{nk} \Big[(\hat\Psi_{0,\overline{a}}^{(n,k)} - \mathbb{E}_{nk}[\hat\Psi_{0,\overline{a}}^{(n,k)}])^2 \Big].
		\end{equation}
	\end{algorithm}

	Following the same logic as in Theorem~\ref{thm: longitudinal AIV}, we derive the following proposition, which enables identification using the adaptively selected weights. 
	\begin{theorem}\label{thm: longitudinal AIV'}
		Under Assumptions~\ref{as: consistency'}--\ref{as: strong IV relevance'}, let \(0 \le s \le T+1\), \(r \ge 0\), and \(s + r \le T+1\). Suppose that, for each \(t=0,\ldots,T\), \(Z_t\) serves as an AIV for \(A_t\). Then, the mean potential outcomes \(\mathbb{E}\bigl[Y(\underline{a}_{s})\bigr]\) can be expressed as
		\begin{align*}
			\mathbb{E}\left[
			\left(\prod_{t=s}^{T-r}
			\frac{\pi_{t,a_t}^o(Z_t,H_t) - \delta_{t,a_t}^o(H_t)}
			{\mathrm{Var}\!\{\pi_{t,a_t}^o(Z_t,H_t) \mid H_t \}}A_t^{(a_t)}\right)
			\times \gamma_{T+1-r,\underline{a}_{T+1-r}}^o(H_{T+1-r})
			\right].
		\end{align*}
	\end{theorem}
	The proof for this theorem is similar to that of Theorem~\ref{thm: longitudinal AIV}, thus omitted.
	Specifically, one can identify the mean of potential outcomes $\mathbb{E}[Y(\overline{a})]$ by
	\begin{align}
		\psi_{\overline{a},*}^o :=
		\mathbb{E}\left[
		\prod_{t=0}^{T}
		\frac{\pi_{t,a_t}^o(Z_t,H_t) - \delta_{t,a_t}^o(H_t)}
		{\mathrm{Var}\!\{\pi_{t,a_t}^o(Z_t,H_t) \mid H_t \}}A_t^{(a_t)}
		\times Y
		\right]. \label{eq: identification AIV longitudinal' 2}
	\end{align}
	The next theorem derives the EIF for $\psi_{\overline{a},*}^o$.
	\begin{theorem}\label{thm: EIF AIV longitudinal'}
		Under Assumptions~\ref{as: consistency'}--\ref{as: strong IV relevance'}, suppose that for each \(t=0,\ldots,T\), \(Z_t\) serves as an AIV for \(A_t\). Then, the EIF for $\psi_{\overline{a},*}^o$ is $\varphi_{\overline{a},*}(O;\psi_{\overline{a},*}^o,\beta_{\overline{a}}^o)$, where 
		\begin{align*}
			&\varphi_{\overline{a},*}(O;\psi_{\overline{a},*},\beta_{\overline{a}}):=
			\prod_{t=0}^{T}
			\frac{\pi_{t,a_t}(Z_t,H_t) - \delta_{t,a_t}(H_t)}
			{\kappa_{t,a_t}(H_t)}A_t^{(a_t)}
			\times Y - \psi_{\overline{a},*}\\
			&+\sum_{t=0}^T\left(\displaystyle\prod_{s=0}^{t-1}
			\dfrac{\pi_{s,a_s}(Z_s,H_s) - \delta_{s,a_s}(H_s)}
			{\kappa_{s,a_s}(H_s)}A_s^{(a_s)}\right)\times \dfrac{1}{\kappa_{t,a_t}(H_t)}\\
			&\times
			\left\{\begin{array}{l}
				\{A_t^{(a_t)} - \pi_{t,a_t}(Z_t,H_t)\}
				\xi_{t,\underline{a}_t}(Z_t,H_t)
				-\{A_t^{(a_t)} - \delta_{t,a_t}(H_t)\}
				\eta_{t,\underline{a}_t}(H_t)\\
				+\gamma_{t,\underline{a}_t}(H_t)
				\times\left(\kappa_t(H_t)+\{A_t^{(a_t)} - \pi_{t,a_t}(Z_t,H_t)\}^2-\{A_t^{(a_t)} - \delta_{t,a_t}(H_t)\}^2\right)
			\end{array}\right\}.
		\end{align*}
	\end{theorem}
	Algorithm~\ref{alg: longitudinal'} demonstrates the backward cross-fitting procedure for training a debiased estimator for $\psi_{\overline{a},*}^o$, which is the solution to
	\begin{align*}
		\dfrac{1}{n}\sum_{k=1}^K\sum_{i\in I_k}\varphi_{\overline{a},*}(O_i;\hat\psi_{\overline{a},*}^{(n)},\hat\beta_{\overline{a}}^{(n,k)})=0.
	\end{align*}
	Finally, we establish the asymptotic normality of $\hat\psi_{\overline{a},*}^{(n)}$ in Equation~\eqref{eq: identification AIV longitudinal' 2}.
	\begin{theorem}[Asymptotic normality of $\hat\psi_{\overline{a},*}^{(n)}$ in Algorithm~\ref{alg: longitudinal'}]\label{thm: asymptotic normality unknown weight longitudinal setting}
		Under Assumptions~\ref{as: consistency'}--\ref{as: strong IV relevance'}, suppose that $Z_t$ is an AIV for $A_t$. Assume that for any $k=1,\ldots,K$ and $t=0,\ldots,T$, $\mathbb{E}[\|\hat \beta_{t,\underline{a}_t}^{(n,k)}-\beta_{t,\underline{a}_t}^o\|_2^2]=o(1)$, and that
		\begin{align*}
			\left\{
			\begin{array}{l}
				\|\hat\gamma_{t,\underline{a}_t}^{(n,k)}-\gamma_{t,\underline{a}_t}^o\|_2 \times \|\hat\kappa_{t,a_t}^{(n,k)}-\kappa_{t,a_t}^o\|_2\\
				+\|\hat\delta_{t,a_t}^{(n,k)}-\delta_{t,a_t}^o\|_2^2+\|\hat\pi_{t,a_t}^{(n,k)}-\pi_{t,a_t}^o\|_2^2\\
				+\|\hat\xi_{t,\underline{a}_t}^{(n,k)}-\xi_{t,\underline{a}_t}^o\|_2\times \|\hat\pi_{t,a_t}^{(n,k)}-\pi_{t,a_t}^o\|_2\\
				+\|\hat\eta_{t,\underline{a}_t}^{(n,k)}-\eta_{t,\underline{a}_t}^o\|_2\times \|\hat\delta_{t,a_t}^{(n,k)}-\delta_{t,a_t}^o\|_2
			\end{array}\right\}
			=o_p(n^{-1/2}).
		\end{align*}
		Then $\sqrt{n}\left(\hat\psi_{\overline{a},*}^{(n)}-\psi_{\overline{a},*}^o\right)/\sigma_{\overline{a},*}^o$ converges in distribution to $\mathcal{N}(0,1)$, where the asymptotic variance is defined as $(\sigma_{\overline{a},*}^o)^2:=\mathbb{E}[\varphi(O;\psi_{\overline{a},*}^o,\beta^o)^2]$. In addition, the variance estimator $(\hat\sigma_{\overline{a},*}^{(n)})^2$ in Equation~\eqref{eq: AUG estimator longitudinal'} converges in probability to $(\sigma_{\overline{a},*}^o)^2$.
	\end{theorem}
	The proof for this theorem is omitted, since its similar to the proof of Theorem~\ref{thm: asymptotic normality known weight}.

	\section{Additional extensions under point-exposure setting}\label{sec: additional extension}
	\subsection{Multiplicative instrumental variable}
	The AIV condition is relatively strong and may fail to hold in some empirical settings. As an alternative, we consider an identification strategy based on a natural generalization of the multiplicative IV framework for binary instruments, originally proposed by \citet{liu2025multiplicativeinstrumentalvariablemodel}. This generalized framework relaxes the strict additive structure required by the AIV condition, while preserving the essential exclusion and relevance properties of a valid instrument. It highlights that the AIV condition is not the only means to ensure the well-posedness of the nonparametric IV problem in Equation~\eqref{eq: npiv}.
	
	\begin{definition}[Multiplicative IV]\label{defn: multiplicative IV}
		For each $a\in\mathcal{A}$, we say that $Z$ is a \emph{multiplicative IV} (MIV) for $A = a$ if there exist functions $b(U,L)$ and $c(Z,L)$ such that
		\begin{equation*}
			\Pr(A\neq a\mid Z,U,L) = b(U,L)\cdot c(Z,L).
		\end{equation*}
	\end{definition}
	
	The MIV condition implies that, conditional on the observed confounders $L$, the instrument–treatment association on the multiplicative scale is unaffected by unmeasured confounding, effectively ruling out any $U$--$Z$ interaction. Accordingly, it relies on the analyst’s ability to observe and adjust for a sufficiently rich set of covariates to ensure that the instrument's effect on treatment remains stable across levels of the hidden confounder. The following proposition presents the resulting identification strategy under an MIV.
	
	\begin{theorem}[MIV identification]\label{thm: MIV identification}
		Under Assumptions~\ref{as: consistency}--\ref{as: IV relevance}, for \( a\in\mathcal{A} \), assume that $Z$ is an MIV for $A=a$, and that $\pi(Z,L)$ is an RWF. Then there exists a unique solution \( f_a^o(A^{(a)}, L) \) to Equation~\eqref{eq: npiv}, which has explicit form
		\begin{equation*}
			\begin{aligned}
				&f_a^o(0,L)=\mathbb{E}[Y(a)\mid L]-\mathbb{E}[Y(a)\mid L,A\neq a],\\
				&f_a^o(1,L)=\mathbb{E}[Y(a)\mid L].
			\end{aligned}
		\end{equation*}
		In particular, for any regular $\pi(Z,L)$ for $A=a$, it holds that
		\begin{align*}
			\mathbb{E}[Y(a)]=\psi_{a,MIV}^o:=\mathbb{E}\left[(1-A^{(a)})\dfrac{\mathrm{Cov}\!\{A^{(a)}Y, \pi(Z,L) \mid L\}}{\mathrm{Cov}\!\{A^{(a)}, \pi(Z,L) \mid L\}}+A^{(a)}Y\right].
		\end{align*}
	\end{theorem}
	
	\begin{example}[Binary IV and treatment]
		Without loss of generality, set $\pi(Z,L)=Z$.
		Under Assumptions~\ref{as: consistency}--\ref{as: IV relevance}, if $Z$ is an MIV for $A$, one can recover a variation of the results in \citet{liu2025multiplicativeinstrumentalvariablemodel}. Specifically, Theorem~\ref{thm: MIV identification} asserts that
		\begin{align*}
			&\mathbb{E}[Y(1)]
			=\mathbb{E}\left[AY+(1-A)\frac{\mathbb{E}[AY\mid Z=1,L] - \mathbb{E}[AY\mid Z=0,L]}{\mathbb{E}[A\mid Z=1,L] - \mathbb{E}[A\mid Z=0,L]}\right].
		\end{align*}
	\end{example}
	At this stage, one might regard the MIV condition as a viable alternative to the AIV condition.  
	However, as we demonstrate in the subsequent subsections, the AIV condition possesses several desirable properties that are not, in general, ensured under the MIV condition.  
	This highlights the distinctive advantages of the AIV condition.
	
	Without loss of generality, we just take $\pi(Z,L)=Z$.
	Denote the nuisance functions as 
	\begin{align*}
		&\delta_a^o(L) := \mathbb{E}[A^{(a)} \mid L], 
		&& \eta_a^o(L) := \mathbb{E}[A^{(a)}Y \mid L],\\
		&\kappa_{a}^o(L) := \mathrm{Cov}\!\{A^{(a)}, Z \mid L\}, 
		&& \zeta_{a}^o(L) := \mathbb{E}[A^{(a)}Y Z \mid L],\\
		&\rho^o(L) := \mathbb{E}[Z \mid L], 
		&& \gamma_{a}^o(L) := \dfrac{\mathrm{Cov}\!\{A^{(a)}Y, Z \mid L\}}{\mathrm{Cov}\!\{A^{(a)}, Z \mid L\}}.
	\end{align*}
	We can unify these nuisance functions into a nuisance vector as
	\begin{align*}
		\alpha_{a,MIV}^o = [\delta_a^o,\eta_a^o,\kappa_{a}^o, \zeta_{a}^o,\rho^o,\gamma_{a}^o].
	\end{align*}
	Next, we derive the EIF for $\psi_{a,MIV}^o$.
	\begin{theorem}\label{thm: EIF for MIV}
		Under Assumptions~\ref{as: consistency}--\ref{as: IV relevance}, for \( a\in\mathcal{A} \), assume that $\pi(Z,L)=Z$ is an RWF for $A=a$. Then, the EIF for $\psi_{a,MIV}^o$ is
		\begin{align*}
			\varphi_{a,MIV}(O;\psi_{a,MIV}^o,\alpha_{a,MIV}^o)=&
			(1-A^{(a)})\gamma_{a}^o(L)+A^{(a)}Y-\psi_{a,MIV}^o\\
			&+\dfrac{(1-\delta_a^o(L))}
			{\kappa_{a}^o(L)}
			\left(A^{(a)}Y-\eta_a^o(L)\right)(Z-\rho^o(L))\\
			&-\dfrac{(1-\delta_a^o(L))}{\kappa_a^o(L)}\gamma_a^o(L) 
			\left(A^{(a)}-\delta_a^o(L)\right)(Z - \rho^o(L)).
		\end{align*}
	\end{theorem}

	\subsection{Marginal structural model}
	As an extension of Theorem~\ref{thm: AIV identification}, we identify the parameter of interest in the parametric marginal structural mean model, specified for each $a \in \mathcal{A}$,
	\begin{equation}\label{eq: MSM}
		\mathbb{E}[Y(a) \mid V] = g(a, V; \psi_{MSM}^o),
	\end{equation}
	where \( g(a, V; \psi_{MSM}^o) \) is a known function, \( V \) is a subset of the observed confounders \( L \), and \( \psi_{MSM}^o \in \mathbb{R}^q \) is the finite-dimensional parameter of interest.
	This type of model has been extensively studied in \citet{robins2000marginal, hernan2001marginal,tchetgen2018marginal,michael2024instrumental}.
	Specifically, for any RWF $\pi(Z,L)$ of $A$, we denote the propensity score function as
	\begin{align*}
		\omega_{\pi}^o(a,Z,L):=\dfrac{\pi(Z,L)-\mathbb{E}[\pi(Z,L)\mid L]}{\mathrm{Cov}\!\{I\{A=a\},\pi(Z,L)\mid L\}}.
	\end{align*}	
	\begin{theorem}\label{thm: MSM identification}
		Under Assumptions~\ref{as: consistency}--\ref{as: IV relevance}, suppose that \( Z \) is an AIV for \( A \), and that the potential outcome \( Y(a) \) follows the parametric marginal structural model in Equation~\eqref{eq: MSM}.  
		Then, for any RWF \( \pi(Z, L) \) of \( A \), the parameter \( \psi_{MSM}^o \in\mathbb{R}^q\) satisfies
		\begin{equation*}
			\mathbb{E}\left[\omega_{\pi}^o(A, Z, L) \left\{Y - g(A, V; \psi_{MSM}^o)\right\}\mid A,V\right] = 0.
		\end{equation*}
	\end{theorem}
	Notably, when \( Z \) is binary, Theorem~\ref{thm: MSM identification} reduces to the identification result in Theorem~\ref{thm: AIV identification}. 
	Now, one may construct an estimator for \( \psi_{MSM}^o \) using the GMM framework. A detailed exploration of such estimation procedures is beyond the scope of this article.

	\subsection{Connection to handling continuous instruments via discretization}
	Rather than directly identifying the causal effect as in Theorem~\ref{thm: AIV identification}, 
	some studies adopt the practice of dichotomizing a continuous instrument $Z$ into a binary IV \citep{baiocchi2014instrumental,cui2023instrumental}. 
	A key observation is that if the original continuous $Z$ satisfies the AIV condition for $A$, 
	then its discretized counterpart also satisfies the AIV condition. 
	The following proposition formalizes this result, providing theoretical guarantees for identification based on discretized IVs and thereby enabling valid inference via the bounded IV approach of \citet{wang2018bounded}.
	
	\begin{theorem}[Discretized AIV identification]\label{thm: cut IV}
		Let \( \mathcal{S}=\{S_1, \ldots, S_M\} \) be an arbitrary partition of $\mathcal{Z}$ such that $\Pr(Z\in S_M)>0$. 
		For $m=1,\ldots,M$, define \( Z_{\mathcal{S}} := m \) if \( Z \in S_m \). 
		Assume that for any $a\in\mathcal{A}$ and $l\in \mathcal{L}$, there exist two distinct values $z_1,z_2$ such that
		\begin{equation}\label{eq: IV relevance discrete}
			\Pr(A=a\mid Z_{\mathcal{S}}=z_1,L=l) \neq \Pr(A=a\mid Z_{\mathcal{S}}=z_2,L=l).
		\end{equation}
		Under Assumptions~\ref{as: consistency}--\ref{as: IV independence}, if $Z$ is an AIV for $A$, the nonparametric IV equation
		\begin{equation*}
			\mathbb{E}[A^{(a)}Y\mid Z_{\mathcal{S}},L]=\mathbb{E}[f_{a,\mathcal{S}}^o(A^{(a)},L)\mid Z_{\mathcal{S}},L]
		\end{equation*}
		has the unique solution $f_{a,\mathcal{S}}^o(A^{(a)},L)$ given by
		\begin{align*}
			f_{a,\mathcal{S}}^o(0,L)=&\ \mathrm{Cov}\!\{\mathbb{E}[Y(a)\mid U,L], \Pr(A=a\mid Z_{\mathcal{S}},U,L)\mid L\},\\
			f_{a,\mathcal{S}}^o(1,L)=&\ \mathrm{Cov}\!\{\mathbb{E}[Y(a)\mid U,L], \Pr(A=a\mid Z_{\mathcal{S}},U,L)\mid L\}
			+\mathbb{E}[Y(a)\mid L].
		\end{align*}
	\end{theorem}
	
	This constitutes a suboptimal identification strategy when the original \(Z\) already satisfies the AIV condition.  
	The key drawback is that once \(Z_{\mathcal{S}}\) is discretized into a binary instrument, 
	the relevance condition in Equation~\eqref{eq: IV relevance discrete} becomes more restrictive than its counterpart in Assumption~\ref{as: IV relevance}.  
	Thus, although discretization may simplify estimation, we do not recommend this practice, as it can compromise estimator stability and reduce statistical efficiency.

	\subsection{Continuous treatment}
	So far, we have assumed that the treatment is either binary or multi-categorical. 
	In practice, however, the treatment of interest may be continuous \citep{kennedy2017non}. 
	For completeness, we now extend our theory and derive the identification strategy for a continuous treatment variable $A$. 
	Let $p_{A\mid Z,U,L}(a \mid Z, U, L)$ denote the conditional probability density function of $A$ given $(Z,U,L)$.
	
	\begin{definition}[Additive IV]\label{defn: additive IV continuous}
		For each $a \in \mathcal{A}$, we say that $Z$ is an \emph{additive IV} (AIV) for $A = a$ if there exist functions $b(U,L)$ and $c(Z,L)$ such that
		\begin{equation*}
			p_{A\mid Z,U,L}(a \mid Z, U, L) = b(U, L) + c(Z, L).
		\end{equation*}
		Moreover, we say that $Z$ is an AIV for $A$ if, for every $a\in\mathcal{A}$, $Z$ is an AIV for $A=a$.
	\end{definition}
	
	\begin{definition}[Regular weighting function]\label{defn: regular weighting function continuous}
		For each $a\in\mathcal{A}$, we say that $\pi(Z,L)$ is a \emph{regular weighting function} (RWF) for $A = a$ if there exists a positive constant $\epsilon_0$ such that
		\begin{equation*}
			\big| \mathbb{E}[\pi(Z,L)\mid A=a,L] - \mathbb{E}[\pi(Z,L)\mid L] \big|\geq \epsilon_0 \quad \text{uniformly over $L$.}
		\end{equation*}
	\end{definition}
	
	We make several remarks.
	First, the definitions of continuous AIV and RWF are natural extensions of their multi-categorical counterparts introduced in Section~\ref{sec: identification}. 
	Second, a continuous AIV \( Z \) for a continuous treatment \( A \) can be constructed, for example, via a Gaussian mixture model. 
	We now present an identification strategy for the continuous treatment-response curve $\mathbb{E}[Y(a)]$.
	
	\begin{theorem}[AIV identification]\label{thm: AIV identification continuous}
		Under Assumptions~\ref{as: consistency}--\ref{as: IV relevance}, for each \( a\in\mathcal{A} \), if $Z$ is an AIV for $A=a$, then for any RWF $\pi(Z,L)$, it holds that
		\begin{align*}
			\mathbb{E}[Y(a)] 
			= \mathbb{E}\left[ 
			\dfrac{ \mathbb{E}[Y\,\pi(Z,L)\mid A=a,L] - \mathbb{E}[Y\mid A=a,L] \, \mathbb{E}[\pi(Z,L)\mid L]}{ \mathbb{E}[\pi(Z,L)\mid A=a,L] - \mathbb{E}[\pi(Z,L)\mid L]} 
			\right].
		\end{align*}
	\end{theorem}
	
	\begin{example}[Binary IV with a continuous treatment]
		Assume that \( Z \) takes values in \( \{0,1\} \) and the treatment variable \( A \) is continuous. 
		Let $\pi(Z,L) = Z$, which serves as an RWF for \( A = a \). 
		Under Assumptions~\ref{as: consistency}--\ref{as: IV relevance}, if \( Z \) is an AIV for \( A \), Theorem~\ref{thm: AIV identification continuous} implies
		\begin{align*}
			\mathbb{E}[Y(a)] =
			\mathbb{E}\left[\dfrac{\mathbb{E}[YZ \mid A=a,L] - \mathbb{E}[Y \mid A=a,L]\mathbb{E}[Z \mid L]}
			{\mathbb{E}[Z \mid A=a,L] - \mathbb{E}[Z \mid L]}\right].
		\end{align*}
	\end{example}
	
	Notably, a continuous IV can be discretized into a binary (or multi-categorical) IV while preserving the AIV property, as established in Theorem~\ref{thm: cut IV}, even when the treatment variable is continuous. 
	This allows the identification strategy to be naturally simplified in settings with discretized instruments. 
	Moreover, semiparametric techniques, analogous to those developed in \citet{kennedy2017non}, can be applied to the result in Theorem~\ref{thm: AIV identification continuous} to construct a multiply robust and efficient estimator, which lies beyond the scope of this article.
	
	\section{Additional Empirical illustration}
	\begin{table}[htbp]
		\centering
		\caption{Bootstrap results for the college distance data based on 500 replications.}
		\label{tbl: college distance}
		\begin{tabular}{c|ccc}
			\toprule
			& Treated & Control & ATE \\
			\midrule
			EST & .4291 & .1074 & .1450 \\
			SE  & .2610 & .1199 & .2184 \\
			SD  & .4572 & .1450 & .2276 \\
			\bottomrule
		\end{tabular}
	\end{table}
	The \texttt{CollegeDistance} dataset, available in the \texttt{AER} package in \textsf{R}, originates from the High School and Beyond survey conducted by the U.S. Department of Education in 1980, with a follow-up survey in 1986. The survey includes data on 4,739 students from approximately 1,100 high schools. This dataset was originally analyzed in \citet{rouse1995democratization}, who studied the impact of community colleges on educational attainment.
	
	The dataset contains demographic, socioeconomic, and geographic variables commonly used in applied econometrics, particularly in instrumental variable (IV) analyses of educational attainment. Key variables include parental education (\texttt{fcollege}, \texttt{mcollege}), family characteristics (\texttt{home}, \texttt{income}), local economic conditions (\texttt{unemp}, \texttt{wage}), and measures of college accessibility such as the \texttt{distance} to the nearest four-year college and average state \texttt{tuition}. We adaptively estimate the weighting function $\pi_*^o(Z,L)$ and use the R package \texttt{mgcv} to estimate nuisance functions.
	
	The treatment variable is \texttt{education}, measured as the number of years of schooling completed by 1986, ranging from 12 years (high school completion) to 18 years (graduate degree). For analytical convenience, we dichotomize \texttt{education} into two groups: individuals with 14 or more years of schooling (treated group) and those with 13 years or fewer (control group). The outcome of interest is \texttt{income}, a binary indicator of whether the annual family income in 1980 exceeded \$25,000 (in 1980 U.S. dollars). We use \texttt{distance} and \texttt{tuition} as instrumental variables, and include \texttt{fcollege} and \texttt{mcollege} as baseline covariates $L$.
	
	Table~\ref{tbl: college distance} reports the bootstrap results based on 500 replications. The estimated mean outcome for the treated group is approximately 0.43, whereas that for the control group is around 0.11, yielding an average treatment effect (ATE) of roughly 0.15. Reported standard errors range from 0.12 to 0.26, depending on the subgroup, while the empirical standard deviations are somewhat larger, particularly for the treated group. Overall, the results indicate a moderate positive effect of educational attainment on family income.

	\section{Proof for Theorems}	
	The proof for Theorems~\ref{thm: asymptotic normality known weight},~\ref{thm: asymptotic normality unknown weight},~\ref{thm: asymptotic normality longitudinal},~\ref{thm: asymptotic normality longitudinal DTR}~and~\ref{thm: asymptotic normality unknown weight longitudinal setting} are analogous. Thus, we only provide the proof for Theorem~\ref{thm: asymptotic normality known weight}.
	\subsection{Proof for Theorem~\ref{thm: uniqueness}}	
	Suppose there exists two solution $f_a^o(1,L)$ and $f_a'(1,L)$ to Equation~\eqref{eq: npiv}. Then it holds that
	\begin{align*}
		0=&\mathbb{E}[A^{(a)}Y-A^{(a)}Y\mid Z,L]=\mathbb{E}[f_a'(A^{(a)},L)-f_a^o(A^{(a)},L)\mid Z,L]\\
		=&\{f_a'(1,L)-f_a^o(1,L)\}\mathbb{E}[A^{(a)}\mid Z,L]
		+\{f_a'(0,L)-f_a^o(0,L)\}\mathbb{E}[1-A^{(a)}\mid Z,L]\\
		=&\{f_a'(1,L)-f_a'(0,L)-f_a^o(1,L)+f_a^o(0,L)\}\mathbb{E}[A^{(a)}\mid Z,L]
		+f_a'(0,L)-f_a^o(0,L)
	\end{align*}
	From Assumption~\ref{as: IV relevance}, for any $l$, there exists two distinct $z_1,z_2$, such that
	$$\mathbb{E}[A^{(a)}\mid Z=z_1,L=l]\neq \mathbb{E}[A^{(a)}\mid Z=z_2,L=l].$$
	Then we take difference to get
	\begin{align*}
		&\{f_a'(1,l)-f_a'(0,l)-f_a^o(1,l)+f_a^o(0,l)\}\\
		&\cdot \{\mathbb{E}[A^{(a)}\mid Z=z_1,L=l]-\mathbb{E}[A^{(a)}\mid Z=z_2,L=l]\}=0.
	\end{align*}
	Now we know that 
	\begin{align*}
		&f_a'(1,L)-f_a'(0,L)-f_a^o(1,L)+f_a^o(0,L)\equiv 0,\\
		&f_a^o(0,L)-f_a^o(0,L)\equiv 0.
	\end{align*}
	Therefore $f_a^o(A^{(a)},L)\equiv f_a'(A^{(a)},L)$. This is equivalent to say that the solution to Equation~\eqref{eq: npiv} is unique. Second,
	\begin{align*}
		&\mathrm{Cov}\!\{A^{(a)}Y, \pi(Z,L) \mid L\}\\
		=&\mathbb{E}[A^{(a)}Y \pi(Z,L)\mid L]-\mathbb{E}[A^{(a)}Y\mid L]\mathbb{E}[\pi(Z,L)\mid L]\\
		=&\mathbb{E}[\mathbb{E}[A^{(a)}Y\mid Z,L] \pi(Z,L)\mid L]-\mathbb{E}[\mathbb{E}[A^{(a)}Y\mid Z,L]\mid L]\mathbb{E}[\pi(Z,L)\mid L]\\
		=&\mathbb{E}[\mathbb{E}[f_a^o(A^{(a)},Y)\mid Z,L] \pi(Z,L)\mid L]-\mathbb{E}[\mathbb{E}[f_a^o(A^{(a)},Y)\mid Z,L]\mid L]\mathbb{E}[\pi(Z,L)\mid L]\\
		=&\mathbb{E}[f_a^o(A^{(a)},L)\pi(Z,L)\mid L]-\mathbb{E}[f_a^o(A^{(a)},L)\mid L]\mathbb{E}[\pi(Z,L)\mid L]\\
		=&f_a^o(1,L)\mathbb{E}[A^{(a)}\pi(Z,L)\mid L]+f_a^o(0,L)\mathbb{E}[(1-A^{(a)})\pi(Z,L)\mid L]\\
		&-\{f_a^o(1,L)\mathbb{E}[A^{(a)}\mid L]+f_a^o(0,Y)\mathbb{E}[(1-A^{(a)})\mid L]\}\mathbb{E}[\pi(Z,L)\mid L]\\
		=&f_a^o(1,L)\mathrm{Cov}\!\{A^{(a)},\pi(Z,L)\mid L\}
		+f_a^o(0,L)\mathrm{Cov}\!\{1-A^{(a)},\pi(Z,L)\mid L\}\\
		=&\{f_a^o(1,L)-f_a^o(0,L)\}\cdot\mathrm{Cov}\!\{A^{(a)},\pi(Z,L)\mid L\}.
	\end{align*}
	This show that 
	\begin{align*}
		f_a^o(1,L)-f_a^o(0,L)=\dfrac{\mathrm{Cov}\!\{A^{(a)}Y, \pi(Z,L) \mid L\}}{\mathrm{Cov}\!\{A^{(a)},\pi(Z,L)\mid L\}}.
	\end{align*}
	In addition, by Equation~\eqref{eq: npiv}, we know that
	\begin{align*}
		&\mathbb{E}[A^{(a)}Y\mid Z,L]=\mathbb{E}[f_a^o(A^{(a)},L)\mid Z,L]\\
		=&\{f_a^o(1,L)-f_a^o(0,L)\}\mathbb{E}[A^{(a)}\mid Z,L]+f_a^o(0,L)\\
		=&\dfrac{\mathrm{Cov}\!\{A^{(a)}Y, \pi(Z,L) \mid L\}}{\mathrm{Cov}\!\{A^{(a)},\pi(Z,L)\mid L\}}\mathbb{E}[A^{(a)}\mid Z,L]+f_a^o(0,L).
	\end{align*}
	Now we can deduce the results in Equation~\eqref{eq: explicit form}, finishing the proof for Theorem~\ref{thm: uniqueness}.

	\subsection{Proof for Theorem~\ref{thm: AIV identification}}
	Define $g_a(U,L):=\mathbb{E}[Y(a)\mid U,L]$.
	We can observe that the left side of Equation~\eqref{eq: npiv} equals to
	\begin{align}
		&\mathbb{E}[A^{(a)}Y\mid Z,L]=\mathbb{E}[\mathbb{E}[A^{(a)}Y\mid U,Z,L]\mid Z,L]\notag\\
		=&\mathbb{E}\left[\mathbb{E}[Y\mid U,L,Z,A=a]\Pr(A=a\mid Z,U,L)\middle| Z,L\right]\notag\\
		=&\mathbb{E}\left[\mathbb{E}[Y(a)\mid U,L,Z,A=a]\Pr(A=a\mid Z,U,L)\middle| Z,L\right]\label{exp:4}\\
		=&\mathbb{E}\left[g_a(U,L)\Pr(A=a\mid Z,U,L)\middle| Z,L\right].\label{exp:5}
	\end{align}
	Equation~\eqref{exp:4} follows from Assumption~\ref{as: consistency} that $Y=Y(a)$.
	Equation~\eqref{exp:5} follows from Assumption~\ref{as: latent ignorability} that $Y(a)\indep \{Z,A\}\mid L,U$.
	Similarly, the right side of Equation~\eqref{eq: npiv} equals to
	\begin{align*}
		&\mathbb{E}[f_a^o(A^{(a)},L)\mid Z,L]=\mathbb{E}[\mathbb{E}[f_a^o(A^{(a)},L)\mid U,Z,L]\mid Z,L]\\
		=&\mathbb{E}\left[f_a^o(1,L)\Pr(A=a\mid Z,U,L)\middle| Z,L\right]+\mathbb{E}\left[f_a^o(0,L)\{1-\Pr(A=a\mid Z,U,L)\}\middle| Z,L\right]\\
		=&\mathbb{E}\left[\{f_a^o(1,L)-f_a^o(0,L)\}\Pr(A=a\mid Z,U,L)\middle| Z,L\right]+f_a^o(0,L).
	\end{align*}
	Next, from Assumption~\ref{as: IV independence} that $Z\indep U\mid L$, we see that for any $z\in \mathcal{Z}$,
	\begin{equation}\label{exp:6}
		\begin{aligned}
			&\mathbb{E}\left[g_a(U,L)\Pr(A=a\mid Z=z,U,L)\middle| L\right]\\
			=&\mathbb{E}\left[\{f_a^o(1,L)-f_a^o(0,L)\}\Pr(A=a\mid Z=z,U,L)\middle| L\right]
			+f_a^o(0,L).
		\end{aligned}
	\end{equation}	
	Then from Equation~\eqref{exp:6} and Assumption~\ref{defn: additive IV} that there exists $b(U,L)$ and $c(Z,L)$, such that $$\Pr(A=a\mid Z,U,L)=b(U,L)+c(Z,L).$$
	Then for any $z$, it holds that
	\begin{align}
		f_a^o(0,L)
		=&\mathbb{E}\left[\{g_a(U,L)-f_a^o(1,L)+f_a^o(0,L)\}\{b(U,L)+c(z,L)\}\middle| L\right]\notag
		\\=&\mathbb{E}\left[\{g_a(U,L)-f_a^o(1,L)+f_a^o(0,L)\}b(U,L)\middle| L\right]\label{exp:1}\\
		&+\mathbb{E}\left[g_a(U,L)-f_a^o(1,L)+f_a^o(0,L)\middle| L\right]c(z,L).\notag
	\end{align}
	From Assumption~\ref{as: IV relevance}, for any $l$, there exists $z_1,z_2$, such that $c(z_1,l)\neq c(z_2,l)$, and we can take difference to get
	\begin{align*}
		&\mathbb{E}[g_a(U,L)-f_a^o(1,L)+f_a^o(0,L)\mid L=l]\{c(z_1,l)-c(z_2,l)\}= 0.\\
		\Rightarrow&
		\mathbb{E}[g_a(U,L)-f_a^o(1,L)+f_a^o(0,L)\mid L=l]= 0.
	\end{align*}
	Then we see that
	$\mathbb{E}[g_a(U,L)-f_a^o(1,L)+f_a^o(0,L)\mid L]\equiv 0.$
	Next, note that for $a\in\mathcal{A}$,
	\begin{equation*}
		\mathbb{E}[g_a(U,L)\mid L]=\mathbb{E}[\mathbb{E}[Y(a)\mid L,U]\mid L]=\mathbb{E}[Y(a)\mid L].
	\end{equation*}
	Now we know that $f_a^o(1,L)-f_a^o(0,L)=\mathbb{E}[Y(a)\mid L].$ We can now substitute this into Equation~\eqref{exp:1} to deduce that
	\begin{align*}
		f_a^o(0,L)=&\mathbb{E}\left[\{\mathbb{E}[Y(a)\mid U,L]-\mathbb{E}[Y(a)\mid L]\}b(U,L)\middle| L\right]\\
		=&\mathrm{Cov}\!\{\mathbb{E}[Y(a)\mid U,L], \Pr(A=a\mid Z,U,L)\mid L\},\\
		f_a^o(1,L)=&\mathrm{Cov}\!\{\mathbb{E}[Y(a)\mid U,L], \Pr(A=a\mid Z,U,L)\mid L\}
		+\mathbb{E}[Y(a)\mid L].
	\end{align*}
	This step follows from Assumption~\ref{as: IV independence} that $\mathrm{Cov}\!\{\mathbb{E}[Y(a)\mid U,L],c(Z,L)\mid L\}=0$.
	This is equivalent to say that there only exist one solution $f_a^o(A^{(a)},L)$ to Equation~\eqref{eq: npiv},
	finishing the proof for Theorem~\ref{thm: AIV identification}.

	\subsection{Proof for Theorem~\ref{thm: Y(1)-Y(0)}}
	We start from solving Equation~\eqref{eq: npiv Y(1)-Y(0)}. 
	First, for a fixed value $z\in\mathcal{Z}$, the right side of Equation~\eqref{eq: npiv Y(1)-Y(0)} equals
	\begin{align*}
		\Pr(A=1\mid Z=z,L)\{f(1,L)-f(0,L)\}+f(0,L).
	\end{align*}
	Meanwhile, we repeatedly apply Assumptions~\ref{as: latent ignorability}~and~\ref{as: IV independence} to deduce that the left side of Equation~\eqref{eq: npiv Y(1)-Y(0)} equals
	\begin{align}
		&\mathbb{E}[A(Y(1)-Y(0))|Z=z,L]+\mathbb{E}[Y(0)\mid Z=z,L]\notag\\
		=&\mathbb{E}[\mathbb{E}[A(Y(1)-Y(0))|Z=z,U,L]|A=1,Z=z,L]\notag\\& +\mathbb{E}[\mathbb{E}[Y(0)\mid U,L]\mid Z=z,L]\notag\\
		=&\mathbb{E}[\Pr(A=1\mid Z=z,U,L)\mathbb{E}[Y(1)-Y(0)\mid A=1,Z=z,U,L]|Z=z,L]\notag\\& +\mathbb{E}[\mathbb{E}[Y(0)\mid U,L]\mid L]\notag\\
		=&\mathbb{E}[\Pr(A=1\mid Z=z,U,L)\mathbb{E}[Y(1)-Y(0)\mid U,L]|L] +\mathbb{E}[Y(0)\mid L]\notag\\
		=&\mathbb{E}[\left\{\Pr(A=1\mid Z=z,U,L)-\Pr(A=1\mid U,L)\right\}\mathbb{E}[Y(1)-Y(0)\mid U,L]|L] \notag\\
		&+\mathbb{E}[\Pr(A=1\mid U,L)\mathbb{E}[Y(1)-Y(0)\mid U,L]|L]+\mathbb{E}[Y(0)\mid L]\notag\\
		=&\mathbb{E}[\Pr(A=1\mid Z=z,U,L)-\Pr(A=1\mid U,L)|L] \times\mathbb{E}[\mathbb{E}[Y(1)-Y(0)\mid U,L]\mid L]\label{exp: add1}\\
		&+\mathbb{E}[\Pr(A=1\mid U,L)\mathbb{E}[Y(1)-Y(0)\mid U,L]|L]+\mathbb{E}[Y(0)\mid L]\notag\\
		=&\left\{\mathbb{E}[\Pr(A=1\mid Z=z,U,L)|L]-\Pr(A=1\mid L)\right\} \times\mathbb{E}[Y(1)-Y(0)\mid L]\notag\\
		&+\mathbb{E}[\Pr(A=1\mid U,L)\mathbb{E}[Y(1)-Y(0)\mid U,L]|L]+\mathbb{E}[Y(0)\mid L]\notag\\
		=&\left\{\mathbb{E}[\Pr(A=1\mid Z=z,U,L)|Z=z,L]-\Pr(A=1\mid L)\right\} \times\mathbb{E}[Y(1)-Y(0)\mid L]\notag\\
		&+\mathbb{E}[\Pr(A=1\mid U,L)\mathbb{E}[Y(1)-Y(0)\mid U,L]|L]+\mathbb{E}[Y(0)\mid L]\notag\\
		=&\left\{\Pr(A=1\mid Z=z,L)-\Pr(A=1\mid L)\right\} \times\mathbb{E}[Y(1)-Y(0)\mid L]\notag\\
		&+\mathbb{E}[\Pr(A=1\mid U,L)\mathbb{E}[Y(1)-Y(0)\mid U,L]|L]+\mathbb{E}[Y(0)\mid L]\notag\\
		=&\Pr(A=1\mid Z=z,L)\mathbb{E}[Y(1)-Y(0)\mid L]\notag\\
		&+\mathbb{E}[\{\Pr(A=1\mid U,L)-\Pr(A=1\mid L)\}\mathbb{E}[Y(1)-Y(0)\mid U,L]|L]+\mathbb{E}[Y(0)\mid L]\notag\\
		=&\Pr(A=1\mid Z=z,L)\mathbb{E}[Y(1)-Y(0)\mid L]\notag\\
		&+\mathrm{Cov}\{\Pr(A=1\mid U,L), \mathbb{E}[Y(1)-Y(0)\mid U,L]\mid L\}+\mathbb{E}[Y(0)\mid L].\notag
	\end{align}
	Equation~\eqref{exp: add1} holds by the no unmeasured common effect modifier assumption that
	\begin{align}\label{exp: add2}
		\mathrm{Cov}\!\{Y(1)-Y(0),\mathbb{E}[A\mid Z=z,U,L]-\mathbb{E}[A\mid U,L]\mid L\}=0.    
	\end{align}
	
	Now from Equation~\eqref{eq: npiv Y(1)-Y(0)}, we know that
	\begin{align*}
		&\Pr(A=1\mid Z=z,L)\{f(1,L)-f(0,L)-\mathbb{E}[Y(1)-Y(0)\mid L]\}\\
		&+\mathrm{Cov}\{\Pr(A=1\mid U,L), \mathbb{E}[Y(1)-Y(0)\mid U,L]\mid L\}+\mathbb{E}[Y(0)\mid L]-f(0,L).
	\end{align*}
	Since this equality holds for all $z\in\mathcal{Z}$, from Assumption~\ref{as: IV relevance}, we know that for any $l\in\mathcal{L}$ there exists two constant $z_1,z_2$, such that $\Pr(A=1|Z=z_1,L=l)\neq\Pr(A=1|Z=z_0,L=l)$. Then it is easy to show that
	\begin{align*}
		&\{\Pr(A=1\mid Z=z_1,L=l)-\Pr(A=1\mid Z=z_2,L=l)\}\\
		&\times\{f(1,L)-f(0,L)-\mathbb{E}[Y(1)-Y(0)\mid L=l]\}=0\\
		\Rightarrow&\left\{\begin{array}{l}
			f(1,L)-f(0,L)\equiv \mathbb{E}[Y(1)-Y(0)\mid L],\\
			f(0,L)\equiv\mathrm{Cov}\{\Pr(A=1\mid U,L), \mathbb{E}[Y(1)-Y(0)\mid U,L]\mid L\}+\mathbb{E}[Y(0)\mid L],\\
			f(1,L)\equiv\mathrm{Cov}\{\Pr(A=1\mid U,L), \mathbb{E}[Y(1)-Y(0)\mid U,L]\mid L\}+\mathbb{E}[Y(1)\mid L].
		\end{array}\right.
	\end{align*}
	
	The proof of Equation~\eqref{eq: identification AIV Y(1)-Y(0)} follows analogously to the arguments in Theorem~\ref{thm: AIV identification}, and is therefore omitted for brevity.  
	This concludes the proof of Theorem~\ref{thm: Y(1)-Y(0)}.
	
	Finally, without the no unmeasured common effect modifier assumption in Equation~\eqref{exp: add2}, we prove the fact that 
	\begin{equation*}
		\mathbb{E}\left[\dfrac{\mathrm{Cov}\{Y,\pi(Z,L)\mid L\}}{\mathrm{Cov}\{A,\pi(Z,L)\mid L\}}\right]
		=\mathbb{E}\left[\dfrac{\mathrm{Cov}\{A,\pi(Z,L)\mid U,L\}\mathbb{E}[Y(1)-Y(0)\mid U,L]}
		{\mathrm{Cov}\{A,\pi(Z,L)\mid L\}}\right].
	\end{equation*}
	which is a weighted average of conditional ATE $\mathbb{E}[Y(1)-Y(0)\mid U,L]$. In fact,
	\begin{align*}
		&\mathrm{Cov}\{Y,\pi(Z,L)\mid L\} = 
		\mathbb{E}\left[\mathbb{E}[Y|Z,L]\pi(Z,L)\middle| L\right] - 
		\mathbb{E}[Y\mid L]\mathbb{E}[\pi(Z,L)\mid L]\\
		=&\mathbb{E}\left[\mathbb{E}[A\{Y(1)-Y(0)\}+Y(0)|Z,L]\pi(Z,L)\middle| L\right] \\&- 
		\mathbb{E}[A\{Y(1)-Y(0)\}+Y(0)\mid L]\mathbb{E}[\pi(Z,L)\mid L]\\
		=&\mathbb{E}\left[\left\{\mathbb{E}[A\mathbb{E}[Y(1)-Y(0)\mid U,L]+\mathbb{E}[Y(0)\mid U,L]\mid Z,L]\right\}\pi(Z,L)\middle| L\right] \\&- 
		\mathbb{E}[A\mathbb{E}[Y(1)-Y(0)\mid U,L]+\mathbb{E}[Y(0)\mid U,L]\mid L]\mathbb{E}[\pi(Z,L)\mid L]\\
		=&\mathbb{E}\left[A\mathbb{E}[Y(1)-Y(0)\mid U,L]\pi(Z,L)\middle| L\right] 
		+\mathbb{E}\left[\mathbb{E}[Y(0)\mid L]\pi(Z,L)\middle| L\right] \\
		&- \mathbb{E}[A\mathbb{E}[Y(1)-Y(0)\mid U,L]\mid L]\mathbb{E}[\pi(Z,L)\mid L]
		- \mathbb{E}[Y(0)\mid L]\mathbb{E}[\pi(Z,L)\mid L]\\
		=&\mathbb{E}\left[\mathbb{E}[A\pi(Z,L)\mid U,L]\mathbb{E}[Y(1)-Y(0)\mid U,L]\middle| L\right] \\
		&- \mathbb{E}[A\mathbb{E}[Y(1)-Y(0)\mid U,L]\mid L]\mathbb{E}[\pi(Z,L)\mid L]\\
		=&\mathbb{E}\left[\mathbb{E}[A\pi(Z,L)\mid U,L]\mathbb{E}[Y(1)-Y(0)\mid U,L]\middle| L\right] \\
		&- \mathbb{E}[\mathbb{E}[\pi(Z,L)\mid L]\mathbb{E}[A\mid U,L]\mathbb{E}[Y(1)-Y(0)\mid U,L]\mid L]\\
		=&\mathbb{E}\left[\mathrm{Cov}\{A,\pi(Z,L)\mid U,L\}\mathbb{E}[Y(1)-Y(0)\mid U,L]\middle| L\right].
	\end{align*}
	Then we can show that
	\begin{align*}
		\dfrac{\mathrm{Cov}\{Y,\pi(Z,L)\mid L\}}{\mathrm{Cov}\{A,\pi(Z,L)\mid L\}}
		=\mathbb{E}\left[\dfrac{\mathrm{Cov}\{A,\pi(Z,L)\mid U,L\}\mathbb{E}[Y(1)-Y(0)\mid U,L]}
		{\mathrm{Cov}\{A,\pi(Z,L)\mid L\}}\middle| L\right].
	\end{align*}

	\subsection{Proof for Theorem~\ref{thm: EIF AIV fixed pi}}
	For any pathwise differentiable parameterization $p_\theta(O)$, we denote \(s(O):=\nabla_{\theta}\log p_\theta(O)\eval_{\theta=0}\) as the score function, $\mathbb{E}_\theta$ as taking expectation with respect to $p_\theta(O)$. Denote
	\begin{align*}
		\psi_{\pi;\theta}=&
		\mathbb{E}_\theta\left[\dfrac{\mathbb{E}_\theta[Y\pi(Z,L)\mid L]-\mathbb{E}_\theta[Y\mid L]\mathbb{E}_\theta[\pi(Z,L)\mid L]}{\mathbb{E}_\theta[A\pi(Z,L)\mid L]-\mathbb{E}_\theta[A\mid L]\mathbb{E}_\theta[\pi(Z,L)\mid L]}\right].
	\end{align*}
	We calculate the path-wise derivative as
	\begin{align*}
		\nabla_{\theta}\psi_{\pi;\theta}\eval_{\theta=0}=&\nabla_{\theta}
		\mathbb{E}_\theta\left[\dfrac{\mathbb{E}_\theta[Y\pi(Z,L)\mid L]-\mathbb{E}_\theta[Y\mid L]\mathbb{E}_\theta[\pi(Z,L)\mid L]}{\mathbb{E}_\theta[A\pi(Z,L)\mid L]-\mathbb{E}_\theta[A\mid L]\mathbb{E}_\theta[\pi(Z,L)\mid L]}\right]\eval_{\theta=0}\\
		=&\mathbb{E}[\{\gamma_{\pi}^o(L)-\psi_{\pi}^o\}s(O)]\\
		&+\mathbb{E}\left[\dfrac{\{Y\pi(Z,L)-\zeta_{\pi}^o(L)\}s(O)}{\mathbb{E}[A\pi(Z,L)\mid L]-\mathbb{E}[A\mid L]\mathbb{E}[\pi(Z,L)\mid L]}\right]\\
		&-\mathbb{E}\left[\dfrac{\rho_{\pi}^o(L)\{Y-\eta^o(L)\}s(O)}{\mathbb{E}[A\pi(Z,L)\mid L]-\mathbb{E}[A\mid L]\mathbb{E}[\pi(Z,L)\mid L]}\right]\\
		&-\mathbb{E}\left[\dfrac{\eta^o(L)\{\pi(Z,L)-\rho_{\pi}^o(L)\}s(O)}{\mathbb{E}[A\pi(Z,L)\mid L]-\mathbb{E}[A\mid L]\mathbb{E}[\pi(Z,L)\mid L]}\right]\\
		&-\mathbb{E}\left[\dfrac{\gamma_{\pi}^o(L)}{\kappa_{\pi}^o(L)}\left\{
		\begin{array}{l}
			A\pi(Z,L)-\kappa_{\pi}^o(L)\\
			-\rho_{\pi}^o(L)\{A-\delta^o(L)\}\\
			-\delta^o(L)\{\pi(Z,L)-\rho_{\pi}^o(L)\}
		\end{array}
		\right\}s(O)\right].
	\end{align*}
	Thus, one influence function for $\psi_{\pi}^o$ is
	\begin{align*}
		&\varphi_{\pi}(O;\psi_{\pi}^o,\alpha_{\pi}^o):=\gamma_{\pi}^o(L)-\psi_{\pi}^o\\
		&+\dfrac{1}{\kappa_{\pi}^o(L)}
		\cdot\left\{\begin{array}{l}
			Y\pi(Z,L)-\zeta_{\pi}^o(L)\\
			-\rho_{\pi}^o(L)\{Y-\eta^o(L)\}\\
			-\eta^o(L)\{\pi(Z,L)-\rho_{\pi}^o(L)\}
		\end{array}\right\}\\
		&-\dfrac{\gamma_{\pi}^o(L)}{\kappa_{\pi}^o(L)}\left\{
		\begin{array}{l}
			A\pi(Z,L)-\kappa_{\pi}^o(L)\\
			-\rho_{\pi}^o(L)\{A-\delta^o(L)\}\\
			-\delta^o(L)\{\pi(Z,L)-\rho_{\pi}^o(L)\}
		\end{array}
		\right\}\\
		=&\dfrac{\left\{\pi(Z,L)-\rho_{\pi}^o(L)\right\}}
		{\kappa_{\pi}^o(L)}
		Y-\psi_{\pi}^o
		\\&+ 
		\left(1-\dfrac{\{A-\delta^o(L)\}\{\pi(Z,L)-\rho_{\pi}^o(L)\}}{\kappa_{\pi}^o(L)} \right)\gamma_{\pi}^o(L) \\
		&-\dfrac{(\pi(Z,L)-\rho_{\pi}^o(L))^o(L)}{\kappa_{\pi}^o(L)}.
	\end{align*}
	Since the tangent space spanned by the score function $s(O)$ equals \(L_2(O)\), any influence function is automatically the EIF.

	\subsection{Proof for Theorem~\ref{thm: EIF AIV}}
	Recall that 
	\begin{align*}
		\gamma^o(L):=\dfrac{\zeta^o(L) - \eta^o(L)\delta^o(L)}{\kappa^o(L)},\qquad
		\gamma(L):=\dfrac{\zeta(L) - \eta(L)\delta(L)}{\kappa(L)}.
	\end{align*}
	For any pathwise differentiable parameterization $p_\theta(O)$ (probability density function), we denote \(s(O):=\nabla_{\theta}\log p_\theta(O)\eval_{\theta=0}\) as the score function, $\mathbb{E}_\theta$ as taking expectation with respect to $p_\theta(O)$.
	For $\zeta^o(L;\theta)=\mathbb{E}_\theta[Y\mathbb{E}_\theta[A\mid Z,L]\mid L]$, we calculate
	\begin{align*}
		&\nabla_{\theta}\zeta(L;\theta)\eval_{\theta=0}=\nabla_{\theta}\mathbb{E}_\theta[Y\mathbb{E}_\theta[A\mid Z,L]\mid L]\eval_{\theta=0}\\
		=&\nabla_{\theta}\mathbb{E}_\theta[Y\mathbb{E}[A\mid Z,L]\mid L]\eval_{\theta=0}
		+\mathbb{E}\left[Y\nabla_{\theta}\mathbb{E}_\theta[A\mid Z,L]\eval_{\theta=0}\mid L\right]\\
		=&\mathbb{E}[\{Y\pi_*^o(Z,L)-\zeta^o(L)\}s(O)\mid L]
		+\mathbb{E}[Y\mathbb{E}[\{A-\pi_*^o(Z,L)\}s(O)\mid Z,L]\mid L]\\
		=&\mathbb{E}[\{Y\pi_*^o(Z,L)-\zeta^o(L)\}s(O)\mid L]
		+\mathbb{E}[\xi^o(Z,L)\{A-\pi_*^o(Z,L)\}s(O)\mid L].
	\end{align*}
	Similarly, for $\kappa(L;\theta)=\mathbb{E}_\theta[A\mathbb{E}_\theta[A\mid Z,L]\mid L]-\mathbb{E}_{\theta}[A\mid L]^2$, it holds that
	\begin{align*}
		&\nabla_{\theta}\kappa(L;\theta)\eval_{\theta=0}=
		\nabla_{\theta}\mathbb{E}_\theta[A\mathbb{E}_\theta[A\mid Z,L]\mid L]\eval_{\theta=0}
		-\nabla_{\theta}\mathbb{E}_\theta[A\mid L]^2\eval_{\theta=0}\\
		=&\mathbb{E}\left[\{A\pi_*^o(Z,L)-\kappa^o(L)-\delta^o(L)^2\}s(O)\middle| L\right]\\
		&+\mathbb{E}\left[\pi_*^o(Z,L)\{A-\pi_*^o(Z,L)\}s(O)\middle| L\right]\\
		&-\mathbb{E}[2\delta^o(L)(A-\delta^o(L))s(O)\mid L].
	\end{align*}
	For $\eta(L;\theta)=\mathbb{E}_\theta[Y\mid L]$, we calculate
	\begin{align*}
		&\nabla_{\theta}\eta(L;\theta)\eval_{\theta=0}
		=\nabla_{\theta}\mathbb{E}_\theta[Y\mid L]\eval_{\theta=0}\\
		=&\mathbb{E}[\{Y-\mathbb{E}[Y\mid L]\}s(O)\mid L]\\
		=&\mathbb{E}\left[\{Y-\eta^o(L)\}s(O)\middle| L\right].
	\end{align*}
	For $\delta(L;\theta)=\mathbb{E}_\theta[A\mid L]$, we calculate
	\begin{align*}
		&\nabla_{\theta}\delta(L;\theta)\eval_{\theta=0}
		=\nabla_{\theta}\mathbb{E}_\theta[A\mid L]\eval_{\theta=0}
		=\mathbb{E}[\{A-\delta^o(L)\}s(O)\mid L].
	\end{align*}
	Now we calculate the EIF as follows. For $$\psi_{ada,\theta}=\mathbb{E}_\theta\left[
	\dfrac{\zeta(L;\theta) - \eta(L;\theta)\delta(L;\theta)}{\kappa(L;\theta)}
	\right],$$
	we deduce the path-wise derivative as
	\begin{align*}
		&\nabla_{\theta}\psi_{ada,\theta}\eval_{\theta=0}
		=\nabla_{\theta}\mathbb{E}_\theta\left[
		\dfrac{\zeta(L;\theta) - \eta(L;\theta)\delta(L;\theta)}{\kappa(L;\theta)}
		\right]\eval_{\theta=0}\notag\\
		=&\mathbb{E}[(\gamma^o(L)-\psi_{*}^o)s(O)]\notag\\
		&+\mathbb{E}\left[\dfrac{\nabla_{\theta}\zeta(L;\theta)\eval_{\theta=0}
			-\nabla_{\theta} (L;\theta) \delta(L;\theta)\eval_{\theta=0}}{
			\kappa^o(L)}\right]\notag\\
		&-\mathbb{E}\left[\gamma^o(L)\dfrac{
			\nabla_{\theta}\kappa(L;\theta)\eval_{\theta=0}			
		}{
			\kappa^o(L)}\right]\notag\\
		=&\mathbb{E}[(\gamma^o(L)-\psi_{*}^o)s(O)]\\
		&+\mathbb{E}\left[\dfrac{1}{\kappa^o(L)}
		\left\{\begin{array}{l}
			\{Y\pi_*^o(Z,L)-\zeta^o(L)\}\\
			+\xi^o(Z,L)\{A-\pi_*^o(Z,L)\}
		\end{array}\right\}s(O)
		\right]\\
		&-\mathbb{E}\left[\dfrac{1}{\kappa^o(L)}\left\{
		\begin{array}{l}
			\delta^o(L)\{Y-\eta^o(L)\}\\
			+^o(L)\{A-\delta^o(L)\}
		\end{array}
		\right\}s(O)\right]\\
		&+\mathbb{E}\left[\dfrac{\gamma^o(L)}{\kappa^o(L)}2\delta^o(L)\left\{
		A-\delta^o(L)
		\right\}s(O)\right]\\
		&-\mathbb{E}\left[\dfrac{\gamma^o(L)}{\kappa^o(L)}\left\{
		\begin{array}{l}
			\{A\pi_*^o(Z,L)-\kappa^o(L)\}\\
			+\pi_*^o(Z,L)\{A-\pi_*^o(Z,L)\}
		\end{array}
		\right\}s(O)\right].
	\end{align*}
	The EIF is
	\begin{align*}
		&\varphi(O;\psi_{*}^o,\beta^o)=\gamma^o(L)-\psi_{*}^o\\
		&+\dfrac{1}{\kappa^o(L)}
		\left\{\begin{array}{l}
			\{Y\pi_*^o(Z,L)-\zeta^o(L)\}\\
			+\xi^o(Z,L)\{A-\pi_*^o(Z,L)\}\\
			-\delta^o(L)\{Y-\eta^o(L)\}\\
			-\eta^o(L)\{A-\delta^o(L)\}
		\end{array}\right\}\\
		&-\dfrac{\gamma^o(L)}{\kappa^o(L)}\left\{
		\begin{array}{l}
			A\pi_*^o(Z,L)-\kappa^o(L)-\delta^o(L)^2\\
			+\pi_*^o(Z,L)\{A-\pi_*^o(Z,L)\}\\
			-2\delta^o(L)\left\{A-\delta^o(L)\right\}
		\end{array}
		\right\}\\
		=&\dfrac{\pi_*^o(Z,L)-\delta^o(L)}{\kappa^o(L)}Y-\psi_{*}^o\\
		&+\dfrac{1}{\kappa^o(L)}
		\left\{
		\xi^o(Z,L)(A-\pi_*^o(Z,L))
		-\eta^o(L)(A-\delta^o(L))
		\right\}\\
		&+\gamma^o(L)\left\{
		1 - \dfrac{(\pi_*^o(Z,L)-\delta^o(L))(2A-\pi_*^o(Z,L)-\delta^o(L))}{\kappa^o(L)}
		\right\}.
	\end{align*}	
	
	\subsection{Proof for Theorem~\ref{thm: asymptotic normality known weight}}
	Define the conditional expectation on the $k$-th fold as $\mathbb{E}_k[O]=\mathbb{E}[O\mid I_{-k}]$.
	From the definition of $\psi_{\pi}^o$ in Equation~\eqref{eq: AUG estimator prespecified weight},
	\begin{align}
		0=&\dfrac{\sqrt{n}}{K}\sum_{k=1}^K\mathbb{E}_{nk}\left[\varphi_{\pi}(O;\hat\psi_{\pi}^{(n)},\hat\alpha_{\pi}^{(n,k)})\right]\notag\\
		=&\dfrac{\sqrt{n}}{K}\sum_{k=1}^K\mathbb{E}_{nk}\left[\varphi_{\pi}(O;\psi_{\pi}^{o},\hat\alpha_{\pi}^{(n,k)})\right]+\sqrt{n}\{\hat\psi_{\pi}^{(n)}-\psi_{\pi}^{o}\}\notag\\
		=&\dfrac{\sqrt{n}}{K}\sum_{k=1}^K\mathbb{E}_{nk}\left[\varphi_{\pi}(O;\psi_{\pi}^{o},\alpha_{\pi}^{o})\right]+\sqrt{n}\{\hat\psi_{\pi}^{(n)}-\psi_{\pi}^{o}\}\notag\\
		&+\dfrac{\sqrt{n}}{K}\sum_{k=1}^K\mathbb{E}_{k}\left[\varphi_{\pi}(O;\psi_{\pi}^{o},\hat\alpha_{\pi}^{(n,k)})-\varphi_{\pi}(O;\psi_{\pi}^{o},\alpha_{\pi}^{o})\right]\label{exp:10}\\
		&+\dfrac{\sqrt{n}}{K}\sum_{k=1}^K\{\mathbb{E}_{nk}-\mathbb{E}_{k}\}\left[\varphi_{\pi}(O;\psi_{\pi}^{o},\hat\alpha_{\pi}^{(n,k)})-\varphi_{\pi}(O;\psi_{\pi}^{o},\alpha_{\pi}^{o})\right].\label{exp:9}
	\end{align}
	From Proposition~\ref{prop: mixed bias property fixed pi}, we can deduce that the quantity in Equation~\eqref{exp:10} is $o_p(1)$, since
	\begin{align*}
		&\mathbb{E}_{k}\left[\varphi_{\pi}(O;\psi_{\pi}^{o},\hat\alpha_{\pi}^{(n,k)})-\varphi_{\pi}(O;\psi_{\pi}^{o},\alpha_{\pi}^{o})\right]\\
		\lesssim&
		\left\{\begin{array}{l}
			\|\hat\kappa_{\pi}^{(n,k)}(L)-\kappa_{\pi}^o(L)\|_2\times \|\hat\gamma_{\pi}^{(n,k)}(L)-\gamma_{\pi}^o(L)\|_2\\
			+\|\hat\rho_{\pi}^{(n,k)}(L)-\rho_{\pi}^o(L)\|_2\times
			\|\hat\delta^{(n,k)}(L)-\delta^o(L)\|_2\\
			+\|\hat\rho_{\pi}^{(n,k)}(L)-\rho_{\pi}^o(L)\|_2\times
			\|\hat\eta^{(n,k)}(L)-\eta^o(L)\|_2
		\end{array}\right\}
		=o_p(n^{-1/2}).
	\end{align*}
	Define the empirical process as $\mathbb{G}_{nk}[f(O)]:=\sqrt{n_k}\{\mathbb{E}_{nk}-\mathbb{E}_{k}\}[f(O)],$ and we have
	\begin{align*}
		&\Pr\left(\mathbb{G}_{nk}\left[\varphi_{\pi}(O;\psi_{\pi}^{o},\hat\alpha_{\pi}^{(n,k)})-\varphi_{\pi}(O;\psi_{\pi}^{o},\alpha_{\pi}^{o})\right]\geq \epsilon_0\middle| I_{-k}\right)\\
		\lesssim&\dfrac{1}{\epsilon_0^2}\mathrm{Var}\left[\mathbb{G}_{nk}\left\{\varphi_{\pi}(O;\psi_{\pi}^{o},\hat\alpha_{\pi}^{(n,k)})-\varphi_{\pi}(O;\psi_{\pi}^{o},\alpha_{\pi}^{o})\right\}\middle| I_{-k}\right]\\
		\lesssim&\dfrac{1}{\epsilon_0^2}\mathrm{Var}\left[\varphi_{\pi}(O;\psi_{\pi}^{o},\hat\alpha_{\pi}^{(n,k)})-\varphi_{\pi}(O;\psi_{\pi}^{o},\alpha_{\pi}^{o})\middle| I_{-k}\right]\\
		\lesssim&\dfrac{1}{\epsilon_0^2}\mathbb{E}\left[\left\{\varphi_{\pi}(O;\psi_{\pi}^{o},\hat\alpha_{\pi}^{(n,k)})-\varphi_{\pi}(O;\psi_{\pi}^{o},\alpha_{\pi}^{o})\right\}^2\middle| I_{-k}\right]\\
		\lesssim&\|\hat\rho_{\pi}(L)-\rho_{\pi}^o(L)\|_2^2+
		\|\hat\delta^{(n,k)}(L)-\delta^o(L)\|_2^2+
		\|\hat\eta^{(n,k)}(L)-\eta^o(L)\|_2^2\\&+
		\|\hat\kappa^{(n,k)}(L)-\kappa^o(L)\|_2^2+
		\|\hat\gamma_{\pi}^{(n,k)}(L)-\gamma_{\pi}^o(L)\|_2^2\\\lesssim&
		\|\hat\alpha_{\pi}^{(n,k)}-\alpha_{\pi}^o\|_2^2.
	\end{align*}
	We can now integrates $I_{-k}$ out to deduce that
	\begin{align*}
		\Pr\left(\mathbb{G}_{nk}\left[\varphi_{\pi}(O;\psi_{\pi}^{o},\hat\alpha_{\pi}^{(n,k)})-\varphi_{\pi}(O;\psi_{\pi}^{o},\alpha_{\pi}^{o})\right]\geq \epsilon_0\right)\lesssim 
		\mathbb{E}[\|\hat\alpha_{\pi}^{(n,k)}-\alpha_{\pi}^o\|_2^2]=o(1).
	\end{align*}
	This is equivalent to say that the quantity in Equation~\eqref{exp:9} is $o_p(1)$. Next, we can see that
	\begin{align*}
		\sqrt{n}\{\hat\psi_{\pi}^{(n)}-\psi_{\pi}^{o}\}=&-\dfrac{\sqrt{n}}{K}\sum_{k=1}^K\mathbb{E}_{nk}\left[\varphi_{\pi}(O;\psi_{\pi}^{o},\alpha_{\pi}^{o})\right]+o_p(1)\\
		=&-\dfrac{1}{\sqrt{n}}\sum_{i = 1}^n\varphi_{\pi}(O_i;\psi_{\pi}^{o},\alpha_{\pi}^{o})+o_p(1).
	\end{align*}
	From the central limit theory and the Slutskys lemma, we know that
	$\sqrt{n}\{\hat\psi_{\pi}^{(n)}-\psi_{\pi}^{o}\}$ converges in distribution to $\mathcal{N}(0,(\sigma_{\pi}^o)^2)$.
	Next, we prove the consistency of the variance estimator $$\left(\hat\sigma_{\pi}^{(n,k)}\right)^2:=\mathbb{E}_{nk}\left[\varphi_{\pi}(O;\hat\psi_{\pi}^{(n)},\hat\alpha_{\pi}^{(n,k)})^2\right].$$
	First, we define
	\(
	\left(\tilde\sigma_{\pi}^{(n,k)}\right)^2:=\mathbb{E}_{nk}\left[\varphi_{\pi}(O;\psi_{\pi}^{o},\alpha_{\pi}^{o})^2\right]=\mathbb{E}_n\left[
	\varphi_{\pi}(O;\psi_{\pi}^{o},\alpha_{\pi}^{o})^2\right].
	\)
	By the law of large number, $\left(\tilde\sigma_{\pi}^{(n,k)}\right)^2$ converges to $\left(\sigma_{\pi}^{o}\right)^2$ in probability. Simultaneously, by the boundness for $\varphi_{\pi}(O;\psi_{\pi},\alpha_{\pi})$, we know that
	\begin{align*}
		&\Pr\left(\left|\left(\hat\sigma_{\pi}^{(n,k)}\right)^2-\left(\tilde\sigma_{\pi}^{(n,k)}\right)^2\right|\geq \epsilon_0\middle| I_{-k}\right)
		\leq \dfrac{1}{\epsilon_0^2}\mathbb{E}_k\left[\left|\left(\hat\sigma_{\pi}^{(n,k)}\right)^2-\left(\tilde\sigma_{\pi}^{(n,k)}\right)^2\right|^2\right]\\
		\lesssim&\dfrac{1}{\epsilon_0^2}
		\mathbb{E}_{k}\left[\left|\mathbb{E}_{nk}\left[\varphi_{\pi}(O;\hat\psi_{\pi}^{(n)},\hat\alpha_{\pi}^{(n,k)})^2-\varphi_{\pi}(O;\psi_{\pi}^{o},\alpha_{\pi}^{o})^2\right]\right|\right]\\
		\lesssim&\dfrac{1}{\epsilon_0^2}\mathbb{E}_{k}\left[\left|\varphi_{\pi}(O;\hat\psi_{\pi}^{(n)},\hat\alpha_{\pi}^{(n,k)})^2-\varphi_{\pi}(O;\psi_{\pi}^{o},\alpha_{\pi}^{o})^2\right|\right]\\
		\lesssim&\dfrac{1}{\epsilon_0^2}\left(\mathbb{E}_{k}\left[\left|\varphi_{\pi}(O;\hat\psi_{\pi}^{(n)},\hat\alpha_{\pi}^{(n,k)})-\varphi_{\pi}(O;\psi_{\pi}^{o},\alpha_{\pi}^{o})\right|^2\right]\right)^{1/2}\\
		\lesssim&\dfrac{1}{\epsilon_0^2}\left\{\left|\hat\psi_{\pi}^{(n,k)}-\psi_{\pi}^{o}\right|^2+
		\|\hat\alpha_{\pi}^{(n,k)}-\alpha_{\pi}^{o}\|_2^2\right\}^{1/2}.
	\end{align*}
	Now we can integrates $I_{-k}$ out to deduce that
	\begin{align*}
		&\Pr\left(\left|\left(\hat\sigma_{\pi}^{(n,k)}\right)^2-\left(\tilde\sigma_{\pi}^{(n,k)}\right)^2\right|\geq \epsilon_0\right)\\
		\lesssim&\dfrac{1}{\epsilon_0^2}\mathbb{E}\left[\left\{\left|\hat\psi_{\pi}^{(n,k)}-\psi_{\pi}^{o}\right|^2+
		\|\hat\alpha_{\pi}^{(n,k)}-\alpha_{\pi}^{o}\|_2^2\right\}^{1/2}\right]\\
		\lesssim&\dfrac{1}{\epsilon_0^2}\left\{\mathbb{E}\left[\left|\hat\psi_{\pi}^{(n,k)}-\psi_{\pi}^{o}\right|^2+
		\|\hat\alpha_{\pi}^{(n,k)}-\alpha_{\pi}^{o}\|_2^2\right]\right\}^{1/2}=o(1).
	\end{align*}
	Now $\left(\hat\sigma_{\pi}^{(n,k)}\right)^2=\left(\tilde\sigma_{\pi}^{(n,k)}\right)^2+o_p(1)=\left(\sigma_{\pi}^{o}\right)^2+o_p(1)$, finishing the proof for Theorem~\ref{thm: asymptotic normality known weight}.

	\subsection{Proof for Theorem~\ref{thm: longitudinal AIV}}
	We provide a proof for the following result, which is stronger than the results in Equation~\eqref{eq: identification AIV longitudinal}. For any \( s \) with \( 0 \leq s \leq T+1 \) and \( r \) with \( 0 \leq r \leq T+1-s \), we have
	\begin{align}
		&\mathbb{E}[Y(\underline{a}_{s}) \mid H_s]\notag\\
		=&\mathbb{E}\left[\prod_{t=s}^{T-r}
		\dfrac{\left(\pi_t(Z_t,H_t)-\mathbb{E}[\pi_t(Z_t,H_t)\mid H_t]\right)I\{A_t=a_t\}}{\mathrm{Cov}\!\{I\{A_t=a_t\},\pi_t(Z_t,H_t)\mid H_t\}}
		\gamma_{T+1-r,\underline{a}_{T+1-r}}^o(H_{T+1-r})\middle| H_s
		\right]\label{exp:13},
	\end{align}
	which is clearly a stronger version of Equation~\eqref{eq: identification AIV longitudinal}.
	We establish the result by induction. When \( s = T+1 \), Equation~\eqref{exp:13} reduces to \( \mathbb{E}[Y(\overline{A}) \mid H_{T+1}] = \mathbb{E}[Y \mid H_{T+1}] \), which follows directly from the consistency assumption (Assumption~\ref{as: consistency'}). 
	Now, assume that Equation~\eqref{exp:13} holds for all \( s_c = T+1, T, \ldots, s+1 \) and \( r_c \) satisfying \( s_c + r_c \leq T+1 \). We verify that the equation also holds for \( s_c = s \) and all \( r_c = 0, \ldots, T+1 - s \). This completes the proof by induction. 
	In particular, we begin by verifying the case \( r_c = T+1 - s \). By the definition of \( \gamma_{s,\underline{a}_s}^o \), this is equivalent to verifying that
	\begin{equation}\label{exp:11}
		\begin{aligned}
			&\mathbb{E}[Y(\underline{a}_{s})|H_s]
			=\gamma_{s,\underline{a}_s}^o(H_{s})
			:=\frac{\mathrm{Cov}\bigl\{ I\{A_s = a_s\}  \gamma_{s+1,\underline{a}_{s+1}}^o(H_{s+1}),  \pi_s(Z_s, H_s) \mid H_s \bigr\}}
			{\mathrm{Cov}\bigl\{ I\{A_s = a_s\},  \pi_s(Z_s, H_s) \mid H_s \bigr\}}.
		\end{aligned}
	\end{equation}
	By induction,
	$\mathbb{E}[Y(\underline{a}_{s+1})|H_{s+1}]=\gamma_{s+1,\underline{a}_{s+1}}^o(H_{s+1})$. Then
	\begin{align*}
		&\mathbb{E}\left[ I\{A_s = a_s\}  \gamma_{s+1,\underline{a}_{s+1}}^o(H_{s+1})\pi_t(Z_s, H_s) \middle| H_s \right]\\
		=&\mathbb{E}\left[ I\{A_s = a_s\}  \mathbb{E}[Y(\underline{a}_{s+1})\mid H_{s+1}]\pi_t(Z_s, H_s) \middle| H_s \right]\\
		=&\mathbb{E}\left[ I\{A_s = a_s\}  Y(\underline{a}_{s+1})\pi_t(Z_s, H_s) \middle| H_s \right]\\
		=&\mathbb{E}\left[ I\{A_s = a_s\}  Y(\underline{a}_{s})\pi_t(Z_s, H_s) \middle| H_s \right]\\
		=&\mathbb{E}\left[ \mathbb{E}[Y(\underline{a}_{s})\mid A_s=a_s,Z_s,H_s]\times
		\mathbb{E}[I\{A_s = a_s\}\mid Z_s,H_s] \pi_t(Z_s, H_s)\middle| H_s \right]\\
		=&\mathbb{E}[Y(\underline{a}_{s})\mid H_s]\times\mathbb{E}\left[ 
		\Pr(A_s = a_s\mid  Z_s,H_s) \pi_t(Z_s, H_s)\middle| H_s \right].
	\end{align*}
	Since $A_s$ is an AIV for $H_s$, there exists functions \(b_{s,a_s}(\overline{U}_s,H_s)\) and \(c_{s,a_s}(Z_s,H_s)\) such that, for any \(a_s \in \mathcal{A}_s\),
	\[
	\Pr(A_s = a_s \mid Z_s, \overline{U}_s, H_s) = b_{s,a_s}(\overline{U}_s, H_s) + c_{s,a_s}(Z_s, H_s).
	\]
	Then we can deduce that
	\begin{align*}
		&\mathbb{E}\left[ I\{A_s = a_s\}  \gamma_{s+1,\underline{a}_{s+1}}^o(H_{s+1})\pi_t(Z_s, H_s) \middle| H_s \right]\\
		=&\mathbb{E}[Y(\underline{a}_{s})\mid H_s]\times\mathbb{E}\left[ 
		\left\{b_{s,a_s}(\overline{U}_s, H_s) + c_{s,a_s}(Z_s, H_s)\right\} \pi_t(Z_s, H_s)\middle| H_s \right]\\
		=&\mathbb{E}[Y(\underline{a}_{s})\mid H_s]\times
		\left\{\begin{array}{l}
			\mathbb{E}\left[ 
			b_{s,a_s}(\overline{U}_s, H_s)\middle| H_s\right]\times\mathbb{E}\left[\pi_t(Z_s, H_s)\middle| H_s \right]\\
			+\mathbb{E}\left[ 
			c_{s,a_s}(Z_s, H_s) \pi_t(Z_s, H_s)\middle| H_s \right]
		\end{array}\right\}.
	\end{align*}
	Similarly, we can calculate
	\begin{align*}
		&\mathbb{E}\left[ I\{A_s = a_s\}  \gamma_{s+1,\underline{a}_{s+1}}^o(H_{s+1}) \middle| H_s \right]
		\times\mathbb{E}\left[\pi_t(Z_s, H_s)\middle| H_s \right]\\
		=&\mathbb{E}[Y(\underline{a}_{s})\mid H_s]\times
		\left\{\begin{array}{l}
			\mathbb{E}\left[ 
			b_{s,a_s}(\overline{U}_s, H_s)\middle| H_s\right]\\
			+\mathbb{E}\left[ 
			c_{s,a_s}(Z_s, H_s) \middle| H_s \right]
		\end{array}\right\}\times\mathbb{E}\left[\pi_t(Z_s, H_s)\middle| H_s \right];\\\\
		&\mathrm{Cov}\bigl\{ I\{A_s = a_s\}  \gamma_{s+1,\underline{a}_{s+1}}^o(H_{s+1}),  \pi_s(Z_s, H_s) \mid H_s \bigr\}\\
		=&\mathbb{E}[Y(\underline{a}_{s})\mid H_s]\times
		\left\{\begin{array}{l}
			\mathbb{E}\left[
			c_{s,a_s}(Z_s, H_s) \pi_t(Z_s, H_s)\middle| H_s \right]\\
			-\mathbb{E}\left[
			c_{s,a_s}(Z_s, H_s)\middle| H_s \right]\times
			\mathbb{E}\left[\pi_t(Z_s, H_s)\middle| H_s \right]
		\end{array}\right\}\\
		=&\mathbb{E}[Y(\underline{a}_{s})\mid H_s]\times
		\mathrm{Cov}\bigl\{c_{s,a_s}(Z_s, H_s),  \pi_s(Z_s, H_s) \mid H_s \bigr\}\\
		=&\mathbb{E}[Y(\underline{a}_{s})\mid H_s]\times
		\mathrm{Cov}\bigl\{ b_{s,a_s}(\overline{U}_s, H_s) +c_{s,a_s}(Z_s, H_s),  \pi_s(Z_s, H_s) \mid H_s \bigr\}\\
		=&\mathbb{E}[Y(\underline{a}_{s})\mid H_s]\times
		\mathrm{Cov}\bigl\{\Pr(A_s = a_s \mid Z_s, \overline{U}_s, H_s),  \pi_s(Z_s, H_s) \mid H_s \bigr\}\\
		=&\mathbb{E}[Y(\underline{a}_{s})\mid H_s]\times
		\mathrm{Cov}\bigl\{I\{A_s = a_s\} ,  \pi_s(Z_s, H_s) \mid H_s \bigr\}.
	\end{align*}
	This finishes the proof for Equation~\eqref{exp:11}, and we finish the proof for $s_c=s$ and $r_c=T+1-s$. For $r_c=T-s$,
	\begin{align}
		&\mathbb{E}[Y(\underline{a}_{s})|H_s]
		=\frac{\mathrm{Cov}\bigl\{ I\{A_s = a_s\}  \gamma_{s+1,\underline{a}_{s+1}}^o(H_{s+1}),  \pi_s(Z_s, H_s) \mid H_s \bigr\}}
		{\mathrm{Cov}\bigl\{ I\{A_s = a_s\},  \pi_s(Z_s, H_s) \mid H_s \bigr\}}\notag\\
		=&\mathbb{E}\left[\frac{
			\left\{\pi_s(Z_s, H_s)-\mathbb{E}\left[\pi_s(Z_s, H_s)\middle| H_s\right]\right\}
			I\{A_s = a_s\}  \gamma_{s+1,\underline{a}_{s+1}}^o(H_{s+1})}
		{\mathrm{Cov}\bigl\{ I\{A_s = a_s\},  \pi_s(Z_s, H_s) \mid H_s \bigr\}}\middle| H_s\right].\label{exp:12}
	\end{align}
	By induction, $\gamma_{s+1,\underline{a}_{s+1}}^o(H_{s+1})$ equals to
	\begin{align*}
		&\mathbb{E}[Y(\underline{a}_{s+1}) \mid H_{s+1}]\notag\\
		=&\mathbb{E}\left[\prod_{t=s+1}^{T-r}
		\dfrac{\left(\pi_t(Z_t,H_t)-\mathbb{E}[\pi_t(Z_t,H_t)\mid H_t]\right)I\{A_t=a_t\}}{\mathrm{Cov}\!\{I\{A_t=a_t\},\pi_t(Z_t,H_t)\mid H_t\}}
		\gamma_{T+1-r,\underline{a}_{T+1-r}}^o(H_{T+1-r})\middle| H_{s+1}
		\right].
	\end{align*}
	We can substitute this quantity into Equation~\eqref{exp:12} to deduce that
	\begin{align*}
		&\mathbb{E}[Y(\underline{a}_{s})|H_s]\\
		=&\mathbb{E}\left[\begin{array}{l}
			\dfrac{
				\left\{\pi_s(Z_s, H_s)-\mathbb{E}\left[\pi_s(Z_s, H_s)\middle| H_s\right]\right\}
				I\{A_s = a_s\}}  
			{\mathrm{Cov}\bigl\{ I\{A_s = a_s\},  \pi_s(Z_s, H_s) \mid H_s \bigr\}}\\
			\times \displaystyle\prod_{t=s+1}^{T-r}
			\dfrac{\left(\pi_t(Z_t,H_t)-\mathbb{E}[\pi_t(Z_t,H_t)\mid H_t]\right)I\{A_t=a_t\}}{\mathrm{Cov}\!\{I\{A_t=a_t\},\pi_t(Z_t,H_t)\mid H_t\}}
			\gamma_{T+1-r,\underline{a}_{T+1-r}}^o(H_{T+1-r})
		\end{array}\middle| H_s\right]\\
		=&\mathbb{E}\left[\prod_{t=s}^{T-r}
		\dfrac{\left(\pi_t(Z_t,H_t)-\mathbb{E}[\pi_t(Z_t,H_t)\mid H_t]\right)I\{A_t=a_t\}}{\mathrm{Cov}\!\{I\{A_t=a_t\},\pi_t(Z_t,H_t)\mid H_t\}}
		\gamma_{T+1-r,\underline{a}_{T+1-r}}^o(H_{T+1-r})\middle| H_{s}
		\right].
	\end{align*}
	This finishes the proof for Equation~\eqref{exp:13}.

	\subsection{Proof for Theorem~\ref{thm: EIF AIV longitudinal}}
	For any pathwise differentiable parameterization $p_\theta(O)$, we denote \(s(O):=\nabla_{\theta}\log p_\theta(O)\eval_{\theta=0}\) as the score function, $\mathbb{E}_\theta$ as taking expectation with respect to $p_\theta(O)$.
	Recall that the true parameter of interest is 
	\begin{align*}
		\psi_{\overline{a}}^o:=\mathbb{E}\left[
		\prod_{t=0}^{T}
		\dfrac{\left(Z_t-\mathbb{E}[Z_t\mid H_t]\right)A_t^{(a_t)}}
		{\mathbb{E}\left[A_t^{(a_t)}Z_t\middle| H_t\right]
			-\mathbb{E}\left[A_t^{(a_t)}\middle| H_t\right]
			\mathbb{E}\left[Z_t\middle| H_t\right]}
		\times Y
		\right].
	\end{align*}
	Concretely,
	\begin{align*}
		&\nabla_{\theta}\psi_{\overline{a},\theta}\eval_{\theta=0}
		\\=&\nabla_{\theta}
		\mathbb{E}_{\theta}\left[
		\prod_{t=0}^{T}
		\dfrac{\left(Z_t-\mathbb{E}_{\theta}[Z_t\mid H_t]\right)A_t^{(a_t)}}
		{\mathbb{E}_{\theta}\left[A_t^{(a_t)}Z_t\middle| H_t\right]
			-\mathbb{E}_{\theta}\left[A_t^{(a_t)}\middle| H_t\right]
			\mathbb{E}_{\theta}\left[Z_t\middle| H_t\right]}
		\times Y
		\right]\eval_{\theta=0}\\
		=&\mathbb{E}\left[
		\left(\prod_{t=0}^{T}
		\dfrac{\left(Z_t-\mathbb{E}[Z_t\mid H_t]\right)A_t^{(a_t)}}
		{\mathbb{E}\left[A_t^{(a_t)}Z_t\middle| H_t\right]
			-\mathbb{E}\left[A_t^{(a_t)}\middle| H_t\right]
			\mathbb{E}\left[Z_t\middle| H_t\right]}
		\times Y-\psi_{\overline{a}}^o\right)s(O)
		\right]\\
		&+\sum_{t=0}^T\mathbb{E}\left[
		\begin{array}{l}
			\displaystyle\prod_{s=0}^{t-1}\dfrac{\left(Z_s-\mathbb{E}[Z_s\mid H_s]\right)A_s^{(a_s)}}
			{\mathbb{E}\left[A_s^{(a_s)}Z_s\middle| H_s\right]
				-\mathbb{E}\left[A_s^{(a_s)}\middle| H_s\right]
				\mathbb{E}\left[Z_s\middle| H_s\right]}\\
			\times\displaystyle\prod_{s=t+1}^{T}\dfrac{\left(Z_s-\mathbb{E}[Z_s\mid H_s]\right)A_s^{(a_s)}}
			{\mathbb{E}\left[A_s^{(a_s)}Z_s\middle| H_s\right]
				-\mathbb{E}\left[A_s^{(a_s)}\middle| H_s\right]
				\mathbb{E}\left[Z_s\middle| H_s\right]}\times Y\\
			\times\left\{\begin{array}{l}
				\nabla_{\theta}\dfrac{\left(Z_t-\mathbb{E}_{\theta}[Z_t\mid H_t]\right)A_t^{(a_t)}}
				{\mathbb{E}\left[A_t^{(a_t)}Z_t\middle| H_t\right]
					-\mathbb{E}\left[A_t^{(a_t)}\middle| H_t\right]
					\mathbb{E}\left[Z_t\middle| H_t\right]}\displaystyle\eval_{\theta=0}\\
				-\dfrac{\left(Z_t-\mathbb{E}[Z_t\mid H_t]\right)A_t^{(a_t)}}
				{\left\{\mathbb{E}\left[A_t^{(a_t)}Z_t\middle| H_t\right]
					-\mathbb{E}\left[A_t^{(a_t)}\middle| H_t\right]
					\mathbb{E}\left[Z_t\middle| H_t\right]\right\}^2}\\
				\times\nabla_{\theta}\left(\begin{array}{l}
					\mathbb{E}_{\theta}\left[A_t^{(a_t)}Z_t\middle| H_t\right]
					-\mathbb{E}_{\theta}\left[A_t^{(a_t)}\middle| H_t\right]
					\mathbb{E}_{\theta}\left[Z_t\middle| H_t\right]
				\end{array}\right)\eval_{\theta=0}
			\end{array}\right\}
		\end{array}
		\right]\\
		=&\mathbb{E}\left[
		\left(\prod_{t=0}^{T}
		\dfrac{\left(Z_t-\mathbb{E}[Z_t\mid H_t]\right)A_t^{(a_t)}}
		{\mathbb{E}\left[A_t^{(a_t)}Z_t\middle| H_t\right]
			-\mathbb{E}\left[A_t^{(a_t)}\middle| H_t\right]
			\mathbb{E}\left[Z_t\middle| H_t\right]}
		\times Y-\psi_{\overline{a}}^o\right)s(O)
		\right]\\
		&+\sum_{t=0}^T\mathbb{E}\left[
		\begin{array}{l}
			\displaystyle\prod_{s=0}^{t-1}\dfrac{\left(Z_s-\mathbb{E}[Z_s\mid H_s]\right)A_s^{(a_s)}}
			{\mathbb{E}\left[A_s^{(a_s)}Z_s\middle| H_s\right]
				-\mathbb{E}\left[A_s^{(a_s)}\middle| H_s\right]
				\mathbb{E}\left[Z_s\middle| H_s\right]}\\
			\times\displaystyle\prod_{s=t+1}^{T}\dfrac{\left(Z_s-\mathbb{E}[Z_s\mid H_s]\right)A_s^{(a_s)}}
			{\mathbb{E}\left[A_s^{(a_s)}Z_s\middle| H_s\right]
				-\mathbb{E}\left[A_s^{(a_s)}\middle| H_s\right]
				\mathbb{E}\left[Z_s\middle| H_s\right]}\times Y\\
			\times\left\{\begin{array}{l}
				-\dfrac{\mathbb{E}[(Z_t-\rho_t^o(H_t))s(O)\mid H_t]A_t^{(a_t)}}
				{\mathbb{E}\left[A_t^{(a_t)}Z_t\middle| H_t\right]
					-\mathbb{E}\left[A_t^{(a_t)}\middle| H_t\right]
					\mathbb{E}\left[Z_t\middle| H_t\right]}\\
				-\dfrac{\left(Z_t-\mathbb{E}[Z_t\mid H_t]\right)A_t^{(a_t)}}
				{\left\{\mathbb{E}\left[A_t^{(a_t)}Z_t\middle| H_t\right]
					-\mathbb{E}\left[A_t^{(a_t)}\middle| H_t\right]
					\mathbb{E}\left[Z_t\middle| H_t\right]\right\}^2}\\
				\times\left(\begin{array}{l}
					\mathbb{E}\left[(A_t^{(a_t)}Z_t-\kappa_{t,a_t}^o(H_t))s(O)\middle| H_t\right]\\
					-\mathbb{E}\left[(A_t^{(a_t)}-\delta_{t,a_t}^o(H_t))s(O)\middle| H_t\right]
					\rho_t^o(H_t)\\
					-\mathbb{E}\left[(Z_t-\rho_t^o(H_t))s(O)\middle| H_t\right]
					\delta_{t,a_t}^o(H_t)
				\end{array}\right)
			\end{array}\right\}
		\end{array}
		\right]\\
		=&\mathbb{E}\left[
		\left(\prod_{t=0}^{T}
		\dfrac{\left(Z_t-\mathbb{E}[Z_t\mid H_t]\right)A_t^{(a_t)}}
		{\mathbb{E}\left[A_t^{(a_t)}Z_t\middle| H_t\right]
			-\mathbb{E}\left[A_t^{(a_t)}\middle| H_t\right]
			\mathbb{E}\left[Z_t\middle| H_t\right]}
		\times Y-\psi_{\overline{a}}^o\right)s(O)
		\right]\\
		&-\sum_{t=0}^T\mathbb{E}\left[
		\begin{array}{l}
			\displaystyle\prod_{s=0}^{t-1}\dfrac{\left(Z_s-\mathbb{E}[Z_s\mid H_s]\right)A_s^{(a_s)}}
			{\mathbb{E}\left[A_s^{(a_s)}Z_s\middle| H_s\right]
				-\mathbb{E}\left[A_s^{(a_s)}\middle| H_s\right]
				\mathbb{E}\left[Z_s\middle| H_s\right]}\\
			\times\mathbb{E}\left[A_t^{(a_t)}\displaystyle\prod_{s=t+1}^{T}\dfrac{\left(Z_s-\mathbb{E}[Z_s\mid H_s]\right)A_s^{(a_s)}}
			{\mathbb{E}\left[A_s^{(a_s)}Z_s\middle| H_s\right]
				-\mathbb{E}\left[ A_s^{(a_s)}\middle| H_s\right]
				\mathbb{E}\left[Z_s\middle| H_s\right]}Y\middle| H_{t}\right]\\
			\times\left\{\begin{array}{l}
				\dfrac{(Z_t-\rho_t^o(H_t))s(O)}
				{\mathbb{E}\left[A_t^{(a_t)}Z_t\middle| H_t\right]
					-\mathbb{E}\left[A_t^{(a_t)}\middle| H_t\right]
					\mathbb{E}\left[Z_t\middle| H_t\right]}
			\end{array}\right\}
		\end{array}
		\right]\\
		&-\sum_{t=0}^T\mathbb{E}\left[
		\begin{array}{l}
			\displaystyle\prod_{s=0}^{t-1}\dfrac{\left(Z_s-\mathbb{E}[Z_s\mid H_s]\right)A_s^{(a_s)}}
			{\mathbb{E}\left[A_s^{(a_s)}Z_s\middle| H_s\right]
				-\mathbb{E}\left[A_s^{(a_s)}\middle| H_s\right]
				\mathbb{E}\left[Z_s\middle| H_s\right]}\\
			\times\mathbb{E}\left[\begin{array}{l}
				\dfrac{\left(Z_t-\mathbb{E}[Z_t\mid H_t]\right)A_t^{(a_t)}}
				{\left\{\mathbb{E}\left[A_t^{(a_t)}Z_t\middle| H_t\right]
					-\mathbb{E}\left[A_t^{(a_t)}\middle| H_t\right]
					\mathbb{E}\left[Z_t\middle| H_t\right]\right\}^2}\\
				\times\displaystyle\prod_{s=t+1}^{T}\dfrac{\left(Z_s-\mathbb{E}[Z_s\mid H_s]\right)A_s^{(a_s)}}
				{\mathbb{E}\left[A_s^{(a_s)}Z_s\middle| H_s\right]
					-\mathbb{E}\left[A_s^{(a_s)}\middle| H_s\right]
					\mathbb{E}\left[Z_s\middle| H_s\right]}\times Y
			\end{array}\middle| H_{t}\right]\\
			\times\left(\begin{array}{l}
				(A_t^{(a_t)}Z_t-\kappa_{t,a_t}^o(H_t))s(O)\\
				-(A_t^{(a_t)}-\delta_{t,a_t}^o(H_t))
				\rho_t^o(H_t)s(O)\\
				-(Z_t-\rho_t^o(H_t))
				\delta_{t,a_t}^o(H_t)s(O)
			\end{array}\right)
		\end{array}
		\right]\\
		=&\mathbb{E}\left[
		\left(\prod_{t=0}^{T}
		\dfrac{\left(Z_t-\mathbb{E}[Z_t\mid H_t]\right)A_t^{(a_t)}}
		{\mathbb{E}\left[A_t^{(a_t)}Z_t\middle| H_t\right]
			-\mathbb{E}\left[A_t^{(a_t)}\middle| H_t\right]
			\mathbb{E}\left[Z_t\middle| H_t\right]}
		\times Y-\psi_{\overline{a}}^o\right)s(O)
		\right]\\
		&-\sum_{t=0}^T\mathbb{E}\left[
		\begin{array}{l}
			\displaystyle\prod_{s=0}^{t-1}\dfrac{\left(Z_s-\mathbb{E}[Z_s\mid H_s]\right)A_s^{(a_s)}}
			{\mathbb{E}\left[A_s^{(a_s)}Z_s\middle| H_s\right]
				-\mathbb{E}\left[A_s^{(a_s)}\middle| H_s\right]
				\mathbb{E}\left[Z_s\middle| H_s\right]}\\
			\times\mathbb{E}\left[A_t^{(a_t)}\gamma_{t+1,\underline{a}_{t+1}}^o(H_{t+1})\middle| H_{t}\right]\\
			\times\left\{\begin{array}{l}
				\dfrac{(Z_t-\rho_t^o(H_t))s(O)}
				{\mathbb{E}\left[A_t^{(a_t)}Z_t\middle| H_t\right]
					-\mathbb{E}\left[A_t^{(a_t)}\middle| H_t\right]
					\mathbb{E}\left[Z_t\middle| H_t\right]}
			\end{array}\right\}
		\end{array}
		\right]\\
		&-\sum_{t=0}^T\mathbb{E}\left[
		\begin{array}{l}
			\displaystyle\prod_{s=0}^{t-1}\dfrac{\left(Z_s-\mathbb{E}[Z_s\mid H_s]\right)A_s^{(a_s)}}
			{\mathbb{E}\left[A_s^{(a_s)}Z_s\middle| H_s\right]
				-\mathbb{E}\left[A_s^{(a_s)}\middle| H_s\right]
				\mathbb{E}\left[Z_s\middle| H_s\right]}\\
			\times\begin{array}{l}
				\dfrac{\mathbb{E}\left[\left(Z_t-\mathbb{E}[Z_t\mid H_t]\right)A_t^{(a_t)}\times\gamma_{t+1,\underline{a}_{t+1}}^o(H_{t+1})\middle| H_{t}\right]}
				{\left\{\mathbb{E}\left[A_t^{(a_t)}Z_t\middle| H_t\right]
					-\mathbb{E}\left[A_t^{(a_t)}\middle| H_t\right]
					\mathbb{E}\left[Z_t\middle| H_t\right]\right\}^2}
			\end{array}\\
			\times\left(\begin{array}{l}
				(A_t^{(a_t)}Z_t-\kappa_{t,a_t}^o(H_t))s(O)\\
				-(A_t^{(a_t)}-\delta_{t,a_t}^o(H_t))
				\rho_t^o(H_t)s(O)\\
				-(Z_t-\rho_t^o(H_t))
				\delta_{t,a_t}^o(H_t)s(O)
			\end{array}\right)
		\end{array}
		\right]\\
		=&\mathbb{E}\left[
		\left(\prod_{t=0}^{T}
		\dfrac{\left(Z_t-\mathbb{E}[Z_t\mid H_t]\right)A_t^{(a_t)}}
		{\mathbb{E}\left[A_t^{(a_t)}Z_t\middle| H_t\right]
			-\mathbb{E}\left[A_t^{(a_t)}\middle| H_t\right]
			\mathbb{E}\left[Z_t\middle| H_t\right]}
		\times Y-\psi_{\overline{a}}^o\right)s(O)
		\right]\\
		&-\sum_{t=0}^T\mathbb{E}\left[
		\begin{array}{l}
			\displaystyle\prod_{s=0}^{t-1}\dfrac{\left(Z_s-\mathbb{E}[Z_s\mid H_s]\right)A_s^{(a_s)}}
			{\mathbb{E}\left[A_s^{(a_s)}Z_s\middle| H_s\right]
				-\mathbb{E}\left[A_s^{(a_s)}\middle| H_s\right]
				\mathbb{E}\left[Z_s\middle| H_s\right]}\\
			\times\left\{\begin{array}{l}
				\dfrac{(Z_t-\rho_t^o(H_t))\times\eta_{t,\underline{a}_t}^o(H_t)s(O)}
				{\mathbb{E}\left[A_t^{(a_t)}Z_t\middle| H_t\right]
					-\mathbb{E}\left[A_t^{(a_t)}\middle| H_t\right]
					\mathbb{E}\left[Z_t\middle| H_t\right]}
			\end{array}\right\}
		\end{array}
		\right]\\
		&-\sum_{t=0}^T\mathbb{E}\left[
		\begin{array}{l}
			\displaystyle\prod_{s=0}^{t-1}\dfrac{\left(Z_s-\mathbb{E}[Z_s\mid H_s]\right)A_s^{(a_s)}}
			{\mathbb{E}\left[A_s^{(a_s)}Z_s\middle| H_s\right]
				-\mathbb{E}\left[A_s^{(a_s)}\middle| H_s\right]
				\mathbb{E}\left[Z_s\middle| H_s\right]}\\
			\times\begin{array}{l}
				\dfrac{\zeta_{t,\underline{a}_t}^o(H_t)-\rho_t^o(H_t)\eta_{t,\underline{a}_t}^o(H_t)}
				{\left\{\mathbb{E}\left[A_t^{(a_t)}Z_t\middle| H_t\right]
					-\mathbb{E}\left[A_t^{(a_t)}\middle| H_t\right]
					\mathbb{E}\left[Z_t\middle| H_t\right]\right\}^2}
			\end{array}\\
			\times\left(\begin{array}{l}
				(A_t^{(a_t)}Z_t-\kappa_{t,a_t}^o(H_t))\\
				-(A_t^{(a_t)}-\delta_{t,a_t}^o(H_t))
				\rho_t^o(H_t)\\
				-(Z_t-\rho_t^o(H_t))
				\delta_{t,a_t}^o(H_t)
			\end{array}\right)s(O)
		\end{array}
		\right]\\
		=&\mathbb{E}\left[
		\left(\prod_{t=0}^{T}
		\dfrac{\left(Z_t-\rho_t^o(H_t)\right)A_t^{(a_t)}}
		{\kappa_{s,a_s}^o(H_s)}
		\times Y-\psi_{\overline{a}}^o\right)s(O)
		\right]\\
		&-\sum_{t=0}^T\mathbb{E}\left[
		\begin{array}{l}
			\displaystyle\prod_{s=0}^{t-1}\dfrac{\left(Z_s-\rho_t^o(H_t)\right)A_s^{(a_s)}}
			{\kappa_{s,a_s}^o(H_s)}\\
			\times\left\{\begin{array}{l}
				\dfrac{(Z_t-\rho_t^o(H_t))\times\eta_{t,\underline{a}_t}^o(H_t)}
				{\kappa_{t,a_t}^o(H_t)}
			\end{array}\right\}s(O)
		\end{array}
		\right]\\
		&-\sum_{t=0}^T\mathbb{E}\left[
		\begin{array}{l}
			\displaystyle\prod_{s=0}^{t-1}\dfrac{\left(Z_s-\rho_t^o(H_t)\right)A_s^{(a_s)}}
			{\kappa_{s,a_s}^o(H_s)}\\
			\times\begin{array}{l}
				\dfrac{\zeta_{t,\underline{a}_t}^o(H_t)-\rho_t^o(H_t)\eta_{t,\underline{a}_t}^o(H_t)}
				{\left\{\kappa_{t,a_t}^o(H_t)\right\}^2}
			\end{array}\\
			\times\left(\begin{array}{l}
				(A_t^{(a_t)}Z_t-\kappa_{t,a_t}^o(H_t))\\
				-(A_t^{(a_t)}-\delta_{t,a_t}^o(H_t))
				\rho_t^o(H_t)\\
				-(Z_t-\rho_t^o(H_t))
				\delta_{t,a_t}^o(H_t)
			\end{array}\right)s(O)
		\end{array}
		\right]\\
		=&\mathbb{E}\left[\varphi_{\overline{a}}(O;\psi_{\overline{a}}^o,\alpha_{\overline{a}}^o)s(O)\right],
	\end{align*}
	where
	\begin{align*}
		&\varphi_{\overline{a}}(O;\psi_{\overline{a}}^o,\alpha_{\overline{a}}^o):=
		\prod_{t=0}^{T}
		\dfrac{\left(Z_t-\rho_t^o(H_t)\right)A_t^{(a_t)}}
		{\kappa_{s,a_s}^o(H_s)}
		\times Y-\psi_{\overline{a}}^o\\
		&-\sum_{t=0}^T\displaystyle\prod_{s=0}^{t-1}\dfrac{\left(Z_s-\rho_t^o(H_t)\right)A_s^{(a_s)}}
		{\kappa_{s,a_s}^o(H_s)}\times
		\dfrac{(Z_t-\rho_t^o(H_t))\times\eta_{t,\underline{a}_t}^o(H_t)}{\kappa_{t,a_t}^o(H_t)}\\
		&+\sum_{t=0}^T\displaystyle\prod_{s=0}^{t-1}\dfrac{\left(Z_s-\rho_t^o(H_t)\right)A_s^{(a_s)}}
		{\kappa_{s,a_s}^o(H_s)}
		\times\gamma_{t,\underline{a}_t}^o(H_t)
		\\&\times \left(1-\dfrac{\{A_t^{(a_t)}-\delta_{t,a_t}^o(H_t)\}\{Z_t-\rho_t^o(H_t)\}}{\kappa_{t,a_t}^o(H_t)} \right).
	\end{align*}
	This finishes the proof for Theorem~\ref{thm: EIF AIV longitudinal}.

	\subsection{Proof for Theorem~\ref{thm: EIF AIV longitudinal'}}
	Analogous to the proof for Theorem~\ref{thm: EIF AIV longitudinal},
	we calculate the path-wise derivative for
	\begin{align*}
		\psi_{\overline{a},\theta} :=
		\mathbb{E}_\theta\left[
		\prod_{t=0}^{T}
		\frac{\pi_{t,a_t,\theta}^o(Z_t,H_t) - \delta_{t,a_t,\theta}^o(H_t)}
		{\mathrm{Var}_\theta\!\{\pi_{t,a_t,\theta}^o(Z_t,H_t) \mid H_t \}}A_t^{(a_t)}
		\times Y
		\right].
	\end{align*}
	Concretely,
	\begin{align*}
		&\nabla_{\theta}\psi_{\overline{a},\theta}\eval_{\theta=0}
		=\nabla_{\theta}\mathbb{E}_\theta\left[
		\prod_{t=0}^{T}
		\frac{\pi_{t,a_t,\theta}^o(Z_t,H_t) - \delta_{t,a_t,\theta}^o(H_t)}
		{\mathrm{Var}_\theta\!\{\pi_{t,a_t,\theta}^o(Z_t,H_t) \mid H_t \}}A_t^{(a_t)}
		\times Y
		\right]\eval_{\theta=0}\\
		=&\mathbb{E}\left[
		\left\{\prod_{t=0}^{T}
		\frac{\pi_{t,a_t}^o(Z_t,H_t) - \delta_{t,a_t}^o(H_t)}
		{\mathrm{Var}\!\{\pi_{t,a_t}^o(Z_t,H_t) \mid H_t \}}A_t^{(a_t)}
		\times Y - \psi_{\overline{a},*}^o\right\}s(O)
		\right]\\
		&+\sum_{t=0}^T\mathbb{E}\left[\begin{array}{l}
			\displaystyle\prod_{s=0}^{t-1}
			\dfrac{\pi_{s,a_s}^o(Z_s,H_s) - \delta_{s,a_s}^o(H_s)}
			{\mathrm{Var}_\theta\!\{\pi_{s,a_s}^o(Z_s,H_s) \mid H_s \}}A_s^{(a_s)}
			\times\displaystyle\prod_{s=t+1}^{T}
			\dfrac{\pi_{s,a_s}^o(Z_s,H_s) - \delta_{s,a_s}^o(H_s)}
			{\mathrm{Var}_\theta\!\{\pi_{s,a_s}^o(Z_s,H_s) \mid H_s \}}A_s^{(a_s)}\\
			\times\nabla_{\theta}
			\dfrac{\pi_{t,a_t,\theta}^o(Z_t,H_t) - \delta_{t,a_t,\theta}^o(H_t)}
			{\mathrm{Var}_\theta\!\{\pi_{t,a_t,\theta}^o(Z_t,H_t) \mid H_t \}}A_t^{(a_t)}\displaystyle\eval_{\theta=0}
			\times Y
		\end{array}
		\right]\\
		=&\mathbb{E}\left[
		\left\{\prod_{t=0}^{T}
		\frac{\pi_{t,a_t}^o(Z_t,H_t) - \delta_{t,a_t}^o(H_t)}
		{\mathrm{Var}\!\{\pi_{t,a_t}^o(Z_t,H_t) \mid H_t \}}A_t^{(a_t)}
		\times Y - \psi_{\overline{a},*}^o\right\}s(O)
		\right]\\
		&+\sum_{t=0}^T\mathbb{E}\left[\begin{array}{l}
			\displaystyle\prod_{s=0}^{t-1}
			\dfrac{\pi_{s,a_s}^o(Z_s,H_s) - \delta_{s,a_s}^o(H_s)}
			{\mathrm{Var}\!\{\pi_{s,a_s}^o(Z_s,H_s) \mid H_s \}}A_s^{(a_s)}
			\times\displaystyle\prod_{s=t+1}^{T}
			\dfrac{\pi_{s,a_s}^o(Z_s,H_s) - \delta_{s,a_s}^o(H_s)}
			{\mathrm{Var}\!\{\pi_{s,a_s}^o(Z_s,H_s) \mid H_s \}}A_s^{(a_s)}\\
			\times
			\dfrac{\mathbb{E}[\{A_t^{(a_t)} - \pi_{t,a_t}^o(Z_t,H_t)\}s(O)\mid Z_t,H_t] - 
				\mathbb{E}[\{A_t^{(a_t)} - \delta_{t,a_t}^o(H_t)\}s(O)\mid H_t]}
			{\mathrm{Var}\!\{\pi_{t,a_t}^o(Z_t,H_t) \mid H_t \}}A_t^{(a_t)}
			\times Y
		\end{array}
		\right]\\
		&-\sum_{t=0}^T\mathbb{E}\left[\begin{array}{l}
			\displaystyle\prod_{s=0}^{t-1}
			\dfrac{\pi_{s,a_s}^o(Z_s,H_s) - \delta_{s,a_s}^o(H_s)}
			{\mathrm{Var}\!\{\pi_{s,a_s}^o(Z_s,H_s) \mid H_s \}}A_s^{(a_s)}
			\times\displaystyle\prod_{s=t+1}^{T}
			\dfrac{\pi_{s,a_s}^o(Z_s,H_s) - \delta_{s,a_s}^o(H_s)}
			{\mathrm{Var}\!\{\pi_{s,a_s}^o(Z_s,H_s) \mid H_s \}}A_s^{(a_s)}\\
			\times
			\dfrac{\pi_{t,a_t}^o(Z_t,H_t) - \delta_{t,a_t}^o(H_t)}
			{\mathrm{Var}\!\{\pi_{t,a_t}^o(Z_t,H_t) \mid H_t \}^2}A_t^{(a_t)}
			\times  Y\\
			\times\mathbb{E}\left[(\{\pi_{t,a_t}^o(Z_t,H_t) - \delta_{t,a_t}^o(H_t)\}\{2A_t^{(a_t)} - \pi_{t,a_t}^o(Z_t,H_t) - \delta_{t,a_t}^o(H_t)\} - \kappa_t^o(H_t))s(O)\middle| H_t\right]
		\end{array}
		\right]\\
		=&\mathbb{E}\left[
		\left\{\prod_{t=0}^{T}
		\frac{\pi_{t,a_t}^o(Z_t,H_t) - \delta_{t,a_t}^o(H_t)}
		{\mathrm{Var}\!\{\pi_{t,a_t}^o(Z_t,H_t) \mid H_t \}}A_t^{(a_t)}
		\times Y - \psi_{\overline{a},*}^o\right\}s(O)
		\right]\\
		&+\sum_{t=0}^T\mathbb{E}\left[\begin{array}{l}
			\displaystyle\prod_{s=0}^{t-1}
			\dfrac{\pi_{s,a_s}^o(Z_s,H_s) - \delta_{s,a_s}^o(H_s)}
			{\mathrm{Var}\!\{\pi_{s,a_s}^o(Z_s,H_s) \mid H_s \}}A_s^{(a_s)}
			\times\displaystyle\prod_{s=t+1}^{T}
			\dfrac{\pi_{s,a_s}^o(Z_s,H_s) - \delta_{s,a_s}^o(H_s)}
			{\mathrm{Var}\!\{\pi_{s,a_s}^o(Z_s,H_s) \mid H_s \}}A_s^{(a_s)}\\
			\times
			\dfrac{\mathbb{E}[\{A_t^{(a_t)} - \pi_{t,a_t}^o(Z_t,H_t)\}s(O)\mid Z_t,H_t]}
			{\mathrm{Var}\!\{\pi_{t,a_t}^o(Z_t,H_t) \mid H_t \}}A_t^{(a_t)}
			\times Y
		\end{array}
		\right]\\
		&-\sum_{t=0}^T\mathbb{E}\left[\begin{array}{l}
			\displaystyle\prod_{s=0}^{t-1}
			\dfrac{\pi_{s,a_s}^o(Z_s,H_s) - \delta_{s,a_s}^o(H_s)}
			{\mathrm{Var}\!\{\pi_{s,a_s}^o(Z_s,H_s) \mid H_s \}}A_s^{(a_s)}
			\times\displaystyle\prod_{s=t+1}^{T}
			\dfrac{\pi_{s,a_s}^o(Z_s,H_s) - \delta_{s,a_s}^o(H_s)}
			{\mathrm{Var}\!\{\pi_{s,a_s}^o(Z_s,H_s) \mid H_s \}}A_s^{(a_s)}\\
			\times
			\dfrac{\mathbb{E}[\{A_t^{(a_t)} - \delta_{t,a_t}^o(H_t)\}s(O)\mid H_t]}
			{\mathrm{Var}\!\{\pi_{t,a_t}^o(Z_t,H_t) \mid H_t \}}
			\times A_t^{(a_t)}Y
		\end{array}
		\right]\\
		&-\sum_{t=0}^T\mathbb{E}\left[\begin{array}{l}
			\displaystyle\prod_{s=0}^{t-1}
			\dfrac{\pi_{s,a_s}^o(Z_s,H_s) - \delta_{s,a_s}^o(H_s)}
			{\mathrm{Var}\!\{\pi_{s,a_s}^o(Z_s,H_s) \mid H_s \}}A_s^{(a_s)}
			\times\displaystyle\prod_{s=t+1}^{T}
			\dfrac{\pi_{s,a_s}^o(Z_s,H_s) - \delta_{s,a_s}^o(H_s)}
			{\mathrm{Var}\!\{\pi_{s,a_s}^o(Z_s,H_s) \mid H_s \}}A_s^{(a_s)}\\
			\times
			\dfrac{\pi_{t,a_t}^o(Z_t,H_t) - \delta_{t,a_t}^o(H_t)}
			{\mathrm{Var}\!\{\pi_{t,a_t}^o(Z_t,H_t) \mid H_t \}^2}
			\times A_t^{(a_t)}Y\\
			\times\mathbb{E}\left[(\{\pi_{t,a_t}^o(Z_t,H_t) - \delta_{t,a_t}^o(H_t)\}\{2A_t^{(a_t)} - \pi_{t,a_t}^o(Z_t,H_t) - \delta_{t,a_t}^o(H_t)\} - \kappa_t^o(H_t))s(O)\middle| H_t\right]
		\end{array}
		\right]\\
		=&\mathbb{E}\left[
		\left\{\prod_{t=0}^{T}
		\frac{\pi_{t,a_t}^o(Z_t,H_t) - \delta_{t,a_t}^o(H_t)}
		{\mathrm{Var}\!\{\pi_{t,a_t}^o(Z_t,H_t) \mid H_t \}}A_t^{(a_t)}
		\times Y - \psi_{\overline{a},*}^o\right\}s(O)
		\right]\\
		&+\sum_{t=0}^T\mathbb{E}\left[\begin{array}{l}
			\displaystyle\prod_{s=0}^{t-1}
			\dfrac{\pi_{s,a_s}^o(Z_s,H_s) - \delta_{s,a_s}^o(H_s)}
			{\mathrm{Var}\!\{\pi_{s,a_s}^o(Z_s,H_s) \mid H_s \}}A_s^{(a_s)}
			\times
			\dfrac{\{A_t^{(a_t)} - \pi_{t,a_t}^o(Z_t,H_t)\}s(O)}
			{\mathrm{Var}\!\{\pi_{t,a_t}^o(Z_t,H_t) \mid H_t \}}\\
			\mathbb{E}\left[\displaystyle\prod_{s=t+1}^{T}
			\dfrac{\pi_{s,a_s}^o(Z_s,H_s) - \delta_{s,a_s}^o(H_s)}
			{\mathrm{Var}\!\{\pi_{s,a_s}^o(Z_s,H_s) \mid H_s \}}A_s^{(a_s)}\times A_t^{(a_t)} Y\middle| Z_t,H_t\right]
		\end{array}
		\right]\\
		&-\sum_{t=0}^T\mathbb{E}\left[\begin{array}{l}
			\displaystyle\prod_{s=0}^{t-1}
			\dfrac{\pi_{s,a_s}^o(Z_s,H_s) - \delta_{s,a_s}^o(H_s)}
			{\mathrm{Var}\!\{\pi_{s,a_s}^o(Z_s,H_s) \mid H_s \}}A_s^{(a_s)}
			\times\dfrac{\{A_t^{(a_t)} - \delta_{t,a_t}^o(H_t)\}s(O)}
			{\mathrm{Var}\!\{\pi_{t,a_t}^o(Z_t,H_t) \mid H_t \}}\\
			\mathbb{E}\left[\displaystyle\prod_{s=t+1}^{T}
			\dfrac{\pi_{s,a_s}^o(Z_s,H_s) - \delta_{s,a_s}^o(H_s)}
			{\mathrm{Var}\!\{\pi_{s,a_s}^o(Z_s,H_s) \mid H_s \}}A_s^{(a_s)}\times A_t^{(a_t)} Y\middle| H_t\right]
		\end{array}
		\right]\\
		&-\sum_{t=0}^T\mathbb{E}\left[\begin{array}{l}
			\displaystyle\prod_{s=0}^{t-1}
			\dfrac{\pi_{s,a_s}^o(Z_s,H_s) - \delta_{s,a_s}^o(H_s)}
			{\mathrm{Var}\!\{\pi_{s,a_s}^o(Z_s,H_s) \mid H_s \}}A_s^{(a_s)}\\
			\times\mathbb{E}\left[\displaystyle\prod_{s=t}^{T}
			\dfrac{\pi_{s,a_s}^o(Z_s,H_s) - \delta_{s,a_s}^o(H_s)}
			{\mathrm{Var}\!\{\pi_{s,a_s}^o(Z_s,H_s) \mid H_s \}}A_s^{(a_s)}
			\times Y\middle| H_t\right]\\
			\times\dfrac{\{\pi_{t,a_t}^o(Z_t,H_t) - \delta_{t,a_t}^o(H_t)\}\{2A_t^{(a_t)} - \pi_{t,a_t}^o(Z_t,H_t) - \delta_{t,a_t}^o(H_t)\} - \kappa_t^o(H_t)}
			{\mathrm{Var}\!\{\pi_{t,a_t}^o(Z_t,H_t) \mid H_t \}}s(O)
		\end{array}
		\right]\\
		=&\mathbb{E}\left[
		\left\{\prod_{t=0}^{T}
		\frac{\pi_{t,a_t}^o(Z_t,H_t) - \delta_{t,a_t}^o(H_t)}
		{\mathrm{Var}\!\{\pi_{t,a_t}^o(Z_t,H_t) \mid H_t \}}A_t^{(a_t)}
		\times Y - \psi_{\overline{a},*}^o\right\}s(O)
		\right]\\
		&+\sum_{t=0}^T\mathbb{E}\left[\begin{array}{l}
			\displaystyle\prod_{s=0}^{t-1}
			\dfrac{\pi_{s,a_s}^o(Z_s,H_s) - \delta_{s,a_s}^o(H_s)}
			{\mathrm{Var}\!\{\pi_{s,a_s}^o(Z_s,H_s) \mid H_s \}}A_s^{(a_s)}
			\times
			\dfrac{\{A_t^{(a_t)} - \pi_{t,a_t}^o(Z_t,H_t)\}s(O)}
			{\mathrm{Var}\!\{\pi_{t,a_t}^o(Z_t,H_t) \mid H_t \}}
			\xi_{t,\underline{a}_t}^o(Z_t,H_t)
		\end{array}
		\right]\\
		&-\sum_{t=0}^T\mathbb{E}\left[\begin{array}{l}
			\displaystyle\prod_{s=0}^{t-1}
			\dfrac{\pi_{s,a_s}^o(Z_s,H_s) - \delta_{s,a_s}^o(H_s)}
			{\mathrm{Var}\!\{\pi_{s,a_s}^o(Z_s,H_s) \mid H_s \}}A_s^{(a_s)}
			\times\dfrac{\{A_t^{(a_t)} - \delta_{t,a_t}^o(H_t)\}s(O)}
			{\mathrm{Var}\!\{\pi_{t,a_t}^o(Z_t,H_t) \mid H_t \}}
			\eta_{t,\underline{a}_t}^o(H_t)
		\end{array}
		\right]\\
		&-\sum_{t=0}^T\mathbb{E}\left[\begin{array}{l}
			\displaystyle\prod_{s=0}^{t-1}
			\dfrac{\pi_{s,a_s}^o(Z_s,H_s) - \delta_{s,a_s}^o(H_s)}
			{\mathrm{Var}\!\{\pi_{s,a_s}^o(Z_s,H_s) \mid H_s \}}A_s^{(a_s)}
			\times\gamma_{t,\underline{a}_t}^o(H_t)\\
			\times\dfrac{\{\pi_{t,a_t}^o(Z_t,H_t) - \delta_{t,a_t}^o(H_t)\}\{2A_t^{(a_t)} - \pi_{t,a_t}^o(Z_t,H_t) - \delta_{t,a_t}^o(H_t)\} - \kappa_t^o(H_t)}
			{\mathrm{Var}\!\{\pi_{t,a_t}^o(Z_t,H_t) \mid H_t \}}s(O)
		\end{array}
		\right]
	\end{align*}
	In summary, we see that the influence function of $\psi_{\overline{a}_t}$ is
	\begin{align*}
		&\prod_{t=0}^{T}
		\frac{\pi_{t,a_t}^o(Z_t,H_t) - \delta_{t,a_t}^o(H_t)}
		{\mathrm{Var}\!\{\pi_{t,a_t}^o(Z_t,H_t) \mid H_t \}}A_t^{(a_t)}
		\times Y - \psi_{\overline{a},*}^o\\
		&+\sum_{t=0}^T\left(\displaystyle\prod_{s=0}^{t-1}
		\dfrac{\pi_{s,a_s}^o(Z_s,H_s) - \delta_{s,a_s}^o(H_s)}
		{\mathrm{Var}\!\{\pi_{s,a_s}^o(Z_s,H_s) \mid H_s \}}A_s^{(a_s)}\right)\times \dfrac{1}{\mathrm{Var}\!\{\pi_{t,a_t}^o(Z_t,H_t) \mid H_t \}}\\
		&\times
		\left\{\begin{array}{l}
			\{A_t^{(a_t)} - \pi_{t,a_t}^o(Z_t,H_t)\}
			\xi_{t,\underline{a}_t}^o(Z_t,H_t)
			-\{A_t^{(a_t)} - \delta_{t,a_t}^o(H_t)\}
			\eta_{t,\underline{a}_t}^o(H_t)\\
			+\gamma_{t,\underline{a}_t}^o(H_t)
			\times\left(\kappa_t^o(H_t)+\{A_t^{(a_t)} - \pi_{t,a_t}^o(Z_t,H_t)\}^2-\{A_t^{(a_t)} - \delta_{t,a_t}^o(H_t)\}^2\right)
		\end{array}\right\}\\
		=&\prod_{t=0}^{T}
		\frac{\pi_{t,a_t}^o(Z_t,H_t) - \delta_{t,a_t}^o(H_t)}
		{\kappa_{t,a_t}^o(H_t)}A_t^{(a_t)}
		\times Y - \psi_{\overline{a},*}^o\\
		&+\sum_{t=0}^T\left(\displaystyle\prod_{s=0}^{t-1}
		\dfrac{\pi_{s,a_s}^o(Z_s,H_s) - \delta_{s,a_s}^o(H_s)}
		{\kappa_{s,a_s}^o(H_s)}A_s^{(a_s)}\right)\times \dfrac{1}{\kappa_{t,a_t}^o(H_t)}\\
		&\times
		\left\{\begin{array}{l}
			\{A_t^{(a_t)} - \pi_{t,a_t}^o(Z_t,H_t)\}
			\xi_{t,\underline{a}_t}^o(Z_t,H_t)
			-\{A_t^{(a_t)} - \delta_{t,a_t}^o(H_t)\}
			\eta_{t,\underline{a}_t}^o(H_t)\\
			+\gamma_{t,\underline{a}_t}^o(H_t)
			\times\left(\kappa_t^o(H_t)+\{A_t^{(a_t)} - \pi_{t,a_t}^o(Z_t,H_t)\}^2-\{A_t^{(a_t)} - \delta_{t,a_t}^o(H_t)\}^2\right)
		\end{array}\right\}.
	\end{align*}
	This finishes the proof that $\varphi_{\overline{a},*}(O;\psi_{\overline{a},*}^o,\beta_{\overline{a}}^o)$ is the influence function for $\psi_{\overline{a},*}^o$. 
	Since the tangent space spanned by the score functions equals \(L_2(O)\), we know that the EIF is just $\varphi_{\overline{a},*}(O;\psi_{\overline{a},*}^o,\beta_{\overline{a}}^o)$, finishing the proof for Theorem~\ref{thm: EIF AIV longitudinal'}.

	\subsection{Proof for Theorem~\ref{thm: MIV identification}}
	From Equation~\eqref{exp:6}, we see that
	\begin{equation*}
		\begin{aligned}
			f_a^o(0,L)=&\mathbb{E}\left[g_a(U,L)\{ 1-\Pr(A\neq a\mid Z=z,U,L)\}\middle| L\right]\\
			&-\mathbb{E}\left[\{f_a^o(1,L)-f_a^o(0,L)\}\{1-\Pr(A\neq a\mid Z=z,U,L)\}\middle| L\right].
		\end{aligned}
	\end{equation*}	
	If there exists $b(U,L)$ and $c(Z,L)$, such that $\Pr(A\neq a\mid Z=z,U,L)=b(U,L)c(Z,L)$, and
	\begin{align}
		f_a^o(0,L)=&\mathbb{E}\left[\{g_a(U,L)-f_a^o(1,L)+f_a^o(0,L)\}\{1-b(U,L)c(z,L)\}\middle| L\right]
		\notag\\
		=&\mathbb{E}\left[g_a(U,L)-f_a^o(1,L)+f_a^o(0,L)\middle| L\right]\label{exp:2}\\
		&-c(z,L)\mathbb{E}\left[\{g_a(U,L)-f_a^o(1,L)+f_a^o(0,L)\}b(U,L)\middle| L\right].\notag
	\end{align}
	If for any $l$, there exists $z_1,z_2$, such that $c(z_1,l)\neq c(z_2,l)$, then we can take difference to deduce that
	\begin{align*}
		&\{c(z_1,l)-c(z_2,l)\}\mathbb{E}\left[\{g_a(U,L)-f_a^o(1,L)+f_a^o(0,L)\}b(U,L)\middle| L=l\right]=0\\
		\Rightarrow&\mathbb{E}\left[\{g_a(U,L)-f_a^o(1,L)+f_a^o(0,L)\}b(U,L)\middle| L\right]=0\\
		\Rightarrow&\mathbb{E}\left[\{g_a(U,L)-f_a^o(1,L)+f_a^o(0,L)\}\Pr(A\neq a\mid U,L)\middle| L\right]=0\\
		\Rightarrow&\mathbb{E}\left[\{g_a(U,L)-f_a^o(1,L)+f_a^o(0,L)\}I(A\neq a)\middle| L\right]=0\\
		\Rightarrow&\mathbb{E}\left[\{g_a(U,L)-f_a^o(1,L)+f_a^o(0,L)\}\middle| L,A\neq a\right]=0.
	\end{align*}
	Now we know that 
	\begin{align*}
		&f_a^o(1,L)-f_a^o(0,L)=\mathbb{E}\left[g_a(U,L)\middle| L,A\neq a\right]\\
		=&\mathbb{E}\left[\mathbb{E}[Y(a)\mid U,L]\middle| L,A\neq a\right]\\
		=&\mathbb{E}\left[\mathbb{E}[Y(a)\mid U,L,A\neq a]\middle| L,A\neq a\right]\\
		=&\mathbb{E}\left[Y(a)\middle| L,A\neq a\right].
	\end{align*}
	We can now substitute this expression into Equation~\eqref{exp:2} to deduce that 
	\begin{align*}
		f_a^o(0,L)=&\mathbb{E}\left[g_a(U,L)-f_a^o(1,L)+f_a^o(0,L)\middle| L\right]\\
		=&\mathbb{E}[Y(a)\mid L]-\mathbb{E}[Y(a)\mid L,A\neq a].\\
		f_a^o(1,L)=&f_a^o(0,L)+\mathbb{E}[Y(a)\mid L,A\neq a]=\mathbb{E}[Y(a)\mid L].
	\end{align*}
	This finishes the proof for Theorem~\ref{thm: MIV identification}.

	\subsection{Proof for Theorem~\ref{thm: EIF for MIV}}
	Analogous to the proof for Theorem~\ref{thm: EIF AIV longitudinal},
	we calculate the path-wise derivative for
	\begin{align*}
		\psi_{a,MIV}^o:= \mathbb E\left[
		(1-A^{(a)})\dfrac{\mathrm{Cov}\{A^{(a)}Y,\pi(Z,L)\mid L\}}{\mathrm{Cov}\{A^{(a)},\pi(Z,L)\mid L\}}
		+A^{(a)}Y
		\right].
	\end{align*}
	The EIF can be derived as
	\begin{align*}
		&\nabla_{\theta}
		\mathbb E_\theta\left[
		(1-A^{(a)})\dfrac{\mathrm{Cov}_\theta\{A^{(a)}Y,Z\mid L\}}{\mathrm{Cov}_\theta\{A^{(a)},Z\mid L\}}
		+A^{(a)}Y
		\right]\eval_{\theta=0}\\
		=&\mathbb E\left[
		\begin{array}{l}
			\left\{
			(1-A^{(a)})\gamma_{a}^o(L)
			+A^{(a)}Y-\psi_{a,MIV}^o
			\right\}s(O)\\
			+(1-A^{(a)})\nabla_{\theta}\left(\dfrac{\mathbb{E}_\theta[A^{(a)}Y(Z-\mathbb{E}_\theta[Z\mid L])\mid L]}{\mathrm{Cov}\{A^{(a)},Z\mid L\}}\right)\displaystyle\eval_{\theta=0}\\
			+(1-A^{(a)})\nabla_{\theta}\left(\dfrac{\mathrm{Cov}\{A^{(a)}Y,Z\mid L\}}
			{\mathrm{Cov}_\theta\{A^{(a)},Z\mid L\}}\right)\displaystyle\eval_{\theta=0}
		\end{array}
		\right]\\
		=&\mathbb E\left[
		\begin{array}{l}
			\left\{
			(1-A^{(a)})\gamma_{a}^o(L)
			+A^{(a)}Y-\psi_{a,MIV}^o
			\right\}s(O)\\
			+\dfrac{(1-A^{(a)})}
			{\kappa_{a}^o(L)}
			\left\{
			\begin{array}{l}
				\mathbb{E}[\left\{A^{(a)}Y (Z - \mathbb{E}[Z\mid L]) - \gamma_{a}^o(L)\kappa_{a}^o(L)\right\} s(O)\mid L]\\
				-\mathbb{E}[A^{(a)}Y \mathbb{E}[ (Z-\rho^o(L))s(O)\mid L ]\mid L]
			\end{array}
			\right\}\\
			-\dfrac{(1-A^{(a)})\gamma_a^o(L)}{\kappa_{a}^o(L)}
			\left\{\begin{array}{l}
				\mathbb{E}[\left\{A^{(a)} (Z - \mathbb{E}[Z\mid L]) - \kappa_{a}^o(L)\right\} s(O)\mid L]\\
				-\mathbb{E}[A^{(a)} \mathbb{E}[ (Z-\rho^o(L))s(O)\mid L ]\mid L]
			\end{array}\right\}
		\end{array}
		\right]\\
		=&\mathbb E\left[
		\begin{array}{l}
			\left\{
			(1-A^{(a)})\gamma_{a}^o(L)
			+A^{(a)}Y-\psi_{a,MIV}^o
			\right\}s(O)\\
			+\dfrac{(1-\delta_a^o(L))}
			{\kappa_{a}^o(L)}
			\left\{
			\begin{array}{l}
				\left\{A^{(a)}Y (Z - \mathbb{E}[Z\mid L]) - \gamma_{a}^o(L)\kappa_{a}^o(L)\right\} s(O)\\
				-A^{(a)}Y \mathbb{E}[ (Z-\rho^o(L))s(O)\mid L ]
			\end{array}
			\right\}\\
			-\dfrac{(1-\delta_a^o(L))\gamma_a^o(L)}{\kappa_a^o(L)} 
			\left\{\begin{array}{l}
				\left\{A^{(a)} (Z - \mathbb{E}[Z\mid L]) - \kappa_{a}^o(L)\right\} s(O)\\
				-A^{(a)} \mathbb{E}[ (Z-\rho^o(L))s(O)\mid L ]
			\end{array}\right\}
		\end{array}
		\right]\\
		=&\mathbb E\left[
		\begin{array}{l}
			\left\{
			(1-A^{(a)})\gamma_{a}^o(L)
			+A^{(a)}Y-\psi_{a,MIV}^o
			\right\}s(O)\\
			+\dfrac{(1-\delta_a^o(L))}
			{\kappa_{a}^o(L)}
			\left\{
			\begin{array}{l}
				\left\{A^{(a)}Y (Z - \mathbb{E}[Z\mid L]) - \gamma_{a}^o(L)\kappa_{a}^o(L)\right\} s(O)\\
				-\eta_a^o(L)  (Z-\rho^o(L))s(O)
			\end{array}
			\right\}\\
			-\dfrac{(1-\delta_a^o(L))\gamma_a^o(L)}{\kappa_a^o(L)} 
			\left\{\begin{array}{l}
				\left\{A^{(a)} (Z - \mathbb{E}[Z\mid L]) - \kappa_{a}^o(L)\right\} s(O)\\
				-\delta_a^o(L)  (Z-\rho^o(L))s(O)
			\end{array}\right\}
		\end{array}
		\right].
	\end{align*}
	The EIF corresponds to be
	\begin{align*}
		&(1-A^{(a)})\gamma_{a}^o(L)+A^{(a)}Y-\psi_{a,MIV}^o\\
		&+\dfrac{(1-\delta_a^o(L))}
		{\kappa_{a}^o(L)}
		\left\{
		\begin{array}{l}
			A^{(a)}Y (Z - \rho^o(L)) - \gamma_{a}^o(L)\kappa_{a}^o(L) \\
			-\eta_a^o(L)  (Z-\rho^o(L))
		\end{array}
		\right\}\\
		&-\dfrac{(1-\delta_a^o(L))\gamma_a^o(L)}{\kappa_a^o(L)} 
		\left\{\begin{array}{l}
			A^{(a)} (Z - \rho^o(L)) - \kappa_{a}^o(L) \\
			-\delta_a^o(L)  (Z-\rho^o(L))
		\end{array}\right\}\\
		=&(1-A^{(a)})\gamma_{a}^o(L)+A^{(a)}Y-\psi_{a,MIV}^o\\
		&+\dfrac{(1-\delta_a^o(L))}
		{\kappa_{a}^o(L)}
		\left\{
		\begin{array}{l}
			A^{(a)}Y (Z - \rho^o(L)) \\
			-\eta_a^o(L)  (Z-\rho^o(L))
		\end{array}
		\right\}\\
		&-\dfrac{(1-\delta_a^o(L))\gamma_a^o(L)}{\kappa_a^o(L)} 
		\left\{\begin{array}{l}
			A^{(a)} (Z - \rho^o(L)) \\
			-\delta_a^o(L)  (Z-\rho^o(L))
		\end{array}\right\}\\
		=&(1-A^{(a)})\gamma_{a}^o(L)+A^{(a)}Y-\psi_{a,MIV}^o\\
		&+\dfrac{(1-\delta_a^o(L))}
		{\kappa_{a}^o(L)}
		\left(A^{(a)}Y-\eta_a^o(L)\right)(Z-\rho^o(L))\\
		&-\dfrac{(1-\delta_a^o(L))}{\kappa_a^o(L)}\gamma_a^o(L) 
		\left(A^{(a)}-\delta_a^o(L)\right)(Z - \rho^o(L)).
	\end{align*}
	This finishes the proof for Theorem~\ref{thm: EIF for MIV}.

	\subsection{Proof for Theorem~\ref{thm: MSM identification}}
	For any measurable and bounded $f(A,V)$, we directly calculate
	\begin{align*}
		&\mathbb{E}\left[\omega_{\pi}^o(A, Z, L) f(A, V) \left\{Y - g(A, V; \psi_{MSM}^o)\right\}\middle| L\right]\\
		=&\sum_{a\in\mathcal{A}}f(a, V)\mathbb{E}\left[I\{A=a\}\omega_{\pi}^o(a, Z, L)  \left\{Y - g(a, V; \psi_{MSM}^o)\right\}\middle| L\right]\\
		=&\sum_{a\in\mathcal{A}}f(a, V)\mathbb{E}\left[I\{A=a\} \dfrac{\pi(Z,L)-\mathbb{E}[\pi(Z,L)\mid L]}{\mathrm{Cov}\!\{I\{A=a\},\pi(Z,L)\mid L\}} \left\{Y - g(a, V; \psi_{MSM}^o)\right\}\middle| L\right]\\
		=&\sum_{a\in\mathcal{A}}f(a, V)\mathbb{E}\left[ \dfrac{\pi(Z,L)-\mathbb{E}[\pi(Z,L)\mid L]}{\mathrm{Cov}\!\{I\{A=a\},\pi(Z,L)\mid L\}} A^{(a)}Y\middle| L\right]\\
		&-\sum_{a\in\mathcal{A}}f(a, V)g(a, V; \psi_{MSM}^o)\mathbb{E}\left[ \dfrac{\pi(Z,L)-\mathbb{E}[\pi(Z,L)\mid L]}{\mathrm{Cov}\!\{I\{A=a\},\pi(Z,L)\mid L\}} A^{(a)}\middle| L\right]\\
		=&\sum_{a\in\mathcal{A}}f(a, V)\dfrac{\mathrm{Cov}\!\{I\{A=a\}Y,\pi(Z,L)\mid L\}}{\mathrm{Cov}\!\{I\{A=a\},\pi(Z,L)\mid L\}} \\
		&-\sum_{a\in\mathcal{A}}f(a, V)g(a, V; \psi_{MSM}^o)\dfrac{\mathrm{Cov}\!\{I\{A=a\},\pi(Z,L)\mid L\}}
		{\mathrm{Cov}\!\{I\{A=a\},\pi(Z,L)\mid L\}}\\
		=&\mathbb{E}\left[\sum_{a\in\mathcal{A}}f(a, V)\{Y(a)-g(a,V;\psi_{MSM}^o)\}\middle| L\right].
	\end{align*}
	The last step follows from Theorem~\ref{thm: AIV identification}.
	Now we can take expectation with respect to $L$ and get 
	\begin{align*}
		&\mathbb{E}\left[\omega_{\pi}^o(A, Z, L) f(A, V) \left\{Y - g(A, V; \psi_{MSM}^o)\right\}\right]\\
		=&\mathbb{E}\left[\sum_{a\in\mathcal{A}}f(a, V)\{Y(a)-g(a,V;\psi_{MSM}^o)\}\right] \\
		=&\mathbb{E}\left[\sum_{a\in\mathcal{A}}f(a, V)\{\mathbb{E}[Y(a)\mid V]-g(a,V;\psi_{MSM}^o)\}\right]=0.
	\end{align*}
	The final step follows from the definition of $g(a,V;\psi_{MSM}^o)$ in Equation~\eqref{eq: MSM}.
	Since this equality holds for any $f(A,V)$, we conclude that
	\begin{align*}
		\mathbb{E}\left[\omega_{\pi}^o(A, Z, L)\left\{Y - g(A, V; \psi_{MSM}^o)\right\}\mid A,V\right]=0.
	\end{align*}

	\subsection{Proof for Theorem~\ref{thm: cut IV}}
	We first verify one important fact, that is, if $Z$ is an AIV for $A=a$, then $Z_\mathcal{S}$ is also an AIV for $A=a$. Since $Z$ is an AIV for $A=a$, there exists $b(U,L)$ and $c(Z,L)$, such that
	\begin{equation*}
		\Pr(A = a \mid Z, U, L) = b(U, L) + c(Z, L).
	\end{equation*}
	We define $\tilde c(Z_{\mathcal{S}},L)=\int c(z,L)\text d P_{Z\mid Z_{\mathcal{S}},L}(z\mid Z_\mathcal{S},L)$. 
	Then we can use the law of total probability to verify that
	\begin{align}
		&\Pr(A = a \mid Z_{\mathcal{S}}=z_\mathcal{S}, U, L)\notag\\
		=&\int\Pr(A = a \mid  Z=z, Z_{\mathcal{S}}=z_\mathcal{S}, U, L)\text d P_{Z\mid Z_{\mathcal{S}},U,L}(z\mid z_\mathcal{S},U,L)\notag\\
		=&\int\Pr(A = a \mid  Z=z, Z_{\mathcal{S}}=z_\mathcal{S}, U, L)\text d P_{Z\mid Z_{\mathcal{S}},L}(z\mid z_\mathcal{S},L)\label{exp:7}\\
		=&\int\Pr(A = a \mid  Z=z, U, L)\text d P_{Z\mid Z_{\mathcal{S}},L}(z\mid z_\mathcal{S},L)\label{exp:8}\\
		=&\int b(U,L)+c(z,L)\text d P_{Z\mid Z_{\mathcal{S}},L}(z\mid z_\mathcal{S},L)\notag\\
		=&b(U,L)+\int c(z,L)\text d P_{Z\mid Z_{\mathcal{S}},L}(z\mid z_\mathcal{S},L)\notag\\
		=&b(U,L)+\tilde c(z_{\mathcal{S}},L).\notag
	\end{align}
	Equation~\eqref{exp:7} follows by Assumption~\ref{as: IV independence} that $Z\indep U\mid L$.
	Equation~\eqref{exp:8} follows by when $Z=z$ and $P_{Z\mid Z_{\mathcal{S}},L}(z\mid z_\mathcal{S},L)>0$, it must holds that $Z_{\mathcal{S}}=z_\mathcal{S}$.
	
	This is equivalent to say that $Z_\mathcal{S}$ is an AIV for $A=a$. Furthermore, Assumptions~\ref{as: consistency}--\ref{as: IV relevance} still holds for $Z_\mathcal{S}$. Now we adopt Theorem~\ref{thm: AIV identification} for $Z_\mathcal{S}$ to deduce the results.

	\subsection{Proof for Theorem~\ref{thm: AIV identification continuous}}
	For any fixed $a\in\mathcal{A}$, denote $g(u,L):=\mathbb{E}[Y(a)\mid U,L]$. We first observe that
	\begin{align*}
		&\mathbb{E}[Y\pi(Z,L)\mid A=a,L]=\mathbb{E}[\mathbb{E}[Y\mid Z,A=a,U,L]\pi(Z,L)\mid A=a,L]\\
		=&\mathbb{E}[\mathbb{E}[Y(a)\mid U,L]\pi(Z,L)\mid A=a,L]\\
		=&\int g(u,L)\pi(z,L)p_{Z,U\mid A,L}(z,u\mid a,L)\text d\mu_Z(z)\text d\mu_U(u)\\
		=&\dfrac{1}{p_{A\mid L}(a\mid L)}\int g(u,L)\pi(z,L)p_{A,Z,U\mid L}(a,z,u\mid L)\text d\mu_Z(z)\text d\mu_U(u)\\
		=&\dfrac{1}{p_{A\mid L}(a\mid L)}\int g(u,L)\pi(z,L)p_{A\mid Z,U,L}(a\mid z,u,L)p_{Z\mid L}(z\mid L)p_{U\mid L}(u\mid L)\text d\mu_Z(z)\text d\mu_U(u)\\
		=&\dfrac{1}{p_{A\mid L}(a\mid L)}\int g(u,L)\pi(z,L)b(z,L)p_{Z\mid L}(z\mid L)p_{U\mid L}(u\mid L)\text d\mu_Z(z)\text d\mu_U(u)\\
		&+\dfrac{1}{p_{A\mid L}(a\mid L)}\int g(u,L)\pi(z,L)c(u,L)p_{Z\mid L}(z\mid L)p_{U\mid L}(u\mid L)\text d\mu_Z(z)\text d\mu_U(u)\\
		=&\dfrac{1}{p_{A\mid L}(a\mid L)}\int g(u,L)p_{U\mid L}(u\mid L)\text d\mu_U(u)\cdot\int \pi(z,L)b(z,L)p_{Z\mid L}(z\mid L) \text d\mu_Z(z)\\
		&+\dfrac{1}{p_{A\mid L}(a\mid L)}\int g(u,L)c(u,L)p_{U\mid L}(u\mid L)\text d\mu_U(u)\int \pi(z,L)p_{Z\mid L}(z\mid L)\text d\mu_Z(z)\\
		=&\dfrac{1}{p_{A\mid L}(a\mid L)}\mathbb{E}[Y(a)\mid L]\cdot\int \pi(z,L)b(z,L)p_{Z\mid L}(z\mid L) \text d\mu_Z(z)\\
		&+\dfrac{1}{p_{A\mid L}(a\mid L)}\int g(u,L)c(u,L)p_{U\mid L}(u\mid L)\text d\mu_U(u)\cdot\mathbb{E}[\pi(Z,L)\mid L].
	\end{align*}
	Similarly, we can just set $\pi(Z,L)\equiv 1$ to verify that
	\begin{align*}
		\mathbb{E}[Y\mid A=a,L]=&\dfrac{1}{p_{A\mid L}(a\mid L)}\mathbb{E}[Y(a)\mid L]\cdot\int b(z,L)p_{Z\mid L}(z\mid L) \text d\mu_Z(z)\\
		&+\dfrac{1}{p_{A\mid L}(a\mid L)}\int g(u,L)c(u,L)p_{U\mid L}(u\mid L)\text d\mu_U(u).
	\end{align*}
	Finally, we conclude that
	\begin{align*}
		&\mathbb{E}[Y\pi(Z,L)\mid A=a,L]-\mathbb{E}[Y\mid A=a,L]\mathbb{E}[\pi(Z,L)\mid L]\\
		=&\dfrac{1}{p_{A\mid L}(a\mid L)}\mathbb{E}[Y(a)\mid L]\cdot\int \pi(z,L)b(z,L)p_{Z\mid L}(z\mid L) \text d\mu_Z(z)\\
		&-\dfrac{1}{p_{A\mid L}(a\mid L)}\mathbb{E}[Y(a)\mid L]\cdot\int b(z,L)p_{Z\mid L}(z\mid L) \text d\mu_Z(z)\mathbb{E}[\pi(Z,L)\mid L].
	\end{align*}
	Similarly, we can verify that
	\begin{align*}
		&\mathbb{E}[1\cdot\pi(Z,L)\mid A=a,L]-\mathbb{E}[1\mid A=a,L]\mathbb{E}[\pi(Z,L)\mid L]\\
		=&\dfrac{1}{p_{A\mid L}(a\mid L)}\cdot\int \pi(z,L)b(z,L)p_{Z\mid L}(z\mid L) \text d\mu_Z(z)\\
		&-\dfrac{1}{p_{A\mid L}(a\mid L)}\cdot\int b(z,L)p_{Z\mid L}(z\mid L) \text d\mu_Z(z)\mathbb{E}[\pi(Z,L)\mid L].
	\end{align*}
	Now we get the result that
	\begin{align*}
		&\mathbb{E}[Y\pi(Z,L)\mid A=a,L]-\mathbb{E}[Y\mid A=a,L]\mathbb{E}[\pi(Z,L)\mid L]\\
		=&\{\mathbb{E}[\pi(Z,L)\mid A=a,L]-\mathbb{E}[\pi(Z,L)\mid L]\}\mathbb{E}[Y(a)\mid L].
	\end{align*}
	From this equality, we know that
	\begin{align*}
		\mathbb{E}[Y(a)]=\mathbb{E}\left[\dfrac{\mathbb{E}[Y\pi(Z,L)\mid A=a,L]-\mathbb{E}[Y\mid A=a,L]\mathbb{E}[\pi(Z,L)\mid L]}
		{\mathbb{E}[\pi(Z,L)\mid A=a,L]-\mathbb{E}[\pi(Z,L)\mid L]}\right].
	\end{align*}

	\section{Proof for Propositions}	
	
	\subsection{Proof for Proposition~\ref{prop: existence of RWFs}}
	Without loss of generality, we only consider the case that $A=1$.
	If Assumption~\ref{as: strong IV relevance} holds, we can take 
	\(\pi^{o}(Z,L) = \Pr(A=1 \mid Z,L)\)
	to verify that $\pi^{o}(Z,L)$ is an RWF. Indeed, $\pi^{o}(Z,L)$ is uniformly bounded by one, and  
	\[
	\mathrm{Cov}\!\{A,\pi^{o}(Z,L) \mid L\} 
	= \mathrm{Var}\!\{\Pr(A=1 \mid Z,L) \mid L\}
	\]
	is uniformly bounded below by some positive constant $\epsilon_{0} > 0$.
	
	Conversely, suppose there exists an RWF $\pi(Z,L)$. By the Cauchy–Schwarz inequality, we have  
	\[
	\mathrm{Var}\!\{\Pr(A=1 \mid Z,L) \mid L\} 
	\ge \frac{\left|\mathrm{Cov}\!\{\Pr(A=1 \mid Z,L), \pi(Z,L) \mid L\}\right|^2}{\mathrm{Var}\!\{\pi(Z,L) \mid L\}}.
	\]
	Noting that $\mathrm{Cov}\!\{\Pr(A=1 \mid Z,L), \pi(Z,L) \mid L\} = \mathrm{Cov}\!\{A, \pi(Z,L) \mid L\}$, the RWF property yields  
	\[
	\mathrm{Var}\!\{\Pr(A=1 \mid Z,L) \mid L\} 
	\ge \frac{\epsilon_{0}^2}{\sup_{L \in \mathcal{L}} \mathrm{Var}\!\{\pi(Z,L) \mid L\}}.
	\]
	
	The uniform boundedness of $\pi(Z,L)$ implies that  
	$\mathrm{Var}\!\{\Pr(A=1 \mid Z,L) \mid L\}$ is uniformly bounded away from zero.  
	Thus, $\pi_*^o(Z,L):=\Pr(A=1 \mid Z,L)$ is an RWF for $A$.
	This completes the proof of the statement in Assumption~\ref{as: strong IV relevance}.

	\subsection{Proof for Proposition~\ref{prop: AIV equivalent form}}
	If $Z$ is an AIV for $A = a$, then there exists $b(U,L)$ and $c(Z,L)$, such that 
	\begin{align*}
		\mathbb{E}[A^{(a)}\pi(Z,L)\mid U,L] =& \mathbb{E}[\mathbb{E}[A^{(a)}\mid Z,U,X]\pi(Z,L)\mid U,L]\\
		=&\mathbb{E}[\{b(U,L)+c(Z,L)\}\pi(Z,L)\mid U,L]\\
		=&b(U,L)\mathbb{E}[\pi(Z,L)\mid U,L] + \mathbb{E}[c(Z,L)\pi(Z,L)\mid U,L]\\
		=&b(U,L)\mathbb{E}[\pi(Z,L)\mid L] + \mathbb{E}[c(Z,L)\pi(Z,L)\mid L];\\
		\mathbb{E}[A^{(a)}\mid U,L]=&b(U,L) + \mathbb{E}[c(Z,L)\mid L];\\
		\mathrm{Cov}\{A^{(a)},\pi(Z,L)\mid U,L\}=&
		\mathbb{E}[c(Z,L)\pi(Z,L)\mid L]-
		\mathbb{E}[c(Z,L)\mid L]\mathbb{E}[\pi(Z,L)\mid L]\\
		=&\mathrm{Cov}\{c(Z,L),\pi(Z,L)\mid L\}.
	\end{align*}
	Analogously, one can verify that
	\begin{align}\label{exp:100}
		\mathrm{Cov}\{A^{(a)},\pi(Z,L)\mid L\} = \mathrm{Cov}\{c(Z,L),\pi(Z,L)\mid L\}
		=\mathrm{Cov}\{A^{(a)},\pi(Z,L)\mid U,L\}.
	\end{align}
	Reversely, if Equation~\eqref{exp:100} holds true for any $\pi(Z,L)$, then we know that
	\begin{align*}
		&\mathbb{E}[A^{(a)}\pi(Z,L)\mid U,L]-\mathbb{E}[A^{(a)}\pi(Z,L)\mid L]\\
		=&\{\mathbb{E}[A^{(a)}\mid U,L]-\mathbb{E}[A^{(a)}\mid L]\}\mathbb{E}[\pi(Z,L)\mid L].
	\end{align*}
	
	For any subset $\mathcal{S}\subseteq \mathcal{Z}$, we choose $\delta>0$ and $\pi(Z,L)=I(Z\in \mathcal{S})$
	\begin{align*}
		&\mathbb{E}[A^{(a)}I(Z\in \mathcal{S})\mid U,L]-\mathbb{E}[A^{(a)}I(Z\in \mathcal{S})\mid L]\\
		=&\{\mathbb{E}[A^{(a)}\mid U,L]-\mathbb{E}[A^{(a)}\mid L]\}\mathbb{E}[I(Z\in \mathcal{S})\mid L].\\
		\Rightarrow\quad &
		\mathbb{E}[A^{(a)}\mid Z\in \mathcal{S}, U,L]-\mathbb{E}[A^{(a)}\mid Z\in \mathcal{S},L]=
		\mathbb{E}[A^{(a)}\mid U,L]-\mathbb{E}[A^{(a)}\mid L].
	\end{align*}
	This implies that 
	\begin{align*}
		\mathbb{E}[A^{(a)}\mid Z, U,L]-\mathbb{E}[A^{(a)}\mid Z,L]=
		\mathbb{E}[A^{(a)}\mid U,L]-\mathbb{E}[A^{(a)}\mid L].
	\end{align*}
	We can take $b(U,L) = \mathbb{E}[A^{(a)}\mid U,L]-\mathbb{E}[A^{(a)}\mid L]$ and $c(Z,L) = \mathbb{E}[A^{(a)}\mid Z,L]$, and verify that
	\begin{align*}
		\mathbb{E}[A^{(a)}\mid Z, U,L] = b(U,L)+c(Z,L),
	\end{align*}
	finishing the proof for Proposition~\ref{prop: AIV equivalent form}.

	\subsection{Proof for Proposition~\ref{prop: mixed bias property fixed pi}}
	The proof of this theorem follows from the same approach as Proposition~\ref{prop: mixed bias property longitudinal}, which establishes a parallel result in the longitudinal setting.

	\subsection{Proof for Proposition~\ref{prop: lower efficiency bound}}
	Recall that
	\begin{align*}
		&\pi_*^o(Z,L)=\Pr(A=1\mid Z,L),
		\\ &\gamma^o(L):=\dfrac{\mathrm{Cov}\!\{Y, \pi(Z,L) \mid L\}}{\mathrm{Cov}\!\{A, \pi(Z,L) \mid L\}}=\dfrac{\zeta_{\pi}^o(L)-\eta^o(L)\rho_{\pi}^o(L)}
		{\kappa_{\pi}^o(L)}.
	\end{align*}
	We can see from Theorem~\ref{thm: Y(1)-Y(0)} that $\gamma^o(L)$ is a function that do not related to the choose of $\pi(Z,L)$ and it solve the nonparametric IV problem in Equation~\eqref{eq: npiv Y(1)-Y(0)} as
	\begin{align*}
		\mathbb{E}[Y\mid Z,L]=\mathbb{E}[f^o(A,L)\mid Z,L]=\mathbb{E}[f^o(0,L)+A\gamma^o(L)\mid Z,L].
	\end{align*}
	From this equality, we can see that $\mathbb{E}[Y-A\gamma^o(L)\mid Z,L]=f^o(0,L)$.
	\begin{align*}
		\mathrm{Var}\!\{\mathbb{E}[\varphi_{\pi}(O;\psi_{*}^o,\alpha_{\pi}^o)\mid L]\}=
		\mathrm{Var}\left\{\dfrac{\zeta_{\pi}^o(L)-\eta^o(L)\rho_{\pi}^o(L)}
		{\kappa_{\pi}^o(L)}\right\}
		=\mathrm{Var}\left\{\gamma^o(L)\right\},
	\end{align*}
	which is a constant that not related to $\pi(Z,L)$.
	\begin{align*}
		&\mathrm{Var}\!\{\varphi_{\pi}(O;\psi_{*}^o,\alpha_{\pi}^o) \mid L\}\\
		=&\dfrac{1}{\{\kappa_{\pi}^o(L)\}^2}\mathrm{Var}\left\{
		\begin{array}{l}
			Y\{\pi(Z,L)-\rho_{\pi}^o(L)\}-\eta^o(L)\pi(Z,L)\\
			-\gamma^o(L)\cdot \left\{A(\pi(Z,L)-\rho_{\pi}^o(L))-\delta^o(L)\pi(Z,L)\right\}
		\end{array}\middle| L
		\right\}\\
		=&\dfrac{1}{\{\kappa_{\pi}^o(L)\}^2}\mathrm{Var}\left\{
		\begin{array}{l}
			\{Y-A\gamma^o(L)\}\{\pi(Z,L)-\rho_{\pi}^o(L)\}\\
			+\{\gamma^o(L)\delta^o(L)-\eta^o(L)\}\pi(Z,L)
		\end{array}\middle| L
		\right\}\\
		=&\dfrac{1}{\{\kappa_{\pi}^o(L)\}^2}\mathrm{Var}\left\{
		\begin{array}{l}
			\{Y-A\gamma^o(L)\}\{\pi(Z,L)-\rho_{\pi}^o(L)\}\\
			-\mathbb{E}[Y-A\gamma^o(L)\mid L]\pi(Z,L)
		\end{array}\middle| L
		\right\}\\
		=&\dfrac{1}{\{\kappa_{\pi}^o(L)\}^2}\mathrm{Var}\left\{
		\begin{array}{l}
			\left(\{Y-A\gamma^o(L)\}-\mathbb{E}[Y-A\gamma^o(L)\mid L]\right)\\
			\cdot\{\pi(Z,L)-\rho_{\pi}^o(L)\}
		\end{array}\middle| L
		\right\}.
	\end{align*}
	Notably, $\mathbb{E}[Y|Z,L]=\mathbb{E}[\gamma^o(L)|Z,L]$
	We denote 
	\begin{align*}
		W=&Y-A\gamma^o(L)-\mathbb{E}[\{Y-A\gamma^o(L)\}\mid L]\\
		=&Y-A\gamma^o(L)-f^o(0,L)=Y-f^o(A,L).
	\end{align*}
	From the assumption that $\mathbb{E}[W^2\mid Z,L]\indep Z\mid L$,
	we adopt the Cauchy Schwartz inequality to deduce that
	\begin{align*}
		&\mathrm{Var}\!\{\varphi_{\pi}(O;\psi_{*}^o,\alpha_{\pi}^o) \mid L\}
		=\dfrac{\mathbb{E}[W^2\{\pi(Z,L)-\rho_{\pi}^o(L)\}^2\mid L]}{\{\kappa_{\pi}^o(L)\}^2}\\
		=&\dfrac{\mathbb{E}[W^2\{\pi(Z,L)-\rho_{\pi}^o(L)\}^2\mid L]}
		{\{\mathbb{E}[\{A-\delta^o(L)\}\{\pi(Z,L)-\rho_{\pi}^o(L)\}\mid L]\}^2}\\
		=&\dfrac{\mathbb{E}[\mathbb{E}[W^2\mid Z,L]\{\pi(Z,L)-\rho_{\pi}^o(L)\}^2\mid L]}
		{\{\mathbb{E}[\{\pi_*^o(Z,L)-\rho_{\pi_*^o}^o(L)\}\{\pi(Z,L)-\rho_{\pi}^o(L)\}\mid L]\}^2}\\
		=&\dfrac{\mathbb{E}[W^2\mid L]\mathbb{E}[\{\pi(Z,L)-\rho_{\pi}^o(L)\}^2\mid L]}
		{\{\mathbb{E}[\{\pi_*^o(Z,L)-\rho_{\pi_*^o}^o(L)\}\{\pi(Z,L)-\rho_{\pi}^o(L)\}\mid L]\}^2}\\
		\geq&\mathbb{E}[W^2\mid L]\left\{\mathbb{E}\left[\{\pi_*^o(Z,L)-\rho_{\pi_*^o}^o(L)\}^2\middle| L\right]\right\}^{-1}.
	\end{align*}
	The inequality holds only when there exists $f(L)$, such that
	\begin{align*}
		\pi(Z,L)-\rho_{\pi}^o(L)=f(L)\{\pi_*^o(Z,L)-\rho_{\pi_*^o}^o(L)\}.
	\end{align*}
	Apparently, when $\pi(Z,L)=\pi_*^o(Z,L)$, this lower bound is achieved.
	Now we verify the fact that
	\begin{align*}
		&\mathrm{Var}\!\{\varphi_{\pi}(O;\psi_{*}^o,\alpha_{\pi}^o)\}\\
		=&\mathrm{Var}\!\{\mathbb{E}[\varphi_{\pi}(O;\psi_{*}^o,\alpha_{\pi}^o)\mid L]\}
		+\mathbb{E}\left[\mathrm{Var}\!\{\varphi_{\pi}(O;\psi_{*}^o,\alpha_{\pi}^o)\mid L\}\right]\\
		\geq&\mathrm{Var}\left\{\gamma^o(L)\right\}+\mathbb{E}\left[
		\mathbb{E}[W^2\mid L]\left\{\mathbb{E}\left[\{\pi_*^o(Z,L)-\rho_{\pi_*^o}^o(L)\}^2\middle| L\right]\right\}^{-1}\right]\\
		=&\mathrm{Var}\!\{\varphi_{\pi_*^o}(O;\psi_{*}^o,\alpha_{\pi_*^o}^o)\},
	\end{align*}
	finishing the proof for Proposition~\ref{prop: lower efficiency bound}.

	\subsection{Proof for Proposition~\ref{prop: mixed bias property}}
	
	We finish the proof by calculating
	\begin{align*}
		&\mathbb{E}\left[\varphi(O;\psi_{*}^o,\beta)\right]\\
		=&\mathbb{E}\left[\begin{array}{l}
			\dfrac{\pi(Z,L)\xi^o(Z,L)-\delta(L)^o(L)}{\kappa(L)}-\psi_{*}^o\\
			+\gamma(L)\left\{
			1 - \dfrac{(A-\delta(L))^2-(A-\pi(Z,L))^2}{\kappa(L)}
			\right\}\\
			+\dfrac{1}{\kappa(L)}
			\left\{\xi(Z,L)(\pi_*^o(Z,L)-\pi(Z,L))-\eta(L)(\delta^o(L)-\delta(L))\right\}
		\end{array}
		\right]\\
		=&\mathbb{E}\left[\begin{array}{l}
			\dfrac{\pi_*^o(Z,L)\xi^o(Z,L)-\delta^o(L)^o(L)}{\kappa(L)}-\psi_{*}^o\\
			+\dfrac{\gamma(L)}{\kappa(L)}\left\{
			\begin{array}{l}
				\kappa(L) - (A-\delta^o(L)+\delta^o(L)-\delta(L))^2\\
				+(A-\pi_*^o(Z,L)+\pi_*^o(Z,L)-\pi(Z,L))^2
			\end{array}
			\right\}\\
			+\dfrac{1}{\kappa(L)}
			\left\{\begin{array}{l}
				(\xi(Z,L)-\xi^o(Z,L))(\pi_*^o(Z,L)-\pi(Z,L))\\
				-(\eta(L)-\eta^o(L))(\delta^o(L)-\delta(L))
			\end{array}\right\}
		\end{array}
		\right]\\
		=&\mathbb{E}\left[\begin{array}{l}
			\dfrac{\kappa^o(L)}{\kappa(L)}\gamma^o(L)-\psi_{*}^o\\
			+\dfrac{\gamma(L)}{\kappa(L)}\left\{
			\begin{array}{l}
				\kappa(L) - \kappa^o(L) - (\delta^o(L)-\delta(L))^2\\
				+(\pi_*^o(Z,L)-\pi(Z,L))^2
			\end{array}
			\right\}\\
			+\dfrac{1}{\kappa(L)}
			\left\{\begin{array}{l}
				(\xi(Z,L)-\xi^o(Z,L))(\pi_*^o(Z,L)-\pi(Z,L))\\
				-(\eta(L)-\eta^o(L))(\delta^o(L)-\delta(L))
			\end{array}\right\}
		\end{array}
		\right]\\
		=&\mathbb{E}\left[\begin{array}{l}
			\dfrac{\kappa^o(L)}{\kappa(L)}\gamma^o(L)
			+\dfrac{\gamma(L)}{\kappa(L)}\{\kappa(L) - \kappa^o(L)\}
			-\psi_{*}^o\\
			+\dfrac{\gamma(L)}{\kappa(L)}\left\{
			\begin{array}{l}
				- (\delta^o(L)-\delta(L))^2\\
				+(\pi_*^o(Z,L)-\pi(Z,L))^2
			\end{array}
			\right\}\\
			+\dfrac{1}{\kappa(L)}
			\left\{\begin{array}{l}
				(\xi(Z,L)-\xi^o(Z,L))(\pi_*^o(Z,L)-\pi(Z,L))\\
				-(\eta(L)-\eta^o(L))(\delta^o(L)-\delta(L))
			\end{array}\right\}
		\end{array}
		\right]\\
		=&\mathbb{E}\left[\dfrac{1}{\kappa(L)}\left\{\begin{array}{l}
			\{\gamma(L)-\gamma^o(L)\}\{\kappa(L) - \kappa^o(L)\}\\
			+\gamma(L)\left\{
			\begin{array}{l}
				- (\delta^o(L)-\delta(L))^2\\
				+(\pi_*^o(Z,L)-\pi(Z,L))^2
			\end{array}
			\right\}\\
			-(\xi(Z,L)-\xi^o(Z,L))(\pi(Z,L)-\pi_*^o(Z,L))\\
			+(\eta(L)-\eta^o(L))(\delta(L)-\delta^o(L))
		\end{array}\right\}
		\right].
	\end{align*}

	\subsection{Proof for Proposition~\ref{prop: mixed bias property longitudinal}}
	We first calculate that
	\begin{align*}
		&\varphi_{\overline{a}}(O;\psi_{\overline{a}}^o,\overline{\alpha}_{t,\overline{a}},\underline{\alpha}_{t+1,\overline{a}}^o)
		-\varphi_{\overline{a}}(O;\psi_{\overline{a}}^o,\overline{\alpha}_{t-1,\overline{a}},\underline{\alpha}_{t,\overline{a}}^o)\\
		=&\displaystyle\prod_{s=0}^{t-1}\dfrac{\left\{Z_s-\rho_s(H_s)\right\}A_s^{(a_s)}}
		{\kappa_{s,a_s}(H_s)}
		\times\left(
		\dfrac{\left\{Z_t-\rho_t(H_t)\right\}A_t^{(a_t)}}
		{\kappa_{t,a_t}(H_t)}
		-\dfrac{\left\{Z_t-\rho_t^o(H_t)\right\}A_t^{(a_t)}}
		{\kappa_{t,a_t}^o(H_t)}
		\right)\\
		&\times\left\{
		\begin{array}{l}
			\displaystyle\prod_{s=t+1}^{T}\dfrac{\left\{Z_s-\rho_t^o(H_t)\right\}A_s^{(a_s)}}
			{\kappa_{s,a_s}^o(H_s)}Y-\gamma_{t+1,\underline{a}_{t+1}}^o(H_{t+1})\\
			+\displaystyle\sum_{s=t+1}^T\left(\displaystyle\prod_{r=t+1}^{s-1}
			\dfrac{\left\{Z_r-\rho_r^o(H_r)\right\}A_r^{(a_r)}}
			{\kappa_{r,a_r}^o(H_r) - \delta_{r,a_r}^o(H_r)\rho_r^o(H_r)}\right)\\
			\times\left\{
			\left(1-\dfrac{\{A_s^{(a_s)}-\delta_{s,a_s}^o(H_s)\}\{Z_s-\rho_s^o(H_s)\}}{\kappa_{s,a_s}^o(H_s)} \right)\gamma_{s,\underline{a}_s}^o(H_s) - 
			\dfrac{(Z_s-\rho_s^o(H_s))\times\eta_{s,\underline{a}_s}^o(H_s)}{\kappa_{s,a_s}^o(H_s)}\right\}
		\end{array}
		\right\}\\
		&+\displaystyle\prod_{s=0}^{t-1}\dfrac{\left\{Z_s-\rho_s(H_s)\right\}A_s^{(a_s)}}
		{\kappa_{s,a_s}(H_s)}\\
		&\times\left\{\begin{array}{l}
			\left(
			\dfrac{\left\{Z_t-\rho_t(H_t)\right\}A_t^{(a_t)}}
			{\kappa_{t,a_t}(H_t)}
			-\dfrac{\left\{Z_t-\rho_t^o(H_t)\right\}A_t^{(a_t)}}
			{\kappa_{t,a_t}^o(H_t)}
			\right)\gamma_{t+1,\underline{a}_{t+1}}^o(H_{t+1})\\
			+\left(1-\dfrac{\{A_t^{(a_t)}-\delta_{t,a_t}(H_t)\}\{Z_t-\rho_t(H_t)\}}{\kappa_{t,a_t}(H_t)} \right)\gamma_{t,\underline{a}_t}(H_t) - 
			\dfrac{(Z_t-\rho_t(H_t))\times\eta_{t,\underline{a}_t}(H_t)}{\kappa_{t,a_t}(H_t)}\\
			-\left(1-\dfrac{\{A_t^{(a_t)}-\delta_{t,a_t}^o(H_t)\}\{Z_t-\rho_t^o(H_t)\}}{\kappa_{t,a_t}^o(H_t)} \right)\gamma_{t,\underline{a}_t}^o(H_t) - 
			\dfrac{(Z_t-\rho_t^o(H_t))\times\eta_{t,\underline{a}_t}^o(H_t)}{\kappa_{t,a_t}^o(H_t)}
		\end{array}\right\}\\
		=&\displaystyle\prod_{s=0}^{t-1}\dfrac{\left\{Z_s-\rho_s(H_s)\right\}A_s^{(a_s)}}
		{\kappa_{s,a_s}(H_s)}
		\times\left(
		\dfrac{\left\{Z_t-\rho_t(H_t)\right\}A_t^{(a_t)}}
		{\kappa_{t,a_t}(H_t)}
		-\dfrac{\left\{Z_t-\rho_t^o(H_t)\right\}A_t^{(a_t)}}
		{\kappa_{t,a_t}^o(H_t)}
		\right)Q_{11}\\
		&+\displaystyle\prod_{s=0}^{t-1}\dfrac{\left\{Z_s-\rho_s(H_s)\right\}A_s^{(a_s)}}
		{\kappa_{s,a_s}(H_s)}
		\left\{
		Q_{12}+Q_{13}-Q_{14}
		\right\}.
	\end{align*}
	Since the nuisance functions in $Q_{11}$ all equal to the corresponding true form, it is straight forward to verify that $\mathbb{E}[Q_{11}\mid H_{t+1}]\equiv 0$. Similarly, one can verify that $\mathbb{E}[Q_{14}\mid H_t]\equiv 0$. From
	\begin{align*}
		&\mathbb{E}\left[\dfrac{\left\{Z_t-\rho_t(H_t)\right\}A_t^{(a_t)}\gamma_{t+1,\underline{a}_{t+1}}^o(H_{t+1})}
		{\kappa_{t,a_t}(H_t)}\middle| H_t\right]\\
		=&\dfrac{\mathbb{E}\left[\left\{
			Z_t-\rho_t(H_t)\right\}A_t^{(a_t)}\gamma_{t+1,\underline{a}_{t+1}}^o(H_{t+1})\middle| H_t\right]}
		{\kappa_{t,a_t}(H_t)}\\
		=&\dfrac{\zeta_{t,\underline{a}_t}^o(H_t)-\rho_t(H_t)\eta_{t,\underline{a}_t}^o(H_t)}
		{\kappa_{t,a_t}(H_t)}
		=\dfrac{\gamma_{t,\underline{a}_t}^o(H_t)\kappa_{t,a_t}^o(H_t)+\{\rho_t^o(H_t)-\rho_t(H_t)\}
			\eta_{t,\underline{a}_t}^o(H_t)}
		{\kappa_{t,a_t}(H_t)},
	\end{align*}
	we know that
	\begin{align*}
		&\mathbb{E}[Q_{12}\mid H_t]=\gamma_{t,\underline{a}_t}^o(H_t)\left(\dfrac{\kappa_{t,a_t}^o(H_t)}{\kappa_{t,a_t}(H_t)}-1\right)
		+\dfrac{\{\rho_t^o(H_t)-\rho_t(H_t)\}
			\eta_{t,\underline{a}_t}^o(H_t)}
		{\kappa_{t,a_t}(H_t)};\\
		&\mathbb{E}[Q_{13}\mid H_t]\\=&
		\mathbb{E}\left[\left(1-\dfrac{\{A_t^{(a_t)}-\delta_{t,a_t}(H_t)\}\{Z_t-\rho_t(H_t)\}}{\kappa_{t,a_t}(H_t)} \right)\gamma_{t,\underline{a}_t}(H_t) - 
		\dfrac{(Z_t-\rho_t(H_t))\eta_{t,\underline{a}_t}(H_t)}{\kappa_{t,a_t}(H_t)}\middle| H_t\right]\\=&
		\left(1-\dfrac{
			\mathbb{E}\left[\{A_t^{(a_t)}-\delta_{t,a_t}(H_t)\}\{Z_t-\rho_t(H_t)\}\middle| H_t\right]
		}{\kappa_{t,a_t}(H_t)} \right)\gamma_{t,\underline{a}_t}(H_t) - 
		\dfrac{(\rho_t^o(H_t)-\rho_t(H_t))\eta_{t,\underline{a}_t}(H_t)}{\kappa_{t,a_t}(H_t)}\\=&
		\left(1-\dfrac{
			\kappa_{t,a_t}^o(H_t)
		}{\kappa_{t,a_t}(H_t)} \right)\gamma_{t,\underline{a}_t}(H_t)
		+\dfrac{1}{\kappa_{t,a_t}(H_t)}(\delta_{t,a_t}(H_t)-\delta_{t,a_t}^o(H_t))(\rho_t(H_t) - \rho_t^o(H_t)) \gamma_{t,\underline{a}_t}(H_t)\\&- 
		\dfrac{(\rho_t^o(H_t)-\rho_t(H_t))\times\eta_{t,\underline{a}_t}(H_t)}{\kappa_{t,a_t}(H_t)};\\
		&\mathbb{E}[Q_{12}+Q_{13}\mid H_t]\\=&
		\dfrac{1}{\kappa_{t,a_t}(H_t)}\left\{\kappa_{t,a_t}(H_t)- \kappa_{t,a_t}^o(H_t)\right\} \left\{\gamma_{t,\underline{a}_t}(H_t)-\gamma_{t,\underline{a}_t}^o(H_t)\right\}\\
		&+\dfrac{1}{\kappa_{t,a_t}(H_t)}(\rho_t(H_t)-\rho_t^o(H_t))\left\{\eta_{t,\underline{a}_t}(H_t)-\eta_{t,\underline{a}_t}^o(H_t)\right\}\\
		&+\dfrac{1}{\kappa_{t,a_t}(H_t)}(\delta_{t,a_t}(H_t)-\delta_{t,a_t}^o(H_t))(\rho_t(H_t) - \rho_t^o(H_t)) \gamma_{t,\underline{a}_t}(H_t).
	\end{align*}
	In summary, we can deduce that
	\begin{align*}
		&\mathbb{E}\left[\varphi_{\overline{a}}(O;\psi_{\overline{a}}^o,\overline{\alpha}_{t,\overline{a}},\underline{\alpha}_{t+1,\overline{a}}^o)
		-\varphi_{\overline{a}}(O;\psi_{\overline{a}}^o,\overline{\alpha}_{t-1,\overline{a}},\underline{\alpha}_{t,\overline{a}}^o)\middle| H_t\right]
		=\displaystyle\prod_{s=0}^{t-1}\dfrac{\left\{Z_s-\rho_s(H_s)\right\}A_s^{(a_s)}}
		{\kappa_{s,a_s}(H_s)}\\
		&\times \dfrac{1}{\kappa_{t,a_t}(H_t)}\left\{\begin{array}{l}
			\left\{\kappa_{t,a_t}(H_t)- \kappa_{t,a_t}^o(H_t)\right\} \left\{\gamma_{t,\underline{a}_t}(H_t)-\gamma_{t,\underline{a}_t}^o(H_t)\right\}\\
			+(\rho_t(H_t)-\rho_t^o(H_t))\left\{\eta_{t,\underline{a}_t}(H_t)-\eta_{t,\underline{a}_t}^o(H_t)\right\}\\
			+(\delta_{t,a_t}(H_t)-\delta_{t,a_t}^o(H_t))(\rho_t(H_t) - \rho_t^o(H_t)) \gamma_{t,\underline{a}_t}(H_t).
		\end{array}\right\}.
	\end{align*}
	Finally, we deduce that
	\begin{align*}
		&\mathbb{E}\left[\varphi_{\overline{a}}(O;\psi_{\overline{a}}^o,\alpha_{\overline{a}})\right]\\
		=&\sum_{t=0}^T\mathbb{E}\left[\varphi_{\overline{a}}(O;\psi_{\overline{a}}^o,\overline{\alpha}_{t,\overline{a}},\underline{\alpha}_{t+1,\overline{a}}^o)
		-\varphi_{\overline{a}}(O;\psi_{\overline{a}}^o,\overline{\alpha}_{t-1,\overline{a}},\underline{\alpha}_{t,\overline{a}}^o)\right]\\
		=&\mathbb{E}\left[\begin{array}{l}
			\displaystyle\sum_{t=0}^T\left(\displaystyle\prod_{s=0}^{t-1}\dfrac{\left\{Z_s-\rho_s(H_s)\right\}A_s^{(a_s)}}
			{\kappa_{s,a_s}(H_s)}\right)
			\times \dfrac{1}{\kappa_{t,a_t}(H_t)}\\
			\times\left\{\begin{array}{l}
				\left\{\kappa_{t,a_t}(H_t)- \kappa_{t,a_t}^o(H_t)\right\} \left\{\gamma_{t,\underline{a}_t}(H_t)-\gamma_{t,\underline{a}_t}^o(H_t)\right\}\\
				+(\rho_t(H_t)-\rho_t^o(H_t))\left\{\eta_{t,\underline{a}_t}(H_t)-\eta_{t,\underline{a}_t}^o(H_t)\right\}\\
				+(\delta_{t,a_t}(H_t)-\delta_{t,a_t}^o(H_t))(\rho_t(H_t) - \rho_t^o(H_t)) \gamma_{t,\underline{a}_t}(H_t)
			\end{array}\right\}
		\end{array}\right].
	\end{align*}
	This finishes the proof for Proposition~\ref{prop: mixed bias property longitudinal}.

	% \bibliographystyle{plainnat}
	% \bibliography{bibfile_supp}
    \putbib[bibfile_supp]
	\end{bibunit}

\end{document}